%% file: tcs-infset.tex
\documentclass[3p]{elsarticle}
\include{macros}
 \title{Determinization of $\omega$-automata unified}
 \author[hk]{Hrishikesh Karmarkar}
 \ead{hrishik@cse.iitb.ac.in}
 \author[sc]{Supratik Chakraborty}
 \ead{supratik@cse.iitb.ac.in}
 \address[hk]{Department of Computer Science and Engineering, Indian Institute of Technology Bombay}
 \address[sc]{Department of Computer Science and Engineering, Indian Institute of Technology Bombay}

\begin{document}
 \thispagestyle{empty}
 \begin{abstract}
 We present a uniform construction for converting $\omega$-automata
 with arbitrary acceptance conditions to equivalent deterministic
 parity automata (DPW).  Given a non-deterministic automaton with $n$
 states, our construction gives a DPW with at most $2^{O(n^2 \log n)}$
 states and $O(n^2)$ parity indices.  The corresponding bounds when the
 original automaton is deterministic are $O(n!)$ and $O(n)$,
 respectively.  Our algorithm gives better asymptotic bounds on the
 number of states and parity indices vis-a-vis the best known technique
 when determinizing Rabin or Streett automata with $\Omega{(2^n)}$
 acceptance pairs, where $n > 1$.  We demonstrate this by describing a
 family of Streett (and Rabin) automata with $2^{n}$ non-redundant
 acceptance pairs, for which the best known determinization technique
 gives a DPW with at least $\Omega{(2^{(n^3)})}$ states, while our
 construction constructs a DRW/DPW with $2^{O(n^2\log n)}$ states.
 An easy corollary of our construction is that an
   $\omega$-language with Rabin index $k$ cannot be recognized by any
   $\omega$-automaton (deterministic or non-deterministic) with fewer
   than $O(\sqrt{k})$ states.
 \end{abstract}

\maketitle
\section{Introduction}

The literature contains several interesting constructions for obtaining
deterministic Rabin/parity automata from nondeterministic $\omega$-automata with
different accepting conditions \cite{buchi73, eilenberg74, choueka74, rabin72,
tb73, sch75, thomas81, emerson-sistla, safra, safra06, muller-schupp,
schewe-det, piterman, kaewil, colcombet}. However, all known constructions are
tailor-made to work for nondeterministic automata with a specific kind of
accepting condition. For example, Safra's celebrated B\"{u}chi determinization
construction~\cite{safra, safrathesis} can be used to convert non-deterministic
B\"{u}chi automata over words (NBW) to deterministic Rabin automata over words
(DRW).  Piterman showed that Safra's construction can be augmented with
additional machinery to obtain deterministic parity automata (DPW) over words
from NBW~\cite{piterman, liu-wang}.  It requires the use of a completely
different technique (once again, originally due to
Safra~\cite{safra-stoc92,safra06} and subsequently improved by
Piterman~\cite{piterman}) to convert non-deterministic Streett automata over
words (NSW) to equivalent DRW or DPW.  We are unaware of any construction for
directly converting non-deterministic M\"{u}ller automata over words (NMW) to
DRW or DPW.  A two-step approach would involve first converting an NMW to NBW,
and then using Safra's/Piterman's determinization construction for NBW to obtain
a DRW/DPW.  In this backdrop, we propose a uniform determinization construction
for all $\omega$-automata for which the acceptance condition is based on
\emph{infinity sets}, i.e., the set of states visited infinitely often in a run
of the automaton.  It is worth noting that the acceptance conditions for all
important classes of $\omega$-automata studied in the literature are based on
infinity sets.

We begin by quickly reviewing different acceptance conditions of
$\omega$-automata used in the literature.  Let $\aA = (\Sigma, Q, Q_0,
\delta, \phi)$ be a (possibly non-deterministic) $\omega$-automaton,
where $\Sigma$ is the alphabet, $Q$ is the set of states,
$Q_0\subseteq Q$ is the set of initial states, $\delta: Q \times
\Sigma \rightarrow 2^Q$ is the transition relation, and $\phi$ is the
acceptance condition.
An acceptance condition $\phi$ based on infinity sets specifies properties of
the set of states visited infinitely often in an accepting run of the automaton.
Hence, $\phi$ can be thought of as defining a predicate $P_\phi$ over $2^Q$.
Formally, for every $X \subseteq Q$, we say $P_\phi(X) = \true$ iff $X$, viewed
as the infinity set of a run of $\aA$, satisfies the properties specified by
$\phi$. This is a re-statement of the fact that any $\omega$-automaton with
acceptance condition based on infinity sets can be converted to a Muller
automaton by preserving the transition structure of the automaton and by listing
all subsets of states that satisfy $\phi$ in the Muller acceptance set.  We list
below acceptance conditions of some important classes of $\omega$-automata and
indicate the interpretation of $P_\phi$ in each case.  In all cases, we assume
that $X$ is a subset of $Q$.

\begin{itemize} 
\item {\buchi} condition : $\phi$ is given by $F \subseteq Q$, and
  $P_\phi(X) = \true$ iff $X \cap F \neq \emptyset$.

\item Muller condition : $\phi$ is given by a collection ${\cal
  F} = \{F_1, F_2, \ldots, F_k\}$, where $F_i \subseteq Q$ for all $i
  \in \{1, \ldots, k\}$, and $P_\phi(X) = \true$ iff $X \in
      {\cal F}$.

\item Rabin condition : $\phi$ is given by a table of pairs
  $\mathcal{T} = \{(E_1,F_1), (E_2, F_2), \ldots (E_h, F_h)\}$, where
  $E_i, F_i \subseteq Q$ for all $i \in \{1, \ldots, h\}$, and
  $P_\phi(X) = \true$ iff there exists an $i \in \{1,2, \ldots, h\}$
  such that $X \cap E_i = \emptyset$ and $X \cap F_i \neq \emptyset$.

\item Streett condition: $\phi$ is given by a table of pairs, similar
  to that used for Rabin condition.  However, in this case $P_\phi(X) =
  \true$ iff for all $i \in \{1 \ldots h\}$, $X \cap E_i \neq
  \emptyset$ whenever $X \cap F_i \neq \emptyset$.

\item Parity condition: $\phi$ is given by a sequence of sets
  $\mathcal{F} = \langle F_0, F_1, \ldots F_h \rangle$, where $F_i
  \subseteq Q$ for all $i \in \{0, \ldots h\}$.  Here, $P_\phi(X) =
  \true$ iff for some even number $j \in \{0, \ldots h\}$, $X \cap
  F_j \neq \emptyset$ and for all $m \in \{0, \ldots j-1\}$, $X \cap
  F_m = \emptyset$.

\item Emerson-Lei condition~\cite{emersonlai}: $\phi$ is given by a
  fairness condition, expressed as a Boolean combination $f$ of
  special linear-time temporal logic formulae over atomic propositions
  labeling states of the $\omega$-automaton.  The sub-formulae of $f$
  are such that their truth can be determined simply by knowing the
  set of sets visited infinitely often along a path (or run) of the
  automaton, and from the labels of these states.  Therefore,
  $P_\phi(X) = \true$ iff every run of the automaton with infinity set
  $X$ satisfies the temporal logic formula $f$.
\end{itemize}
It follows from the above discussion that to determine if an
$\omega$-word $\alpha$ is accepted by $\aA$, it suffices to determine
the set of infinity sets for all runs of $\aA$ on $\alpha$, and to
check if $P_\phi$ evaluates to $\true$ for any of these infinity sets.
This observation forms the basis of our construction for determinizing
$\omega$-automata with arbitrary acceptance conditions based on
infinity sets.

The primary contribution of this paper is a uniform construction for
converting $\omega$-automata with arbitrary acceptance conditions
based on infinity sets to deterministic parity automata.  Given a
non-deterministic automaton with $n$ states, our construction gives a
DPW with at most $2^{O(n^2 \log n)}$ states and $O(n^2)$ parity
indices.  The corresponding bounds when the original automaton is
deterministic are $O(n!)$ and $O(n)$, respectively.  Our algorithm
gives better asymptotic bounds on the number of states and parity
indices vis-a-vis the best known technique when determinizing Rabin
or Streett automata with $\Omega(n^k)$ acceptance pairs, where $k >
1$.  We demonstrate this by describing a family of Streett (and Rabin)
automata with $2^{O(n)}$ non-redundant acceptance pairs, for which the
best known determinization technique gives a DPW with at least
$2^{O(n^3)}$ states and $2^{O(n)}$ parity indices.  An easy corollary
  of our construction is that an $\omega$-language with Rabin index
  $k$ cannot be recognized by any $\omega$-automaton (deterministic or
  non-deterministic) with fewer than $O(\sqrt{k})$ states.

The remainder of this paper is organized as follows.
We begin by revisiting Schwoon's version of Safra's NSW determinization construction and
Piterman's optimization of it. We then describe our uniform construction for
determinization of $\omega$-automata along with intuition behind the
construction and an example that demonstrates steps of the construction. We then
prove the correctness of our construction and compute its complexity. Finally,
we demonstrate the existence of a family of NSW for which our construction
provides better upper bounds for determinization than any of the existing
methods.

\section{Determinizing NSW: A Recap of Safra's and Piterman's Constructions}
\label{safransw}
Since our construction is obtained by adapting Safra's determinization
construction for NSW~\cite{safra-stoc92,safra06} and borrows some key
optimization ideas from Piterman's construction~\cite{piterman}, we
provide an overview of Safra's and Piterman's constructions below.
Additional details of Safra's construction can be found
in~\cite{safra-stoc92,safra06,2001automata}, and those of Piterman's
construction can be found in~\cite{piterman}.

Safra's determinization construction for NSW is based on the idea of
\emph{witness sets} and hierarchically related \emph{decompositions}.
Since we will use a different notion of witness sets later in the
paper, we will henceforth call witness sets as defined by Safra as
\emph{Streett Safra witness sets}.  For a Streett automaton $\aS =
(\Sigma, Q, Q_0, \delta, \phi)$, the acceptance condition $\phi$ is
given by a Streett pairs table $\mathcal{T} = \{(E_1, F_1), \ldots,
(E_h, F_h)\}$.  Let $H = \{1, 2, \ldots h\}$ be the set of indices of
Streett pairs in $\mathcal{T}$.  A subset $J$ of $H$ is called a
\emph{Streett Safra witness set} for a run $\rho$ of $\aS$ iff for
every $j \in J$, some state in $E_j$ is visited infinitely often in
$\rho$, and for every $j \not\in J$, no state in $F_j$ is visited
infinitely often in $\rho$.  It is easy to see that every accepting
run of $\aS$ has at least one Streett Safra witness set, and any run
of $\aS$ with a Streett Safra witness set is an accepting run.  Note,
however, that an accepting run of $\aS$ can have multiple Streett Safra
witness sets.  The \emph{decompositions} used in Safra's construction
can be viewed as hierarchically related processes, each of which
tracks a subset of runs of $\aS$ on a given word, and checks if a
certain subset of $H$ is a Streett Safra witness set for all the tracked runs.
While Safra's original exposition~\cite{safra-stoc92,safra06}
represents the hierarchy between decompositions using the notion of
sub-decompositions, Schwoon's exposition of Safra's
construction~\cite{2001automata} explicitly represents the
hierarchical relation between decompositions as a tree.  Each node in
this tree represents a decomposition as defined by Safra, and children
of a node represent sub-decompositions in Safra's terminology.  We
will use the tree representation of decompositions, called $(Q,
H)$-trees by Schwoon~\cite{2001automata}, in the following discussion
for clarity of exposition.

Following the definition given by Schwoon~\cite{2001automata}, a $(Q,
H)$-tree over $\aS$ is a finitely branching rooted tree with the
following properties.
\begin{itemize}
\item Every leaf node is labeled with a non-empty subset of $Q$
  (states of the Streett automaton $\aS$).
\item State labels of leaf nodes are pairwise disjoint.
\item Every node is assigned a name from the set $V=\{1, 2, \ldots
  2\cdot |Q| \cdot (|H|+1)\}$.
\item No two nodes have the same name.
\item Every edge is annotated with an element of $H \cup \{0\}$.
\item No edge annotation other than $0$ occurs more than once on any path from
  the root to a leaf.
\item Every non-leaf node has at least one child connected by an edge
  with a non-zero annotation.
\item The children of every node are ordered from left to right.
\end{itemize}

Every node $v$ in a $(Q, H)$-tree can be thought of as being
associated with a Streett Safra witness set, $W(v)$, defined as
follows.  If $v$ is the root node, then $W(v) = \{1, 2 \ldots h\} =
H$.  Otherwise, if $v'$ is the parent of $v$ and if the edge from $v'$
to $v$ is annotated with $j$, then $W(v) = W(v') \setminus \{j\}$.
Let $\lambda(v)$ denote the set of Streett states labeling the leaves
of the sub-tree rooted at $v$.  Thus, if $v$ is a leaf node,
$\lambda(v)$ is the state label of $v$.  However, if $v$ has children $v_1,
v_2, \ldots v_l$, then $v$ itself does not have a state label but
$\lambda(v)$ is the disjoint union of $\lambda(v_1), \lambda(v_2),
\ldots \lambda(v_l)$.  A node $v$ in a $(Q, H)$-tree represents a
process that tracks the runs represented by states in $\lambda(v)$,
and checks if $W(v)$ is a Streett Safra witness set for all these
runs.  This is done by waiting until all $E_j$ for $j \in W(v)$ are
visited in order along the runs, without visiting any $F_{l}$ for $l
\notin W(v)$.  If this happens, the process represented by $v$ is said
to have ``succeeded''; it is then ``reset'' and the check starts all
over again.  Clearly, if the process represented by $v$ is reset
infinitely often, then $W(v)$ is a Streett Safra witness set for the
runs tracked by this process have, and hence these are accepting runs
of $\aS$.  On the other hand, if some state in $F_{l}$ for $l \notin
W(v)$ is seen in a run being tracked by the process represented by
$v$, then that run is removed from this process, and a new process is
started for that run.  The hierarchical relation between processes is
explicitly represented by the parent-child relation between nodes in a
$(Q, H)$-tree.  Intuitively, if $v'$ is the parent of $v$ and if the
edge from $v'$ to $v$ is annotated with $j$, the process represented
by $v$ tracks a subset of the runs tracked by $v'$ after giving up
hope that it will see a state from $E_{j}$ ever in the future.  While
the parent $v'$ keeps alive the hope that $W(v')$ is the Streett Safra
witness set for all runs tracked by $v'$, the child $v$ refines and
corrects that hope by expecting $W(v) = W(v') \setminus \{j\}$ to be
the Safra Streett witness set for the subset of runs tracked by $v$.

The DRW obtained by applying Safra's construction to a Streett
automaton $\aS = (\Sigma, Q, Q_0, \delta, \phi)$ is given by $\ar =
(\Sigma, Q^r, Q_0^r, \delta^r, \phi^r)$, where $Q^r$ is the set of all
$(Q, H)$-trees over $\aS$, and $Q_0^r$ is a singleton set containing
the $(Q, H)$-tree consisting of only a root node with name $1$ and
labeled with $Q_0$ (set of initial states of $\aS$).  Since $\ar$ is a
deterministic automaton, $\delta^r$ can be thought of as a function
that takes a state (i.e., $(Q, H)$-tree) $t$ and a letter $\sigma \in
\Sigma$ and returns the next state (i.e., $(Q, H)$-tree) $t'$.  The
computation of $t'$ from $t$ and $\sigma$ is detailed in algorithm
\algo{SafraNext} given below (adapted from Schwoon's
exposition~\cite{2001automata} and Piterman's
correction~\cite{piterman} of an erroneous step
in~\cite{safra06,2001automata}).  Note that algorithm \algo{SafraNext}
calls a recursive procedure \algo{SafraNextRecursive} that is
parameterized by the root node of a $(Q, H)$-sub-tree and the
corresponding Streett Safra witness set.  If $|Q| = n$ and $|H| = h$,
the Rabin acceptance condition $\phi^r$ is given by a table
$\mathcal{T}^r = \{(E^{r}_i, F^{r}_i) \mid 1 \le i \le 2\cdot n\cdot
(h +1)\}$, where $E^{r}_i$ is the set of all $(Q, H)$-trees with no
node named $i$, and $F^{r}_i$ is the set of all $(Q, H)$-trees in
which a leaf node named $i$ occurs.

\algoblock{SafraNext}
{$t:$ $(Q, H)$-tree over $\aS$,~~ $\sigma:$ letter in $\Sigma$}
{$t':$ $(Q, H)$-tree over $\aS$} 
{
  \begin{enumerate}
  \item {\bfseries [Initialization]} For every leaf node $u$ of $t$,
    set the state label of $u$ to $\delta(\lambda(u), \sigma)$.
    
  \item {\bfseries [Recursive transformation]} Let $\mathit{root}$ be
    the root node of $t$.  \\Invoke 
    \algo{SafraNextRecursive}$(\mathit{root}, H)$.

  \item Return $t'$ as the $(Q, H)$-tree rooted at $\mathit{root}$.
\end{enumerate}
}

\algoblock{SafraNextRecursive}
{$v:$ root of a $(Q, H)$-sub-tree,~~ $J:$ subset of $H$}
{$t':$ Transformed $(Q, H)$-sub-tree rooted at $v$}
{
  \begin{enumerate}
  \item \label{safra1} If $v$ is a leaf and $J = \emptyset$,
    return $t'$ as the $(Q, H)$-sub-tree rooted at $v$.
    
  \item \label{safra2} If $v$ is a leaf and $J \neq \emptyset$, create a
    new child $v'$ of $v$ with state label $\lambda(v)$, remove $\lambda(v)$ from
    the state label of $v$ (since $v$ is no longer a leaf) and annotate the
    edge from $v$ to $v'$ with $\max W(v)$.  Assign an unused name 
    from $V=\{1, 2, \ldots 2\cdot |Q| \cdot (|H|+1)\}$ to $v'$.
    
  \item \label{safra3} If, after the execution of Steps (\ref{safra1})
    and (\ref{safra2}), $v$ is not a leaf, then let $v_1, \ldots, v_l$
    be the children of $v$ ordered from left to right.  Let the edge
    from $v$ to $v_i$ be annotated with $j_i$ for all $i \in \{1, 2,
    \ldots l\}$.
    
    \begin{enumerate}
      
    \item \label{safra3a} For all $i$ from $1$ to $l$, invoke
      \algo{SafraNextRecursive}$(v_i, J \setminus \{j_i\})$
      
    \item \label{safra3b} For every child $v_i$ of $v$ and every $q \in
      \lambda(v_i)$, do the following
      \begin{enumerate}
        
      \item \label{safra3b1} If $q \in F_{j_i}$, remove $q$ from the
        state labels of all leaves of the sub-tree rooted at $v_i$, create a
        new rightmost child $v'$ of $v$ with state label $\{q\}$, and
        annotate the edge from $v$ to $v'$ with $j_i$.  Assign an
        unused name from $V=\{1, 2, \ldots 2\cdot |Q| \cdot (|H|+1)\}$
        to $v'$.
        
      \item \label{safra3b2} If $q \in E_{j_i}$, create a new
        rightmost child $v'$ of $v$ with state label $\{q\}$ and annotate
        the edge from $v$ to $v'$ with $\max\left((J \cup \{0\}) \cap
        \{0, 1, \ldots j_i-1\}\right)$.  In other words, the edge is
        annotated with the largest integer less than $j_i$ but in $J$,
        if it exists.  Otherwise, it is annotated with $0$.  Assign an
        unused name from $V=\{1, 2, \ldots 2\cdot |Q| \cdot (|H|+1)\}$
        to $v'$.
        
      \end{enumerate}
      
    \end{enumerate}
    
  \item \label{safra4} Let $v_1, v_2, \ldots, v_{l'}$ be the children
    of $v$ after the above steps.  Let $j_1, j_2, \ldots, j_{l'}$ be
    the annotations of the corresponding edges from $v$ to its children.
    For every $q \in \lambda(v_j)\cap \lambda(v_k)$, where $j \neq k$
    and $j, k \in \{1,2, \ldots,l'\}$, do the following.
    \begin{enumerate}
      
    \item If $j_i < j_k$, remove $q$ from the state labels of all leaves of
      the sub-tree rooted at $v_k$.
      
    \item If $j_i = j_k$ and $v_i$ is to the left of $v_k$, remove $q$ 
      from the state labels of all leaves of the sub-tree rooted at $v_k$.
      
    \end{enumerate} 
    
  \item \label{safra5} For every descendant $u$ of $v$ such that
    $\lambda(u) = \emptyset$, delete $u$ and all its descendants.
    
  \item \label{safra6} If, after the previous steps, all edges from
    $v$ to its children are annotated with $0$, then the process
    represented by $v$ has ``succeeded'' and needs to be ``reset''.
    Let $S = \lambda(v)$.  Make $v$ a leaf node by deleting all its
    children and their descendants, and set the state label of $v$ to $S$.
    
  \item \label{safra7} Return $t'$ as the $(Q, H)$-sub-tree rooted at $v$.
  \end{enumerate}
}

It was shown by Safra that given an NSW with $|Q| = n$ and $|H| = h$,
the above construction gives a deterministic Rabin automaton with
$2^{O(n\cdot h \cdot \log(n\cdot h)}$ states and $O(n\cdot h)$ Rabin
acceptance pairs.  Although a proof of correctness of the construction
was provided in~\cite{safra-stoc92,safra06,2001automata}, Piterman
pointed out a minor error in the construction and rectified it
in~\cite{piterman}.  Fortunately, Piterman's correction affects only a
single step of Safra's construction and does not change the asymptotic
count of states or Rabin acceptance pairs.  The fact that this
erroneous step evaded the scrutiny of researchers for almost $14$
years is testimony to the intricate nature of arguments used in
Safra's construction.  Piterman also proposed an adaptation of Safra's
construction that uses only $n \cdot (h+1)$ names (instead of $2\cdot
n \cdots (h+1)$ names used by Safra) and gives a deterministic parity
automaton with $2^{O(n\cdot h \cdot \log(n\cdot h))}$ states and
$2\cdot n\cdot h$ parity indices.  Currently, Piterman's construction
is the best known determinization construction for NSW.

Piterman's adaptation of Safra's construction involves two key ideas:
(i) a new strategy for naming nodes, and (ii) addition of two
integer-valued components, $e$ and $f$, to every state of the
constructed automaton that allows a parity acceptance condition to be
defined.  In the new naming strategy, whenever a new node is created
in steps (\ref{safra2}), (\ref{safra3b1}) or (\ref{safra3b2}) of
algorithm \algo{SafraNextRecursive}, it is assigned the smallest name
larger than all names used so far in the construction of $t'$ from
$t$.  In addition, after algorithm \algo{SafraNext} has finished
computing $t'$, a name-compaction step is performed.  In this step,
for each node $v$ with name $i$ in $t'$, we determine the count,
$\mathit{rem}(v)$, of nodes that were removed during the construction
of $t'$ from $t$ and had names less than $i$.  The name of $v$ is then
reduced from $i$ to $i - \mathit{rem}(v)$.  This ensures that there
are no gaps in the set of names assigned to nodes in a $(Q, H)$-tree
after the name-compaction step.  Piterman's naming strategy also
ensures that the name of a node $v$ is less than that of node $u$ iff
$v$ was created before $u$.  Since the name of a node that stays back
in a run (sequence of $(Q, H)$-trees) can only reduce finitely many
times, it follows that all nodes that eventually stay back in a run
get fixed names that are smaller than the names of all other nodes
that keep getting created and removed.

The new state components $e$ and $f$ in Piterman's construction keep
track of the smallest name of a node removed and the smallest name of
a node that represents a successful process (see step (\ref{safra6})
of algorithm \algo{SafraNextRecursive}) respectively in the
construction of $t'$ from $t$.  A state in the resulting automaton is
therefore a $(Q, H)$-tree coupled with a pair of integers $e, f \in
\{1, \ldots n\cdot (h+1) + 1\}$, with the restriction that the root
node is always named $1$ and all nodes are assigned names from $\{1,
\ldots n\cdot (h+1)\}$.  Piterman calls these states \emph{compact
  Streett Safra trees} over $\aS$, and obtains a deterministic parity
automaton by defining a parity acceptance condition as follows.  Let
$D$ denote the set of all compact Streett Safra trees over $\aS$.
Piterman's parity acceptance condition is given by $\mathcal{F} =
\langle F_0, F_1, \ldots F_{2m - 1}$, where $m = 2\cdot n \cdot (h+1)$
and $F_i$s are defined as follows.
\begin{itemize}
\item $F_0 = \{d \in D \mid f = 1 \mbox{ and } e > 1\}$ 
\item $F_{2i + 1} = \{d \in D \mid e = i + 2 \mbox{ and } f \ge e\}$,
  for all $i \in \{0, \ldots m-1\}$
\item $F_{2i + 2} = \{d \in D \mid f = i + 2 \mbox{ and } e > f\}$, for
  all $i \in \{0, \ldots m-2\}$
\end{itemize}
A proof of correctness of the above construction is given
in~\cite{piterman}.  It is also shown there that the DPW obtained
using this construction has at most $2\cdot n^n \cdot (k+1)^{n\cdot
  (k+1)}\cdot (n\cdot (k+1))!$ states and $2\cdot n \cdot (k+1)$
parity indices.

\section{A uniform determinization construction for $\omega$-automata}

We now describe a construction for converting $\omega$-automata with
arbitrary acceptance conditions based on infinity sets to
deterministic parity automata.  Our construction can be viewed as an
adaptation of Safra's NSW determinization construction that works for
arbitrary acceptance conditions.  As part of our construction, we use
Piterman's naming strategy and his idea of using $e, f$ components of
states to get a parity acceptance condition.  Interestingly, although
our construction is based on Safra's and Piterman's constructions, we
are able to sharpen the asymptotic upper bound for Streett and Rabin
determinization beyond those obtainable by Safra's and Piterman's
constructions.

Let $\aA = (\Sigma, Q, Q_0, \delta, \phi)$ be an $\omega$-automaton,
where $\phi$ is an arbitrary acceptance condition based on infinity
sets.  Let $P_\phi$ denote the predicate corresponding to $\phi$.
Without loss of generality, we will assume that $Q = \{q_1, q_2,
\ldots q_n\}$, where $n = |Q|$. For notational clarity, we will
henceforth refer to states of $\aA$ as $\aA$-states, and use $[p]$ to
denote the set $\{1, 2, \ldots p\}$ for every natural number $p > 0$.
For every $W \subseteq [n]$, we also define $Q_W$ to be the set $\{q_i
\mid q_i \in Q, i \in W\}$.

Motivated by the role played by Streett Safra witness sets in Safra's
NSW determinization construction, we now define \emph{generalized
  witness sets} for $\omega$-automata with arbitrary acceptance
conditions based on infinity sets.
\begin{definition}[Generalized Witness Set]
A set $W \subseteq [n]$ is a generalized witness set for a run $\rho$
of $\aA$ iff $\infi(\rho) = Q_W$ and $P_\phi(Q_W) = \true$.
\end{definition}
Note that Streett Safra witness sets are distinct from generalized
witness sets even when $\aA$ is a Streett automaton.  By definition, a
Streett Safra witness set is a subset of indices of Streett acceptance
pairs, while a generalized witness set is a subset of indices of
$\aA$-states.  Thus, if $\aA$ has $n$ states and $h$ pairs in its
acceptance table, and if $n \ll h$ (examples of NSW with this property
are given in Section~\ref{sec:large-acc-sets}), there can be many more
Streett Safra witness sets than generalized witness sets. The
situation is reversed if $h \ll n$.  It follows from the definition
above that a run $\rho$ of $\aA$ can have at most one generalized
witness set, although it may have multiple Streett Safra witness sets.
Furthermore, the generalized witness set of $\rho$ uniquely determines
$\infi(\rho)$, while a Streett Safra witness set for $\rho$ does not
necessarily determine $\infi(\rho)$ uniquely.  Finally, if $\aA$ is a
Streett automaton and if a run $\rho$ of $\aA$ has a generalized
witness set, then it has at least one (and perhaps more) Streett Safra
witness sets.  Conversely, if $\rho$ has at least one Streett Safra
witness sets, then it has exactly one generalized witness set.

The use of generalized witness sets allows us to adapt Safra's
construction to obtain a uniform determinization construction for
$\omega$-automata with arbitrary acceptance conditions.  We detail
this construction in the following subsections.

\subsection{Intuition}
\label{sec:intuition}
The intuition behind our construction parallels that behind Safra's
NSW determinization construction, with some key differences stemming
from the use of generalized witness sets instead of Streett Safra
witness sets.  The overall idea is to construct a deterministic
automaton that simulates all runs of $\aA$ on an $\omega$-word
$\alpha$, and uses a Rabin acceptance condition to simultaneously
identify the set of state indices in the $\infi$-set of a run and
check if this set is a generalized witness set. The construction of
the Rabin automaton can be adapted to give a deterministic parity
automaton using techniques employed by Piterman~\cite{piterman}.
Although there are an exponential number of potential generalized
witness sets, we use Safra's idea of building a process decomposition
(represented as a tree), in which each process tracks a subset of runs
and checks if a given subset of $\aA$-state indices is a generalized
witness set for these runs.  Using the same reasoning as used by
Safra, we can show that only a polynomial number of generalized
witness sets need to be examined at any time in order to determine if
a run has a generalized witness set.

As in Safra's and Piterman's
constructions~\cite{safra-stoc92,safra06,2001automata,piterman}, each
state of the DPW obtained by our construction is a tree of
hierarchically related processes, with additional book-keeping
information. The process represented by a node in the tree tracks a
subset of runs of the automaton $\aA$.  Each process is also
associated with a set of indices of $\aA$-states, called the
\emph{hope set} for the process.  A process hopes that its hope set
gives the indices of states in the $\infi$-set of all runs tracked by
it.  This is checked by waiting for all states with indices in the
hope set to be visited in turn by every run tracked by the process,
without visiting any state with index outside the hope set.  If this
happens, the process is said to have ``succeeded'' locally; it is then
``reset'' and the check starts all over again.  Clearly, if the
process represented by a node $v$ is reset infinitely often, its hope
set gives the indices of states in the $\infi$-set of all runs
tracked by it.  If, in addition, the set of states with indices in the
hope set causes $P_\phi$ to evaluate to \true, the hope set must be a
generalized witness set of all runs tracked by the process.  In this
case, there exists at least one accepting run of $\aA$ on the input
word.  On the other hand, if some state with an index outside the hope
set is seen in a run tracked by a process, the corresponding run is
removed from the process, and a new process is initiated for that run.
As in Safra's and Piterman's constructions, we use an acceptance
condition that checks for the existence of a node $u$ that is
eventually never deleted in the sequence of trees (states) in an
infinite run of the constructed automaton, but is reset infinitely
often.  Unlike Safra's and Piterman's construction, we also require
that the hope set of the process corresponding to node $u$ be such
that the corresponding set of $\aA$-states renders $P_\phi$ \true.  In
the remainder of the discussion, we will refer to a node and the
process represented by it interchangeably when there is no confusion.

\subsection{The determinization construction}
\label{sec:inf-set}

Piterman used compact Streett Safra trees to represent states of the
deterministic parity automaton in his NSW determinization
construction~\cite{piterman}.  We follow the same approach and use a
variant of compact Streett Safra trees, called compact generalized
Safra trees, or \CGS trees.  Formally, a \CGS tree $t$ over $\aA = (Q,
\Sigma, Q_0, \delta, \phi)$ is a $9$-tuple $(N, M, r, p, \lambda, h,
e, f)$, where
\begin{itemize}

\item $N$  is the set of nodes.

\item $M: N \rightarrow [|Q|^2 + |Q| + 1]$ is the naming function.

\item $r$ is the root node.

\item $p: N \rightarrow N$ is the parenthood function defined for $N
  \setminus \{r\}$.  Thus, $p(v)$ is the parent of $v \in N \setminus
  \{r\}$.

\item $\lambda: N \rightarrow 2^Q$ is a state labeling function that
  associates a subset of $Q$ with each node.  The state label of every
  node is equal to the union of state labels of its
  children. Furthermore, the state labels of two siblings are
  disjoint.

\item $h: N \rightarrow 2^{[|Q|]}$ is an annotation of nodes with a
  subset of $[|Q|]$.  The root is always annotated with $[|Q|]$. The
  annotation of every node is contained in that of its parent and
  differs by atmost one element from the annotation of its parent. Every
  non-leaf node $v$ has at least one child with an annotation that is
  a strict subset of $h(v)$.  For a node $v$ with annotation $J$ and
  child $v'$ with annotation $J' = J \setminus \{j\}$, we will say
  that the edge from $v$ to $v'$ is annotated with $j$.  If $J' = J$,
  we will say that the edge from $v$ to $v'$ is annotated with $0$.

\item $e, f \in [|Q|^2 + |Q| + 2]$ are two integers used to define the
  parity acceptance condition.
\end{itemize}
Note that \CGS trees differ from compact Streett Safra
trees~\cite{piterman} only in the annotation of nodes.  In a compact
Streett Safra tree, each node is annotated with a potential Streett
Safra witness set, while in a \CGS tree, the annotations are potential
generalized witness sets.  As discussed earlier, generalized witness
sets can differ significantly from Streett Safra witness sets even
when $\aA$ is a Streett automaton.  Intuitively, each node $v$ in a
compact generalized Safra tree represents a process that tracks the
runs of $\aA$ currently represented by $\lambda(v)$, and hopes that
$Q_{h(v)}$ is the $\infi$-set of these runs.  The set $h(v)$ may
therefore be viewed as the hope set for the process represented by
$v$.

Given $\aA = (\Sigma, Q, Q_0, \delta, \phi)$, we now construct a
deterministic parity automaton (DPW) $\dD = (\Sigma, T, t_0, \delta^p,
\mathcal{P})$ such that $L(\aA) = L(\dD)$.  In the following, we
assume that $n = |Q|$ and $m = |Q|^2 + |Q| + 1$.  The different components of
$\dD$ are as defined below.
\begin{itemize}
\item $T$ is the set of all \CGS trees over $\aA$.
  
\item $t_0$ is the \CGS tree with a single (root) node $r_0$, with
  $\lambda(r_0)= Q_0$, $M(r_0) = 1$ and $h(r_0)=[n]$. For $t_0$, we
  set $e = f = m+1$.
  
\item The parity acceptance condition $\mathcal{P} = \langle F_0, F_1,
  \ldots, F_{2m-1}\rangle$ is defined in the same manner as done by
  Piterman~\cite{piterman}.  Specifically,
  \begin{itemize}
    
  \item $F_0= \{t \in T \mid f = 1, e > 1\}$
    
  \item $F_{2i+1}=\{t \in T \mid e = i + 2, f \geq e\}$ for $0 \le i < m-1$
    
  \item $F_{2i+2}=\{t \in T \mid f = i + 2, e > f\}$ for $0 \le i < m-1$
    
  \item $F_{2m-1}=\{t \in T \mid e, f > m\}$
    
  \end{itemize}
  \noindent For reasons to be seen later, no \CGS tree that arises
  in our construction can have $e = 1$; hence \CGS trees with $e = 1$
  are excluded from the $F_i$ sets defined above.
  
\item $\delta^p$ is a deterministic transition function that returns a
  unique next state (\CGS tree) $t'$ for every current state $t \in T$
  and input symbol $\sigma \in \Sigma$.  The computation of $t'$ from
  $t$ and $\sigma$ is accomplished by invoking algorithm
  \algo{GeneralizedNext}$(t, \sigma)$, as detailed below.
\end{itemize}


Recall that a \CGS tree has named, state-labeled and annotated nodes
hierarchically arranged as a rooted tree, along with two integer
valued components named $e$ and $f$.  Computing $t'$ from $t$ and
$\sigma$ therefore involves transforming the hierarchical arrangement
of nodes and determining new values for $e$ and $f$, in general.
Component $e$ of $t'$ is intended to record the smallest name of a
node that was deleted during the transformation of the hierarchical
arrangement.  Similarly, component $f$ is meant to record the smallest
name of a node that was ``reset'' (in the sense described in
Section~\ref{sec:intuition}), had a hope set such that the
corresponding set of $\aA$ states satisfies $P_\phi$, and was not
deleted subsequently during the transformation of the hierarchical
arrangement.  Since a node can be deleted in a step after being reset,
algorithm \algo{GeneralizedNext} uses a set $U$ to remember all nodes
that were reset and had hope sets such that the corresponding set of
$\aA$ states satisfies $P_\phi$, in some step during the
transformation.  Finally, component $f$ is set to the smallest name of
a node in $U$ that survives the transformation.  The task of
transforming the hierarchical arrangement of nodes is accomplished by
invoking algorithm \algo{GeneralizedNextRecursive}, as described
below.  As the transformation proceeds through recursive calls to
\algo{GeneralizedNextRecursive} and nodes are reset and/or deleted
from the \CGS tree, component $e$ and the set $U$ described above are
updated.  After the transformation of the hierarchical arrangement is
completed, a name-compaction step is performed on the nodes of the
resulting \CGS tree in the same way as is done in~\cite{piterman}.
Although intermediate steps of algorithm
\algo{GeneralizedNextRecursive} may use names of nodes outside the set
$[m]$, the name-compaction step ensures that all names used in the
final \CGS tree $t'$ are within $[m]$. The pseudocode of algorithms
\algo{GeneralizedNext} and \algo{GeneralizedNextRecursive} are
presented below.

\algoblock{GeneralizedNext}
{$t:$ \CGS tree over $\aA$, ~~ $\sigma:$ letter in $\Sigma$}
{$t':$ \CGS tree over $\aA$}
{
  \begin{enumerate}

  \item \label{Nxt1} {\bfseries [Initialization]} Initialize $e$ and
    $f$ to $m+1$.  Initialize $U$ to $\emptyset$. For every node $u$
    in $t$, set $\lambda(u)$ to $\delta(\lambda(u), \sigma)$.
    
  \item \label{Nxt2} {\bfseries [Recursive transformation]} Let
    $\mathit{root}$ be the root node of $t$.  \\Invoke 
    \algo{GeneralizedNextRecursive}$(\mathit{root})$.
    
  \item \label{Nxt3} {\bfseries [Name-compaction]} Let $\widehat{t}$
    be the \CGS tree rooted at $\mathit{root}$ after Step
    (\ref{Nxt2}).  Let $Z$ be the set of \CGS tree nodes removed
    during the execution of Step (\ref{Nxt2}).  For every node $u$ in
    $\widehat{t}$, let $\mathsf{rem}(u) = |\{u' \in Z \mid M(u') <
    M(u)\}|$ .  Update $M(u)$ to $M(u) - \mathsf{rem}(u)$.
    
  \item \label{Nxt4} {\bfseries [Updation of component $f$]} Let
    $\widetilde{t}$ be the \CGS tree rooted at $\mathit{root}$ that
    results after Step (\ref{Nxt3}).  Let $\widetilde{N}$ be the set
    of nodes in $\widetilde{t}$.  Set $f$ to the minimum of its
    current value and $\min \{M(v') \mid v' \in U \cap
    \widetilde{N}\}$.
  
  \item Return $t'$ as the \CGS tree rooted at $\mathit{root}$ with
    $e$ and $f$ components as calculated above.
  \end{enumerate}
}

\algoblock{GeneralizedNextRecursive}
{$v:$ root of a \CGS sub-tree}
{$\widehat{t}:$ Transformed \CGS sub-tree rooted at $v$, updated values of $e$ and $U$}
{
  \begin{enumerate}

  \item \label{GNR1} If $v$ is a leaf and $h(v) = \emptyset$,
    return $t'$ as the \CGS sub-tree rooted at $v$.
    
  \item \label{GNR2} If $v$ is a leaf and $h(v) \neq \emptyset$, create a
    new child $v'$ of $v$.  Set $\lambda(v') = \lambda(v)$, $h(v') =
    h(v) \setminus \{\max(h(v))\}$ and $M(v')$ to the smallest name
    greater than all names already used.  Note that this may require
    using names not in $[m]$.
    
  \item \label{GNR3} If, after the execution of Steps (\ref{GNR1}) and
    (\ref{GNR2}), $v$ is not a leaf, then let $v_1, \ldots, v_l$ be the
    children of $v$ ordered according to their names.  Let $j_1,
    \ldots j_l$ be indices such that $j_i = \mathsf{max}((h(v) \cup
    \{0\}) \setminus h(v_i))$ \footnote{Note that if $h(v) = h(v_i)$,
      then $j_i=0$.}.  As discussed earlier (in the definition of
    compact generalized Safra trees), we will say that the edge from
    $v$ to $v_i$ is annotated with $j_i$.
    
    \begin{enumerate}
      
    \item \label{GNR31} For all $i$ in $1$ through $l$, invoke \algo{GeneralizedNextRecursive}$(v_i)$
      
      
    \item \label{GNR32} For every child $v_i$ of $v$ and every $q \in \lambda(v_i)$, do
      the following. 
      
      \begin{enumerate}
        
      \item \label{GNR3a} If $q = q_{j_i}$ then 
        create a new child $v'$ of $v$.\\  Set $\lambda(v') = \{q\}$,
        $h(v') = h(v) \setminus \{\max \left((h(v) \cup \{0\})\cap
        \{0, 1, 2, \ldots, j_i-1\}\right)\}$.  The edge from $v$ to $v'$ is
        thus annotated with the largest integer smaller
        than $j_i$ but in $h(v)$, if it exists.  Otherwise, the edge
        is annotated with $0$.  Set $M(v')$ to the smallest
        name greater than all names already used.
        
      \item \label{GNR3b} If $q \neq q_{j_i}$ and $q \notin Q_{h(v_i)} =
        \{q_j \mid q_j \in Q, j \in h(v_i)\}$, remove $q$ from
        $\lambda(v_i)$ and also from $\lambda(u)$ for all descendants
        $u$ of $v_i$.

        
      \end{enumerate}

    \end{enumerate}

  \item \label{GNR4} Let $v_1, v_2, \ldots, v_{l'}$ be the children of
    $v$ after the above steps. Let $j_1, \ldots, j_{l'}$ be the
    annotations of the corresponding edges from $v$ to its children.
    In other words, let $j_i = \mathsf{max}((h(v)\cup \{0\}) \setminus
    h(v_i))$ for $i \in \{1, 2, \ldots l'\}$. Then for every $q \in
    \lambda(v_i) \cap \lambda(v_{k})$, where $i \neq k$ and $i, k \in
    \{1, \ldots l'\}$, do the following.
    
    \begin{enumerate}
    \item If $j_i < j_k$, remove $q$ from $\lambda(v_{k})$ and from
      $\lambda(u)$ for all descendants $u$ of $v_k$.
      
    \item If $j_i = j_k$ and $M(v_i) < M(v_{k}))$, remove $q$ from
      $\lambda(v_{k})$ and from $\lambda(u)$ for all descendants $u$
      of $v_k$.
      
    \end{enumerate}
    
  \item \label{GNR5} For every descendant $u$ of $v$ such that
    $\lambda(u) = \emptyset$, delete $u$ and all its descendants.
    
  \item \label{GNR6} If, after the previous steps, all children of $v$
    have annotation $h(v)$, then the process represented by $v$ is
    said to ``succeed'' locally and needs to be ``reset''.  Delete all
    descendants of $v$, so that $v$ becomes a leaf node.
    Additionally, if $P_\phi(Q_{h(v)}) = \true$, then update $U$ to $U
    \cup \{v\}$.
    
  \item \label{GNR7} Update $e$ to the minimum of its previous value and
    the smallest name among all descendants of $v$ that were deleted.

  \item \label{GNR8} Return $t'$ as the \CGS sub-tree rooted at $v$.
    
    
  \end{enumerate}
}

The similarity of algorithms \algo{GeneralizedNext} and
\algo{GeneralizedNextRecursive} to the corresponding algorithms in
Safra's and Piterman's NSW determinization constructions is striking.
Yet, there are important differences that enable our construction to
achieve something different, and even better Safra's and Piterman's
constructions when the number of Streett pairs is large compared to
the number of Streett states.

The computation of $\delta^p(t, \sigma)$ starts by determining the
successors of all $\aA$-states appearing in state labels of nodes in the
\CGS tree $t$, under the input symbol $\sigma$.  Algorithm
\algo{GeneralizedNextRecursive} is then invoked on the resulting tree
rooted at $\mathit{root}$.  This recursively ``extends'' the tree (in
Steps (\ref{GNR1}), (\ref{GNR2}) and the recursive call in Step
(\ref{GNR3}) of algorithm \algo{GeneralizedNextRecursive}) by adding
new leaf nodes with successively smaller hope sets until each leaf
node has an empty hope set.  As the recursive calls return, algorithm
\algo{GeneralizedNextRecursive}) determines in a bottom-up manner
which nodes in the extended \CGS tree must have their hope sets
invalidated and/or hierarchical relations modified.  We explain below
the reasoning behind this crucial step in the computation of $t'$.

Suppose the hope set of a node $v$ is $h(v)$ and that of its child
$v''$ is $h(v'')$.  Suppose further that the edge from $v$ to $v''$ is
annotated with $j_i$, i.e., $h(v) \setminus h(v'') = \{j_i\}$.  This
represents a situation wherein the process represented by $v$ is
waiting to see $q_{j_i}$ in the subset of runs being tracked by its
child $v''$, but the process represented by $v''$ has given up hope of
seeing any further $q_{j_i}$'s in the runs it is tracking.  Now,
suppose after reading an input symbol $\sigma$, the initialization
step of algorithm \algo{GeneralizedNext} places $q_{j_i}$ in
$\lambda(v'')$ (and hence also in $\lambda(v)$).  This implies that
$v''$ has seen a state along a run it was tracking, such that the
corresponding state index is outside its own hope set but is in the
hope set of its parent.  Since every node expects to see all and only
states with indices in its hope set in all runs being tracked by it,
the above situation warrants two actions: (i)~invalidating the hope
set of $v''$ for the run represented by $q_{j_i}$, and
(ii)~registering progress towards the realization of $v$'s hope set as
the set of state indices in the $\infi$-set of the run represented by
$q_{j_i}$.  Accordingly, $q_{j_i}$ is removed from $\lambda(v'')$ by
the sequence of steps~\ref{GNR3a} and \ref{GNR4} of algorithm
\algo{GeneralizedNextRecursive}.  In addition, step~\ref{GNR3a}
creates a new child $v'$ of $v$ with $\lambda(v'') = \{q_{j_i}\}$, and
annotates the edge from $v$ to $v''$ with the next index (after $j_i$
in decreasing order), say $j_k$, in the hope set of $v$.  This
represents the new situation wherein the process represented by $v$
has seen $q_{j_i}$ and is waiting to see the next $\aA$-state in its
hope set, i.e. $q_{j_k}$, in the run (currently) represented by
$q_{j_i}$.  The new child $v'$ however hopes to see no further
$q_{j_k}$'s in the run represented by $q_{j_i}$; hence its hope set is
set to $h(v) \setminus \{j_k\}$.  A special situation arises if
$q_{j_i}$ is the lowest indexed $\aA$-state in $Q_{h(v)}$.  In this
case, node $v$ has seen all states with indices in its hope set in the
run represented by $q_{j_i}$ since the last time $v$ was ``reset''.
The edge from $v$ to $v'$ is annotated with a special index, i.e. $0$,
to represent this situation.  The newly created child $v'$ retains the
same hope set as $v$, i.e. $h(v))$, and is now delegated the task of
checking if $Q_{h(v)}$ is the $\infi$- set of the run currently
represented by $q_{j_i}$.  Meanwhile, the parent node $v$ continues to
check if all states with indices in its hope set, i.e. $h(v)$, are
seen in the \emph{remaining} runs (other than the one currently
represented by $q_i$) that it was tracking.

A different situation arises if the initialization step of
algorithm \algo{GeneralizedNext} places $q_{j_i}$ in $\lambda(v'')$
for a child $v''$ of $v$, but $j_i$ is neither the annotation of the
edge from $v$ to $v''$, nor is in the hope set of $v''$.  This
represents a situation wherein the process represented by $v$ was
waiting to see some $\aA$-state other than $q_{j_i}$ next in the runs
being tracked by $v''$, and the process represented by $v''$ was
expecting to never see $q_{j_i}$ in any run being tracked by it.
Since $q_{j_i}$ is in $\lambda(v'')$, the hope set of $v''$ must be
invalidated for the run currently represented by $q_{j_i}$.  This is
done in step~\ref{GNR3b} of algorithm \algo{GeneralizedNextRecursive}
by removing $q_{j_i}$ from the state label of $v''$ and all its descendants.
Note, however, that we cannot remove the run represented by $q_{j_i}$
from the state label of $v$ yet.  Although $v$ was not expecting $q_{j_i}$
to be the \emph{next} $\aA$-state in the runs being tracked by $v''$,
the hope set of $v$ may still contain $j_i$.  Therefore, the hope set
of $v$ need not be invalidated yet for the run corresponding to
$q_{j_i}$.  As the recursive calls to algorithm
\algo{GeneralizedNextRecursive} return, the hope set of $v$ will be
examined in turn to determine if a run being tracked by $v$ has
encountered a state with index outside $v$'s hope set.  If so, the run
will then be removed from the set of runs being tracked by $v$.
 
Since runs tracked by different nodes in a \CGS tree may merge, we may
encounter a situation wherein the same $\aA$-state $q$ appears in the
state labels of multiple nodes that are not related as ancestors or
descendants in the tree.  However, by definition, two nodes in a \CGS
tree can have overlapping state labels only if one is an ancestor (or
descendant) of the other.  Algorithm \algo{GeneralizedNextRecursive}
rectifies this situation by ensuring that whenever an $\aA$-state $q$
appears in the state labels of multiple children of a node $v$, at most one
child eventually gets to retain $q$ in its state label.  The chosen child is
the one that represents the maximum progress (since $v$ was last
reset) towards realisation of the hope set of $v$ as the set of state
indices in the $\infi$-set of the run represented by $q$.  This choice
can be made by examining the annotations on the edges from $v$ to the
subset of its children containing $q$ in their state labels.  Specifically,
the child that represents the most progress is the one that has the
smallest annotation on the edge from $v$.  This is because a child
with an edge from $v$ annotated with $i$ represents the situation
wherein all $\aA$-states with indices greater than $i$ and in the hope
set of $v$ have been seen since $v$ was last reset.  In the event that
an $\aA$-state $q$ appears in the state labels of two siblings with the same
annotation on the edges from their parent, we choose to retain $q$ in
the state label of the node that was created earlier, i.e. has a smaller
name.  As the recursive calls to algorithm
\algo{GeneralizedNextRecursive} return, step~\ref{GNR4} examines the
nodes of the \CGS tree in a bottom-up manner and applies the above
criterion to ensure that two nodes not related as ancestor and
descendant do not share any $\aA$-state in their state labels in the final
tree.

Step~\ref{GNR5} of algorithm \algo{GeneralizedNextRecursive} deletes
all nodes with empty state labels from the \CGS tree constructed thus far,
since the processes represented by these nodes no longer track any
runs.  In Step~\ref{GNR6}, we examine the annotations on the edges to
all children of the current node $v$.  If these annotations are all
$0$, we have a situation wherein all runs being tracked by $v$ have
seen all states with indices in $v$'s hope set since the last time $v$
was reset.  This constitutes a step of progress in establishing that
the hope set of $v$ is indeed the set of state indices in the
\infi-set of all runs being tracked by it.  Node $v$ is therefore said
to have ``succeeded'' locally, and is ``reset'' in step~\ref{GNR6} of
algorithm \algo{GeneralizedNextRecursive} by deleting all its
descendants. If, in addition, $Q_{h(v)} \models \phi$ then we have a
step of progress in establishing that $Q_{h(v)}$ is the generalized
Safra witness set of all runs being tracked by $v$.  Step~\ref{GNR6}
of algorithm \algo{GeneralizedNextRecursive} keeps track of this fact
by updating the set $U$.  As explained earlier, $U$ is eventually used
to obtain the value of component $f$ of the \CGS tree $t'$.  Finally,
step~\ref{GNR7} of algorithm \algo{GeneralizedNextRecursive} updates
component $e$ of $t'$ by recording the smallest name of a node deleted
in the recursive transformation of the \CGS tree.

\subsection{An Example}
\label{sec:inf-set-example}
\begin{figure}[ht]
\begin{center}
\includegraphics[scale=0.6]{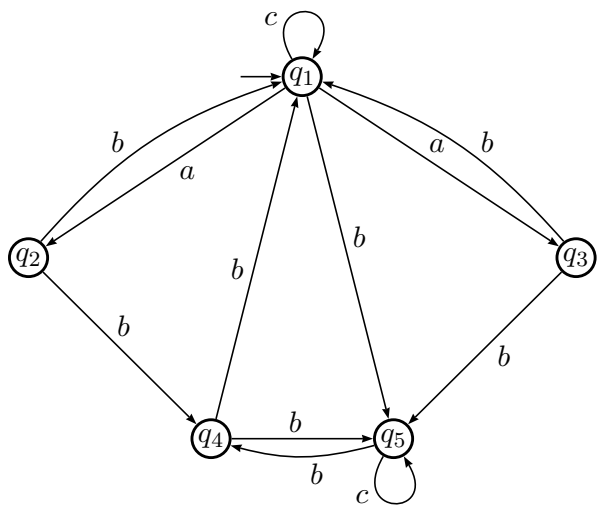}
\end{center}
\caption{Example non-deterministic Muller automaton}
\label{infset1}
\end{figure}

We now illustrate the working of our determinization construction
using the non-determinisic M\"{u}ller automaton (NMW) $\aA$ shown in
Figure (\ref{infset1}).  The M\"{u}ller acceptance condition of this
automaton is given by $\mathcal{F} = \{\{q_1\}\}$.
Let $\dD$ be the corresponding deterministic parity automaton obtained
by our construction.  To see how different states and transitions in
$\dD$ are obtained, we will follow the construction of states
encountered in $\dD$ on reading a short prefix of the word
$bbbc^\omega$ that is accepted by $\aA$.  Since $\aA$ has $5$ states,
we have $n = 5$ and $m = 5^2 + 5 + 1 = 31$.  Thus, every node in the
\CGS tree representing a state of $\dD$ has a name in $[31]$, and a
hope set that is a subset of $[5]$.  Every edge in the tree is
annotated with an element of $\{0, 1, \ldots 5\}$.  Since the hope set
of the root node is always $[5]$, and since the hope set of any other
node $v$ can be obtained by eliminating from $[5]$ the annotations of
edges on the path from the root to $v$, we will simply annotate edges
with elements of $[5]$ and not explicitly represent hope sets.
Similarly, since the state label of every node is the union of the
state labels of its children, we will simply label leaves of the \CGS
tree with subsets of $\aA$-states.  To help illustrate the
intermediate steps of the construction, we will also indicate the
updated values of $e$ and $f$ (components of the \CGS tree) in the
following discussion.
\begin{figure}[ht]
\begin{minipage}[b]{0.48\linewidth}
\begin{center}
\includegraphics[scale=0.63]{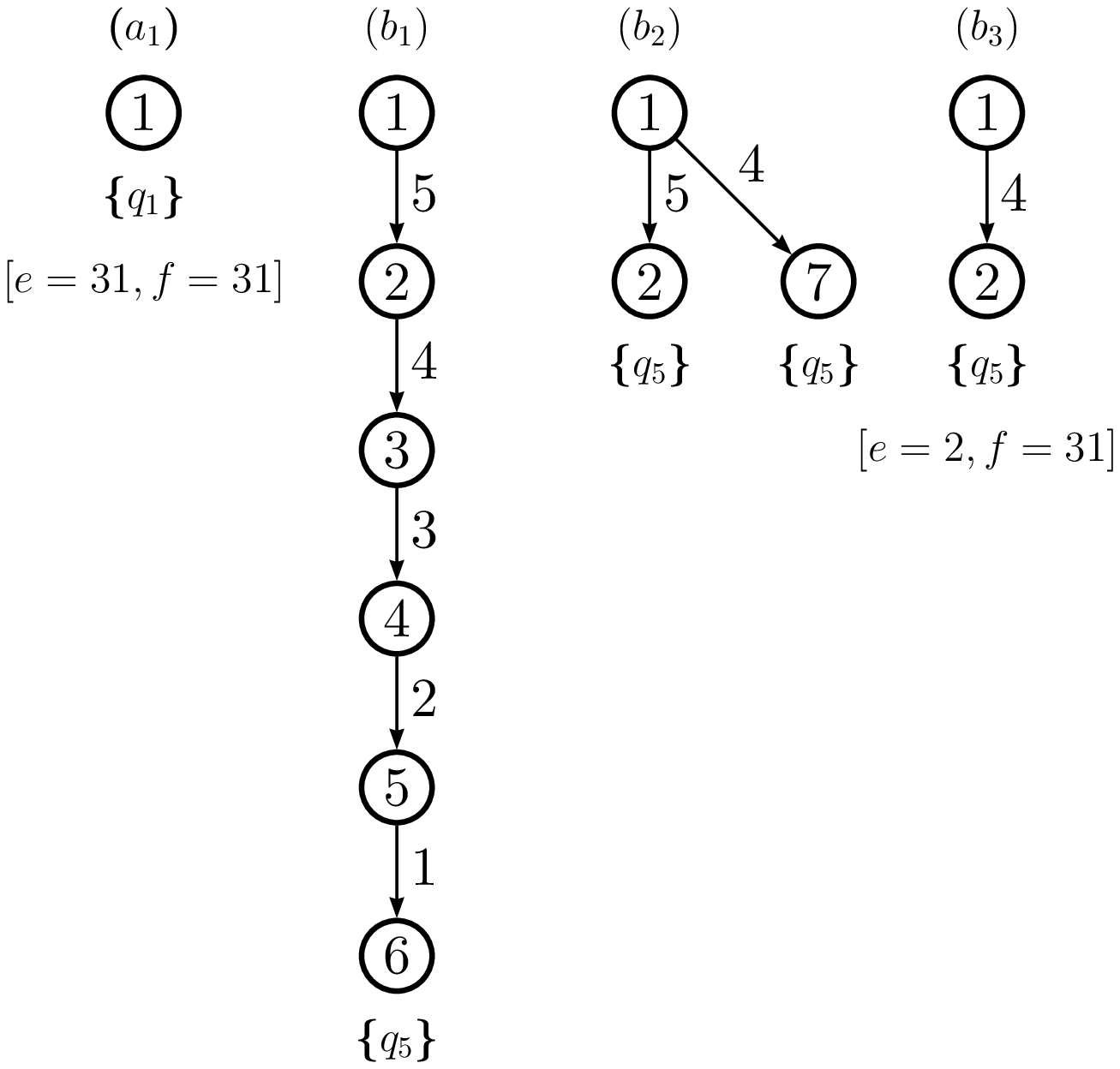}
\end{center}
\label{infset2}
\end{minipage}
\hspace{5mm}
\begin{minipage}[b]{0.48\linewidth}
\begin{center}
\includegraphics[scale=0.63]{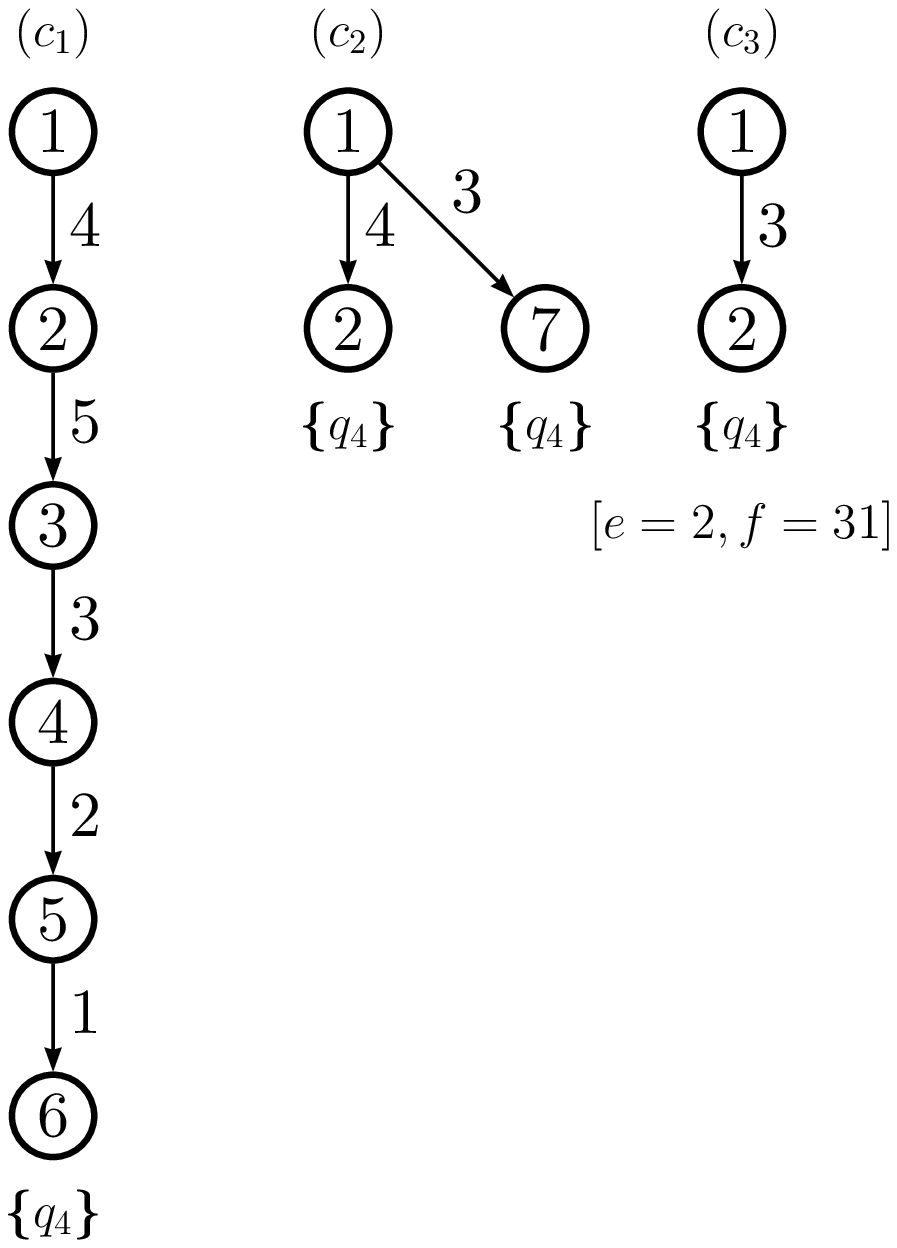}
\end{center}
\label{infset3}
\end{minipage}
\caption{Steps in determinization construction}
\label{infset23}
\end{figure}

\begin{figure}[ht]
\begin{minipage}[t]{0.6\linewidth}
\vspace{0pt}
\begin{center}
\includegraphics[scale=0.63]{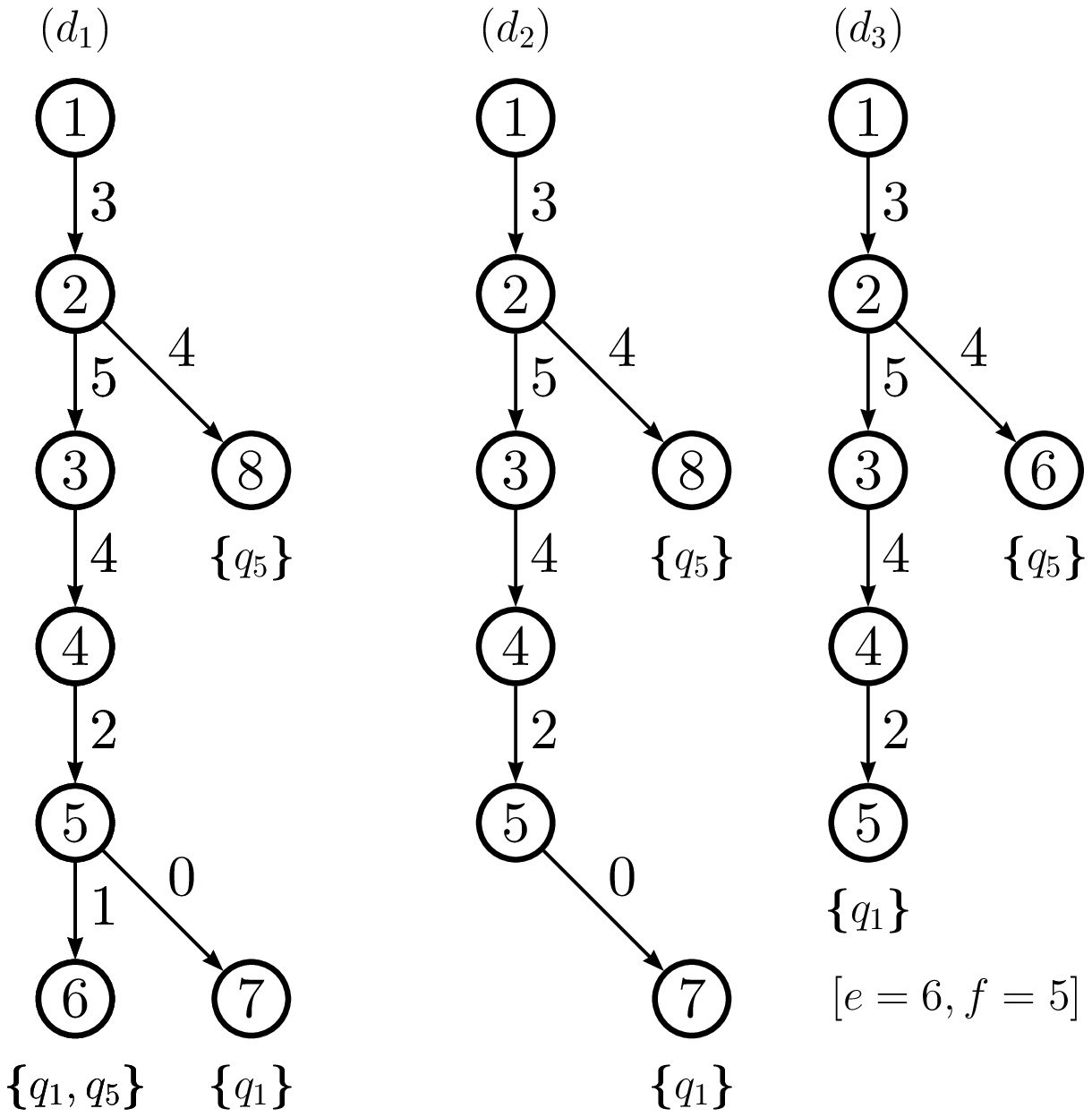}
\end{center}
\label{infset4}
\end{minipage}
\begin{minipage}[t]{0.4\linewidth}
\vspace{0pt}
\begin{center}
\includegraphics[scale=0.63]{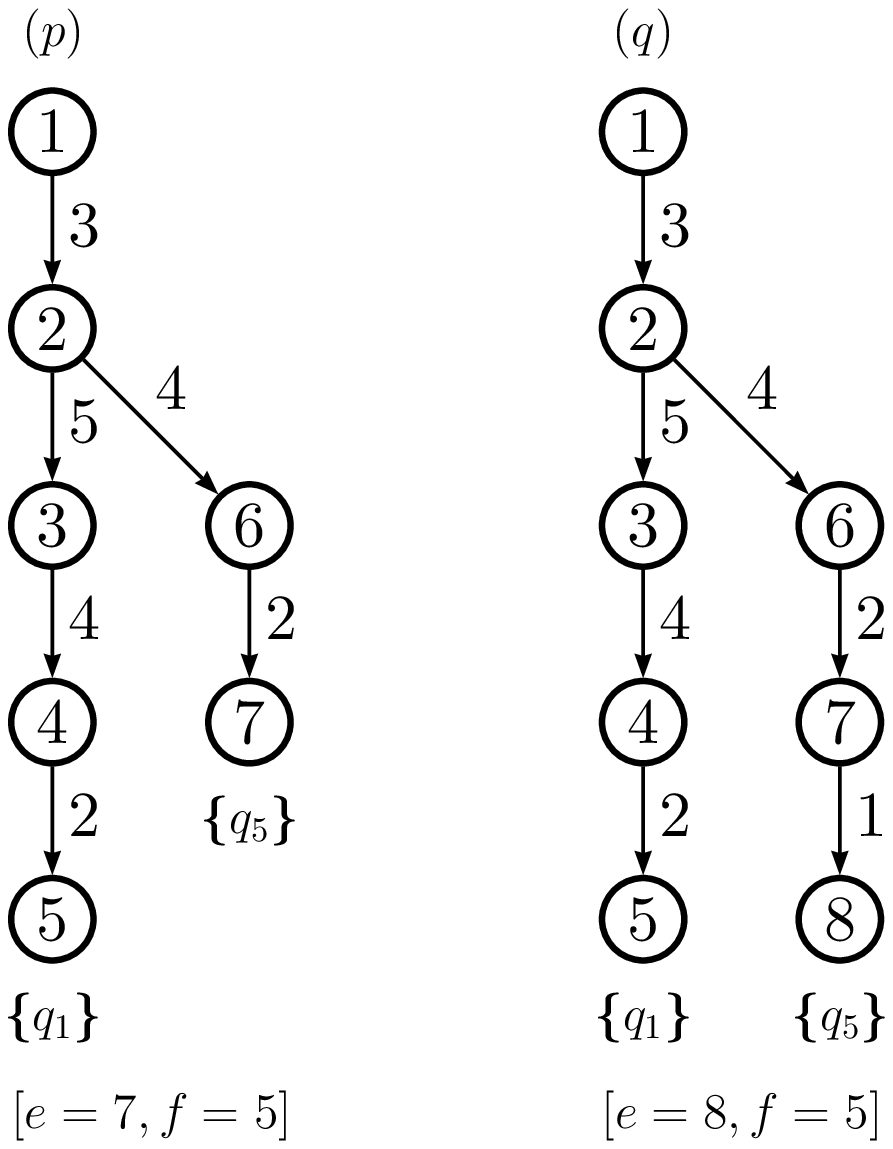}
\end{center}
\end{minipage}
\caption{Steps in determinization construction}
\label{infset45}
\end{figure}

We start in the initial state consisting of a \CGS tree having a
single node named $1$ and labeled $\{q_1\}$, as shown in Figure
(\ref{infset23}-$a_1$).  The values of $e$ and $f$ are both $m + 1 =
32$ in this state.  On reading the letter $b$, the state label of the
node named $1$ (also a leaf in this case) is first changed to
$\{q_5\}$, since $q_1$ transitions to $q_5$ on reading $b$ in
automaton $\aA$.  The \CGS tree consisting of only the root node is
then extended in Steps (\ref{GNR1}), (\ref{GNR2}) and through the
recursion in Step (\ref{GNR31}) of algorithm
\algo{GeneralizedNextRecursive} to give the tree shown in Figure
(\ref{infset23}-$b_1$).  As the recursive calls return in sequence,
all nodes other than the ones named $1$ and $2$ are deleted.  When the
recursion returns to the topmost level with the root named $1$ as the
current node $v$, the condition in Step (\ref{GNR3a}) of algorithm
\algo{GeneralizedNextRecursive} is satisfied.  Consequently, a new
node named $7$ is created as a child of the root, and assigned the
state label $\{q_5\}$.  The edge from the root to this child is
annotated with $4$, as shown in Figure (\ref{infset23}-$b_2$).
Subsequently, Step (\ref{GNR4}) of algorithm
\algo{GeneralizedNextRecursive} removes $q_5$ from the state label of
the leaf named $2$ in Figure (\ref{infset23}-$b_2$).  This is because
the annotation of the edge from the root to this node is larger than
that of the edge from the root to its sibling having the same
$\aA$-state, $q_5$, in its state label.  Removing $q_5$ from its state
label causes the leaf named $2$ in Figure (\ref{infset23}-$b_2$) to
acquire an empty state label; hence this node is deleted in Step
(\ref{GNR5}) of algorithm \algo{GeneralizedNextRecursive}.  This gives
a tree with only two nodes -- a root named $1$ and a leaf named $7$
with state label $\{q_5\}$.  The condition in Step (\ref{GNR6}) is not
satisfied; hence no nodes are ``reset'' and $U$ continues to be the
empty set.  In Step (\ref{GNR7}), the component $e$ finally acquires
the value $2$, since that is the smallest name of a node that is
deleted.  Once we return from algorithm
\algo{GeneralizedNextRecursive} to algorithm \algo{GeneralizedNext},
the name-compaction step assigns the name $2$ to the leaf node that
was named $7$ earlier.  Since no node is reset and the set $U$ is
empty, the updated value of the component $f$ remains at $32$.  The
resulting \CGS tree obtained after reading the first $b$ from the
input word is shown in Figure (\ref{infset23}-$b_3$).

On reading the next $b$, a sequence of transformations similar to that
described above results in a \CGS tree with a root named $1$ and a
leaf named $2$ with state label $\{q_4\}$ and edge annotation $3$.  Here
too, the component $e$ acquires the value $2$ and $U$ remains empty,
causing $f$ to have the value $32$.  Figures (\ref{infset23}-$c_1$) to
(\ref{infset23}-$c_3$) illustrate the steps in the construction of
this \CGS tree.

When the third $b$ in the input word is read, the tree in Figure
(\ref{infset23}-$c_3$) is extended in Steps (\ref{GNR1}), (\ref{GNR2})
and through the recursion in Step (\ref{GNR31}) of algorithm
\algo{GeneralizedNextRecursive} to give the tree shown in Figure
(\ref{infset45}-$d_1$) sans the nodes named $7$ and $8$.  As the
recursive calls to algorithm \algo{GeneralizedNextRecursive} return in
sequence, Step (\ref{GNR3a}) creates two new leaf nodes (albeit in
different recursive calls) named $7$ and $8$, with state labels
$\{q_1\}$ and $\{q_5\}$ respectively.  The edges from the respective
parents to the new leaves named $7$ and $8$ are annotated $0$ and $4$,
respectively.  The resulting tree is as shown in Figure
(\ref{infset45}-$d_1$), except that the node named $6$ no longer has
$q_1$ or $q_5$ in its state label.  In fact, Step (\ref{GNR4}) of
algorithm \algo{GeneralizedNextRecursive} removes both $q_1$ and $q_5$
(once again, in different recursive calls) from the state label of
this node, leaving it with an empty state label.  Subsequently, this
node is removed in Step (\ref{GNR5}), giving the intermediate \CGS
tree shown in Figure (\ref{infset45}-$d_2$).  Observe that the node
named $5$ in this tree has the edge to its sole child annotated $0$.
Therefore, this node is ``reset'' in Step (\ref{GNR6}) of algorithm
\algo{GeneralizedNextRecursive} and the child named $7$ is deleted.
Additionally, since the hope set for the node named $5$ in Figure
(\ref{infset45}-$d_2$) is $\{1, 2, 3, 4, 5\} \setminus \{3, 5, 4, 2\}
= \{1\}$, and since $\{q_1\} \in \mathcal{F}$, we have
$P_\phi(\{q_1\}) = \true$.  Therefore, $5$ is added to the set $U$ in
Step (\ref{GNR6}) of algorithm \algo{GeneralizedNextRecursive}.  Since
the smallest name of a node that is deleted is $6$, component $e$
finally acquires the value $6$ in Step (\ref{GNR7}).  Once we return
to algorithm \algo{GeneralizedNext}, the name-compaction step renames
the leaf node named $8$ to $6$, as shown in Figure
(\ref{infset45}-$d_3$).  The value of $f$ is updated to $\min(32, 5) =
5$.  The final \CGS tree obtained after reading $bbb$ is shown in
Figure (\ref{infset45}-$d_3$).  Figures (\ref{infset45}-$p$) and
(\ref{infset45}-$q$) show the final \CGS trees (states) obtained after
reading $bbbc$ and $bbbcc$ respectively.  For all subsequent $c$'s
that are read from the input word, the \CGS tree in Figure
(\ref{infset45}-$q$) is obtained.  Therefore, the automaton $\dD$
loops infinitely in the state represented by Figure
(\ref{infset45}-$q$) after reading $bbbcc$.  Note that nodes named $5$
and $8$ are deleted only finitely often but appear as leaves
infinitely often in the sequence of \CGS trees (states) visited on
reading the word $bbbc^\omega$.  Interestingly, the hope sets of the
nodes named $5$ and $8$ in Figure (\ref{infset45}-$q$) are precisely
the $\infi$-sets of the runs of $\aA$ on the word $bbbc^\omega$.  As
we will see subsequently, this is not a coincidence, but a consequence
of our construction.

Let $T$ be the set of all \CGS tree over $\aA$.  The parity acceptance
condition for automaton $\dD$ is $\mathcal{P} = \langle F_0, F_1,
\ldots, F_{61}\rangle$, where $F_0= \{t \in T \mid f = 1, e > 1\}$,
$F_{2i+1}=\{t \in T \mid e = i + 2, f \geq e\}$ for $0 \le i < 30$,
$F_{2i+2}=\{t \in T \mid f = i + 2, e > f\}$ for $0 \le i < 30$, and
$F_{61}=\{t \in T \mid e, f > 31\}$.  If we let $\rho$ denote the run
of $\dD$ on the word $bbbc^\omega$, then clearly $\mathit{inf}(\rho)
\cap F_8 \neq \emptyset$, while $\infi(\rho) \cap F_i = \emptyset$ for
$0 \le i < 8$.  Therefore, $bbbc^\omega$ is accepted by $\dD$.


\section{Proof of Correctness}

Let $\aA = (\Sigma, Q, Q_0, \delta, \phi)$ be an $\omega$-automaton
with acceptance condition based on infinity sets, and let $\dD$ be the
corresponding DPW obtained by our construction.  Let $\alpha \in
\Sigma^\omega$ be an $\omega$-word, and let $\rho = t_0 t_1 t_2
\ldots$ be the unique run of $\dD$ on $\alpha$.  Here, $t_i = (N_i,
M_i, r_i, p_i, \lambda_i, h_i, e_i, f_i)$ is the state (tree)
of $\dD$ reached after reading the prefix $\alpha(0, i-1)$ of
$\alpha$.

We will first show that if $t_0$ is a \CGS tree, as defined in Section
(\ref{sec:inf-set}), then every $t_i$, for $i \geq 0$, in $\rho$ is
also a \CGS tree.  From algorithms \algo{GeneralizedNext} and
\algo{GeneralizedNextRecursive}, it is easy to see that if $t_i$ is a
rooted tree with nodes labeled by subsets of $Q$ and annotated with
subsets of $[|Q|]$, then so is $t_{i+1}$, for all $i \geq 0$.  Since
$e_{i+1}$ and $f_{i+1}$ are initialized to $m + 1 = |Q|^2 + |Q| + 2$ and
subsequently updated to the smaller of their respective current value
and the name of a node in $t_{i+1}$, it follows that $e_{i+1}$ and
$f_{i+1}$ are always in $[|Q|^2 + |Q| + 2]$.  Given these observations, it
suffices to show the following three additional properties of
$t_{i+1}$ in order to establish that $t_{i+1}$ is indeed a \CGS tree.
\begin{enumerate}
\item \emph{There are no more than $|Q|^2 + |Q| + 1$ nodes in
  $t_{i+1}$.}  Since the name-compaction step of algorithm
  \algo{GeneralizedNext} ensures the absence of gaps in the set of
  names eventually assigned to nodes of $t_{i+1}$, proving the above
  property guarantees that the range of the naming function $M_{i+1}$
  is indeed $|Q|^2 + |Q| + 1$.  We will defer the proof of this
  property to Section (\ref{sec:inf-set-size}).
\item \emph{The (hope-set) annotation of every node in $t_{i+1}$ is
  contained in the annotation of its parent, and differs by atmost one
  element from that of its parent.  In addition, every non-leaf node
  $v$ in $t_{i+1}$ has at least one child with an annotation that is a
  strict subset of $h_{i+1}(v)$}.  The first property is proved in Lemma
  (\ref{lemmaD}) below.  The second property is a consequence of Lemma
  (\ref{lemmaD}) and Step (\ref{GNR6}) of algorithm
  \algo{GeneralizedNextRecursive}.
\item \emph{The state label of every node in $t_{i+1}$ is the union of
  state labels of its children in $t_{i+1}$.  In addition, the state
  labels of sibling nodes in $t_{i+1}$ are mutually disjoint.}  We will
  prove the first property in Lemma (\ref{lemmaC}) below.  The second
  property is a consequence of Step (\ref{GNR4}) of algorithm
  \algo{GeneralizedNextRecursive} and the fact that no step of
  algorithm \algo{GeneralizedNextRecursive} adds any element to an
  already existing state label of a node.
\end{enumerate}

\begin{lemma}
\label{lemmaD}
For every $i \geq 0$ and for every node $u$ and its child $v$ in
$t_i$, $h_i(v) \subseteq h_i(u)$ and $|h_i(u) \setminus h_i(v)| \leq
1$.
\end{lemma}
\noindent {\bf Proof: } We will prove the lemma by induction on the
indices of $t_0, t_1, \ldots$.

\noindent \emph{Base Case:} For the tree $t_0$ with only the root node
$r_0$, the claim in the lemma holds vacuously since there are no nodes
with children in $t_0$.

\noindent \emph{Hypothesis:} We assume that the claim in the lemma
holds for $t_i$, where $i \geq 0$.

\noindent \emph{Induction:} Consider the tree $t_{i+1}$ obtained by
applying algorithm \algo{GeneralizedNext} to $t_i$.  From the
pseudocode of algorithms \algo{GeneralizedNext} and
\algo{GeneralizedNextRecursive}, we observe that the hope set of a
node in $t_{i+1}$ can be updated only in Step (\ref{GNR2}) or Step
(\ref{GNR3a}) of algorithm \algo{GeneralizedNextRecursive}.  In both
these steps, the node whose hope set is updated is a newly created
node that is added as a child of the current node.

Now let $v$ be an arbitrary node in $t_{i+1}$.  We consider two
cases below.
\begin{itemize}
\item Suppose $v \in N_i \cap N_{i+1}$.  Thus, $v$ was present in
  $t_i$ and was not deleted in the process of transforming $t_i$ to
  $t_{i+1}$.  Since deletion of a node (Step (\ref{GNR5} or Step
  (\ref{GNR6}) of algorithm \algo{GeneralizedNextRecursive}) entails
  deletion of all descendants of the node as well, the fact that $v$
  was not deleted implies that no ancestor of $v$ was deleted either
  in the process of transforming $t_i$ to $t_{i+1}$.  Thus, both $v$
  and its parent, say $u$, in $t_{i+1}$ were present in $t_i$, and
  neither of them was newly created in Step
  (\ref{GNR2}) or Step (\ref{GNR3a}) of algorithm
  \algo{GeneralizedNextRecursive} during the transformation of $t_i$
  to $t_{i+1}$.  Hence, $h_{i+1}(v) = h_{i}(v)$ and $h_{i+1}(u) =
  h_{i}(u)$.  By the induction hypothesis, we already know that
  $h_i(v) \subseteq h_i(u)$ and $|h_i(v) \setminus h_i(u)| \leq 1$.
  The inductive claim now follows immediately.
  
\item Suppose $v$ was newly created in the process of transforming
  $t_i$ to $t_{i+1}$.  Since new nodes can be created only in Step
  (\ref{GNR2}) or Step (\ref{GNR3a}) of the recursive algorithm
  \algo{GeneralizedNextRecursive}, $v$ must have been created in one
  of these steps.  From the pseudocode of algorithm
  \algo{GeneralizedNextRecursive}, it is easy to see that both these
  steps set $h_{i+1}(v)$ to $h_{i+1}(u) \setminus \{k\}$, where $u$ is
  the parent of $v$ in $t_{i+1}$ and $k \in \{0\} \cup h_{i+1}(u)$.  It
    follows that $h_{i+1}(v) \subseteq h_{i+1}(u)$ and $|h_{i+1}(v)
    \setminus h_{i+1}(u)| \leq 1$.
\end{itemize}

Therefore, by the principle of mathematical induction, the claim in
the lemma holds for all $t_i$, where $i \geq 0$. \qed

\begin{lemma}
\label{lemmaC}
For every $i \geq 0$ and for every non-leaf node $v$ in $t_i$,
$\lambda_i(v) = \bigcup_{v' \in V'} \lambda_i(v')$, where $V'=\{v'
\mid v' \in N_i, v = p_i(v')\}$.
\end{lemma}
\noindent {\bf Proof: }
We will prove the lemma by induction on the indices of $t_0, t_1, \ldots$.

\noindent \emph{Base Case:} For the tree $t_0$ with only the root node
$r_0$, the claim in the lemma holds vacuously since there are no
non-leaf nodes in $t_0$.

\noindent \emph{Hypothesis:} We assume that the claim in the lemma
holds for $t_i$, where $i \geq 0$.

\noindent \emph{Induction:} Consider the tree $t_{i+1}$ obtained by
applying algorithm \algo{GeneralizedNext} to $t_i$.  Since the claim
in the lemma holds for $t_i$ (by induction hypothesis), and since the
initialization step of algorithm \algo{GeneralizedNext} replaces the
state label of every node $v$ with $\delta(\lambda_i(v), \sigma)$, it
follows that the state label of every non-leaf node continues to be
the union of state labels of its children even after the
initialization step.  Since no nodes are added or deleted, and the
state labels of no nodes are changed in Steps (\ref{Nxt3}) and
(\ref{Nxt4}) of algorithm \algo{GeneralizedNext} (i.e., during
name-compaction and updation of component $f$), the inductive claim
can be proved by establishing that Step (\ref{Nxt2}) of algorithm
\algo{GeneralizedNext} does not violate the claim.  This amounts to
showing that algorithm \algo{GeneralizedNextRecursive} preserves the
property that the state label of every node is the union of state
labels of its children.  We therefore focus on the steps of algorithm
\algo{GeneralizedNextRecursive} below.

Clearly, Step (\ref{GNR1}) of algorithm
\algo{GeneralizedNextRecursive} preserves the desired property.
Although Step (\ref{GNR2}) results in the creation of a new child $v'$
of $v$, the desired property is preserved, since the state label of
$v'$ is set to that of $v$.  In Step (\ref{GNR3a}), new children may
be created for $v$, but the union of state labels of children of $v$
remains unchanged.  This is because for every new child $v'$ that is
created, Step (\ref{GNR3a}) sets the state label of $v'$ to $\{q\}$,
where $q$ is in the state label of an already existing child of $v$.
Step (\ref{GNR3b}) presents a more interesting situation.  Let $v_k$
be a child of $v$ such that the annotation on the edge from $v$ to
$v_k$ is $j_k$.  From Lemma (\ref{lemmaD}) and from the definition of
edge annotations, we know that $h_{i+1}(v) = h_{i+1}(v_k) \cup
\{j_k\}$.  If a state $q$ in the state label of $v_k$ is such that $q
\neq q_{j_k}$ and $q \notin Q_{h_{i+1}(v_k)}$, Step (\ref{GNR3b}) of
algorithm \algo{GeneralizedNextRecursive} removes $q$ from the state
label of $v_k$ and from the state labels of all its descendants.  This
can give rise to a situation wherein $q$ is in the state label of $v$
(parent of $v_k$) but not in the state label of any child of $v$,
potentially violating the property that the state label of every node
is the union of state labels of its children. However, such a
violation is only temporary and is rectified by the time the recursion
of algorithm \algo{GeneralizedNextRecursive} terminates.  To see why
this is so, notice that since $q \neq q_{j_k}$ and $q \notin
Q_{h_{i+1}(v_k)}$, we must have $q \notin Q_{h_{i+1}(v)} =
Q_{h_{i+1}(v_k)} \cup \{q_{j_k}\}$.  Hence, when the recursion of
algorithm \algo{GeneralizedNextRecursive} returns to the level where
the current node is the parent $u$ of node $v$ in $t_{i+1}$, we have
two possibilities.
\begin{enumerate}
\item Suppose $q = q_{r}$, where $r$ is the annotation of the edge
  from $u$ to $v$ in $t_{i+1}$.  In this case, Step (\ref{GNR3a}) of
  algorithm \algo{GeneralizedNextRecursive} creates a new child $v'$
  of $u$ with state label $\{q\}$, and with an edge annotation that is
  smaller than $r$.  This eventually causes $q$ to be removed from the
  state label of $v$ in Step (\ref{GNR3a}) of algorithm
  \algo{GeneralizedNextRecursive}.

\item Suppose $q \neq q_{r}$, where $r$ is the annotation of the edge
  from $u$ to $v$ in $t_{i+1}$.  Since $q \notin Q_{h_{i+1}(v)}$ as
  well, Step (\ref{GNR3b}) of algorithm
  \algo{GeneralizedNextRecursive} removes $q$ from the state label of
  $v$.
\end{enumerate}
Therefore, if $q$ is removed from the state label of a child $v_k$ of
$v$ by Step (\ref{GNR3b}) of algorithm
\algo{GeneralizedNextRecursive}, then it is also eventually removed
from the state label of $v$.  This ensures that the desired property
of the state label of a node being the union of state labels of its
children is eventually preserved.  Step (\ref{GNR4}) of algorithm
\algo{GeneralizedNextRecursive} can remove a state $q$ from the state
label of a node, but only if $q$ is also present in the state label of
a sibling node.  Hence, Step (\ref{GNR4}) cannot change the union of
state labels of children of a node.  Step (\ref{GNR5}) deletes nodes
with an already empty state label, while Steps (\ref{GNR6}) and (\ref{GNR7})
do not modify the state label of any node.  Step (\ref{GNR6}) can
cause a non-leaf node to turn into a leaf node, but this does not
affect the desired property, which relates only to non-leaf nodes.

Thus, if algorithm \algo{GeneralizedNextRecursive} is invoked on a
tree in which the state label of every node is the union of state
labels of its children, the algorithm preserves this property after it
has transformed the tree recursively.  This, coupled with the
inductive hypothesis, implies that $t_{i+1}$ satisfies the
inductive claim.

Therefore, by the principle of mathematical induction, the claim in
the lemma holds for all $t_i$, where $i \geq 0$.  \qed

\CGS trees encountered along a run of $\dD$ have several interesting
properties that are useful in proving the correctness of our
construction.  We will prove these propeties below by considering an
arbitrary run $\rho = t_0 t_1 t_2 \ldots$ of $\dD$ and by inductively
showing that the respective properties hold for every \CGS tree $t_i$
along $\rho$.

\begin{proposition}\label{initialpath}
For every $i \geq 0$, for every $v \in N_i$ and for every $q \in
\lambda_{i}(v)$, there is a run of the automaton $\aA$ from some $q_0
\in Q_0$ to $q$ on the prefix $\alpha(0, i-1)$.
\end{proposition}
\noindent {\bf Proof:} We will prove this by induction on the indices
of $t_0, t_1, \ldots$.

\noindent \emph{Base Case:} For the tree $t_0$ with only the root node
$r_0$, the claim in the lemma holds trivially, since $\lambda_0(r_0) =
Q_0$ by definition.

\noindent \emph{Hypothesis:} We assume that the claim in the lemma
holds for $t_i$, where $i \geq 0$.

\noindent \emph{Induction:} Consider the tree $t_{i+1}$ obtained by
applying algorithm \algo{GeneralizedNext} to $t_i$.  We know from the
initialization step (Step (\ref{Nxt1}) of algorithm
\algo{GeneralizedNext} that the state label of $r_{i+1}$ is initially
set to $\delta(\lambda_i(r_i), \alpha_i)$.  We also know from the
pseudocode of algorithm \algo{GeneralizedNextRecursive} that invoking
this algorithm on a \CGS tree rooted at a node $v$ does not change the
state label of $v$.  Since Step (\ref{Nxt2}) of algorithm
\algo{GeneralizedNext} invokes algorithm
\algo{GeneralizedNextRecursive} on the \CGS tree rooted at $r_{i+1}$,
the state label of $r_{i+1}$ remains unchanged at
$\delta(\lambda_i(r_i), \alpha_i)$ after the call to
\algo{GeneralizedNextRecursive} returns.  Subsequently, neither Step
(\ref{Nxt3}) nor Step (\ref{Nxt4}) of algorithm \algo{GeneralizedNext}
changes the state label of any node in $t_{i+1}$.  Therefore,
$\lambda_{i+1}(r_{i+1}) = \delta(\lambda_i(r_i), \alpha_i)$.  Now let
$v$ be an arbitrary node in $t_{i+1}$ and let $q \in
\lambda_{i+1}(v)$.  By Lemma (\ref{lemmaC}), we know that $q \in
\lambda_{i+1}(r_{i+1}) = \delta(\lambda_i(r_i), \alpha_i)$.  By the
inductive hypothesis, for every $q' \in \lambda_i(r_i)$, there is a
run of $\aA$ from some $q_0 \in Q_0$ to $q'$ on the prefix $\alpha(0,
i-1)$.  Therefore, there is a run of $\aA$ from some $q_0 \in Q_0$ to
$q \in \delta(\lambda_i(r_i), \alpha_i)$ on the prefix $\alpha(0, i)$.

By the principle of mathematical induction, the claim in the lemma
holds for all $i \geq 0$. \qed

\begin{lemma}\label{lemmaA}
For every $i \geq 0$ and for every $v \in N_i$ such that $v$ is a
non-leaf node of $t_i$, we have $h_i(v) \neq \emptyset$.
\end{lemma}
\noindent {\bf Proof:} From Lemma (\ref{lemmaD}) and Step (\ref{GNR6})
of algorithm \algo{GeneralizedNextRecursive}, it follows that if $v$
is a non-leaf node of $t_i$, it must have a child $v'$ such that
$h_i(v')$ is a \emph{strict subset} of $h_i(v)$.  This immediately
implies that $h_i(v) \neq \emptyset$. \qed

\begin{lemma}
\label{lemmaB}
Let $m = |Q|^2 + |Q| + 1$.  For every $i \geq 0$, if $f_i < m+1$,
there exists a leaf node $v$ in $t_i$ with name $M_i(v) = f_i$ such
that $h_i(v) \neq \emptyset$.
\end{lemma}
\noindent {\bf Proof: } 
We will prove the lemma by induction on the indices of $t_0, t_1 \ldots$.

\noindent \emph{Base Case:} For the \CGS tree $t_0$ with only the root
node $r_0$, the claim in the lemma holds vacuously since $f_0 = m+1$.

\noindent \emph{Hypothesis :} We assume that the claim in the lemma
holds for $t_i$, where $i \geq 0$.

\noindent \emph{Induction :} Consider the \CGS tree $t_{i+1}$ obtained
by applying algorithm \algo{GeneralizedNext} to $t_i$.  The value of
$f_{i+1}$ is set in Step (\ref{Nxt4}) of algorithm
\algo{GeneralizedNext} to the smaller of $m+1$ and the smallest name
of a node added to the set $U$ in Step (\ref{GNR6}) of algorithm
\algo{GeneralizedNextRecursive}.  Therefore, if $f_{i+1} < m+1$, a
node $v$ with $M_{i+1}(v) = f_{i+1}$ must have been added to the set
$U$ in Step (\ref{GNR6}) of a recursive call of algorithm
\algo{GeneralizedNextRecursive}.  Furthermore, $v$ must have been the
root node of the \CGS sub-tree transformed by this specific recursive
call.  The condition in Step (\ref{GNR6}) of algorithm
\algo{GeneralizedNextRecursive} requires that all children of $v$ must
have their hope set equal to $h_{i+1}(v)$ (or alternatively, the
annotations on all edges from $v$ to its children must be $0$).
Therefore, $v$ must have been a non-leaf node prior to being ``reset''
in Step (\ref{GNR6}) of algorithm \algo{GeneralizedNextRecursive}.  We
now consider two cases below depending on whether the node $v$ was
present in $t_i$ or not, and show that $h_{i+1}(v) \neq \emptyset$ in
both cases.
\begin{itemize}
\item Suppose $v \in N_i \cap N_{i+1}$.  By the argument given in the
  proof of Lemma (\ref{lemmaD}), we know that $h_i(v) = h_{i+1}(v)$.
  If $v$ was a non-leaf node in $t_i$, by Lemma (\ref{lemmaA}),
  $h_i(v) \neq \emptyset$.  Hence, $h_{i+1}(v) \neq \emptyset$ as
  well.  If $v$ was a leaf node in $t_i$, we could either have $h_i(v)
  = \emptyset$ or $h_i(v) \neq \emptyset$.  In the latter case, we
  easily get $h_{i+1}(v) = h_i(v) \neq \emptyset$.  In the former
  case, we note that $v$ cannot become a non-leaf node prior to Step
  (\ref{GNR6}) of algorithm \algo{GeneralizedNextRecursive} in the
  process of transforming $t_i$ to $t_{i+1}$.  This is because Step
  (\ref{GNR1}) of algorithm \algo{GeneralizedNextRecursive} prevents
  any children from being added to $v$ if $h_{i}(v) = \emptyset$.
  Therefore, $h_i(v)$ must have been non-empty in $t_i$, and the claim
  in the lemma follows.
\item If $v$ is newly created in the process of transforming $t_i$ to
  $t_{i+1}$, then by the argument used in the proof of Lemma
  (\ref{lemmaD}), $v$ must have been created either in Step
  (\ref{GNR2}) or in Step (\ref{GNR3a}) of algorithm
  \algo{GeneralizedNextRecursive}.  If $v$ was created as a leaf node
  in Step (\ref{GNR3a}), it could not have become a non-leaf node
  prior to execution of Step (\ref{GNR6}). This is because algorithm
  \algo{GeneralizedNextRecursive} is not called recursively on any
  leaf node created in Step (\ref{GNR3a}).  If $v$ was created as a
  leaf node in Step (\ref{GNR2}), the only way it could have become a
  non-leaf node prior to execution of Step (\ref{GNR6}) is by a
  recursive invokation of algorithm \algo{GeneralizedNextRecursive} on
  this node in Step (\ref{GNR3}).  However, Step (\ref{GNR1}) of
  algorithm \algo{GeneralizedNextRecursive} ensures that such a
  recursive invokation adds a child to $v$ only if the hope set of $v$
  is non-empty.  Therefore, we must have $h_{i+1}(v) \neq \emptyset$.
\end{itemize}

Since node $v$ is ``reset'' and all descendants of $v$ are deleted in
Step (\ref{GNR6}) of algorithm \algo{GeneralizedNextRecursive}, $v$
becomes a leaf node at the end of Step (\ref{GNR6}).  Furthermore,
since $t_{i}$ and $t_{i+1}$ are trees, every node has a unique parent
in $t_i$ and $t_{i+1}$, and hence, algorithm
\algo{GeneralizedNextRecursive} is recursively invoked at most once on
a node in Step (\ref{GNR3}).  It follows that after node $v$ is
``reset'' and turned into a leaf by a recursive call of algorithm
\algo{GeneralizedNextRecursive}, there are no subsequent recursive
calls to \algo{GeneralizedNext} with $v$ as the root of a \CGS subtree
to be transformed.  From the pseudocode of algorithm
\algo{GeneralizedNext}, we note that this implies that no child gets
added to $v$ after it is ``reset''.  Therefore, $v$ either remains as
a leaf node in $t_{i+1}$ or is subsequently deleted in the process of
transforming $t_i$ to $t_{i+1}$.  However, since $f_{i+1}$ is set to
$M_{i+1}(v)$, we know from Step (\ref{Nxt4}) of algorithm
\algo{GeneralizedNext} that $v$ is present in $N_{i+1}$.  Therefore,
$v$ is a leaf node in $t_{i+1}$ with $M_{i+1}(v) = f_{i+1}$ and
$h_{i+1}(v) \neq \emptyset$.

By the principle of mathematical induction, the claim in the lemma
holds for all $t_i$, where $i \geq 0$.  \qed

\begin{lemma}\label{claim1} Let $\alpha$ be an $\omega$-word and 
let $\rho = t_0 t_1 \ldots$ be the unique run of $\dD$ on $\alpha$.
Let $i, k$ be indices and let $v$ be a node such that: (i) $i < k$,
(ii) for all $z \in \{i, i+1, \ldots k\}$, node $v$ is present in
$t_z$ and $h_z(v) \neq \emptyset$, and (iii) node $v$ is a leaf in
both $t_i$ and $t_{k}$, and is a non-leaf node in all $t_z$, where $i
< z < k$.  Then the following claims hold.
 \begin{enumerate}
   \item \label{claim1a} Node $v$ is ``reset'' in the process of
     transforming $t_{k-1}$ to $t_{k}$.
   \item \label{claim1b} For every $q' \in \lambda_{k}(v)$, there is a
     $q \in \lambda_i(v)$ such that there is a run $\psi$ of $\aA$ on
     $\alpha(i, k-1)$ with $\psi(0) = q$, $\psi(k-i) = q'$ and
     $\psi(z-i) \in \lambda_z(v)$ for all $z \in \{i, i+1, \ldots
     k\}$.
   \item \label{claim1c} For every run $\psi$ of $\aA$ on the word
     segment $\alpha(i, k-1)$ such that $\psi(z-i) \in \lambda_z(v)$
     for all $z \in \{i, i+1, \ldots k\}$, all states in $Q_{h_i(v)}$
     are visited in $\psi$.
 \end{enumerate}
\end{lemma}
\noindent {\bf Proof:} 
\begin{enumerate}
\item We will prove this claim by contradiction.  If possible, suppose
  $v$ becomes a leaf node in $t_k$ without being ``reset'' in the
  process of transforming $t_{k-1}$ to $t_k$.  Consider the case when
  $k = i+1$.  Since $v$ is a leaf in $t_i$ and $h_i(v) \neq
  \emptyset$, Step (\ref{GNR2}) of algorithm
  \algo{GeneralizedNextRecursive} creates at least one child of $v$
  with the same non-empty state label as that of $v$ when
  \algo{GeneralizedNextRecursive} is invoked with $v$ as the root of
  the \CGS subtree to be transformed.  If $k > i+1$, then since $v$ is
  a non-leaf node in $t_{k-1}$, there is at least one child of $v$
  with a non-empty state label in $t_{k-1}$.  By Lemma (\ref{lemmaC}),
  the state label of $v$ in this case is also the union of state
  labels of its children in $t_{k-1}$.  Thus, in either case, there is
  an intermediate step during the transformation of $t_{k-1}$ to $t_k$
  when $v$ has one or more children with non-empty state labels, and
  the union of state labels of its children equals the state label of
  $v$.  All these children must eventually be deleted before $v$
  becomes a leaf node in $t_{k}$.

  From the pseudocode of algorithm \algo{GeneralizedNextRecursive}, we
  note that the only steps that delete nodes from a \CGS tree are Step
  (\ref{GNR5}) and Step (\ref{GNR6}).  Since $v$ exists in $t_k$ and
  is assumed not to have been ``reset'' in the process of transforming
  $t_{k-1}$ to $t_k$, its children could not have been deleted in Step
  (\ref{GNR6}).  Therefore, all its children must have been deleted in
  Step (\ref{GNR5}) of algorithm \algo{GeneralizedNextRecursive}.
  This requires all children of $v$ to acquire the empty state label.
  We know from above that there exist one or more children of $v$ with
  non-empty state labels in an intermediate step during the
  transformation of $t_{k-1}$ to $t_k$.  The state labels of all such
  children must therefore be emptied before they can be deleted in
  Step (\ref{GNR5}).  From the pseudocode of algorithm
  \algo{GeneralizedNextRecursive}, the only steps that remove states
  from the state labels of nodes are Step (\ref{GNR3b}) and Step
  (\ref{GNR4}).  Unfortunately, State (\ref{GNR4}) simply removes
  duplicates from the state labels of siblings, and cannot render the
  state labels of all children of $v$ empty.  Therefore, Step
  (\ref{GNR3b}) must eventually be responsible for emptying the state
  labels of all children of $v$.  However, we know from the proof of
  Lemma (\ref{lemmaC}) that if a state is removed from the state label
  of a child of $v$ in Step (\ref{GNR3b}) of algorithm
  \algo{GeneralizedNextRecursive}, then that state is eventually
  removed from the state label of $v$ as well.  Since the state label
  of $v$ equals the union of state labels of all its children at an
  intermediate step in the transformation of $t_{k-1}$ to $t_k$, the
  above implies that all states in the state label of $v$ must
  eventually be removed in the process of transforming $t_{k-1}$ to
  $t_k$.  This, in turn, implies that $v$ is removed from $t_k$ in
  Step (\ref{GNR5}) of algorithm \algo{GeneralizedNextRecursive} -- a
  contradiction!

\item Since node $v$ is present in all $t_z$, for $z \in \{i, i+1,
  \ldots k\}$, it follows from Step (\ref{Nxt1}) that $\lambda_r(v)$
  is always initialized to $\delta(\lambda_{r-1}(v), \alpha_{r-1})$,
  for $r \in \{i+1, \ldots k\}$.  Since no other step of algorithm
  \algo{GeneralizedNext} or algorithm \algo{GeneralizedNextRecursive}
  adds states to the state label of an already existing node, the
  claim now follows from an easy induction on $z \in \{i, i+1, \ldots
  k\}$.

\item From the pseudocodes of algorithms \algo{GeneralizedNext} and
  \algo{GeneralizedNextRecursive}, we note that since node $v$ exists
  in $t_z$ for all $z \in \{i, i+1, \ldots k\}$, the hope set of $v$
  must stay unchanged, i.e., $h_i(v) = h_z(v)$ for all $z \in \{i,
  i+1, \ldots k\}$.  Now let $r$ be an arbitrary index such that $ i
  \le r < k$.  Suppose node $v$ has a child $v'$ in a (possibly
  intermediate) step of algorithm \algo{GeneralizedNextRecursive}
  during the transformation of $t_r$ to $t_{r+1}$.  Suppose further
  that the edge from $v$ to $v'$ is annotated with $j$ and the state
  label of $v'$ is $S$ in this step.  We will first prove the
  following claim.

  \emph{{\bfseries Claim 1:} For every run $\psi$ of $\aA$ on
    $\alpha(i, r)$ such that $\psi(z-i) \in \lambda_z(v)$ for all $z
    \in \{i, \ldots r-1\}$ and $\psi(r-i) \in S$, all states in
    $\{q_n, q_{n-1}, \ldots, q_{j+1}\} \cap Q_{h_i(v)}$ are visited
    in $\psi$.}

    The proof is by induction on $r$.

    \noindent \emph{Base Case}: We know that $v$ is a leaf node in
    $t_i$ with $h_i(v) \neq \emptyset$.  Therefore, during the
    transformation of $t_i$ to $t_{i+1}$, Step (\ref{GNR2}) of
    algorithm \algo{GeneralizedNextRecursive} creates a child $v'$ of
    $v$ and adds all states in $\delta(\lambda_i(v), \alpha_i)$ to the
    state label of $v'$.  In addition, the edge from $v$ to $v'$ is
    annotated with $j = \max(h_i(v)) > 0$.  This implies that $\{q_n,
    q_{n-1}, \ldots, q_{j+1} \} \cap Q_{h_i(v)} =\emptyset$.  Hence,
    the claim follows vacuously.

    Suppose additional children of $v$ are subsequently created in
    Step (\ref{GNR3a}) of algorithm \algo{GeneralizedNextRecursive}.
    Since $v$ is a leaf in $t_i$, it can be seen from the pseudocode
    of algorithm \algo{GeneralizedNextRecursive} that prior to
    execution of Step (\ref{GNR3a}), $v$ could have had only a single
    child -- the one created in Step (\ref{GNR2}), with the edge from
    $v$ to this child annotated with $j = \max(h_i(v))$.  In order for
    a new child of $v$, say $v''$, to be created in Step
    (\ref{GNR3a}), we note from the pseudocode of algorithm
    \algo{GeneralizedNextRecursive} that the state label of $v''$ must
    be $\{q_{j}\}$ and the annotation of the edge from $v$ to $v''$
    must be $l = \max(\{0\} \cup (h_i(v) \cap \{1, 2, \ldots j-1\}))$.
    Since $j = \max(h_i(v))$, it follows that $\{q_n, q_{n-1}, \ldots,
    q_{l+1}\} \cap Q_{h_i(v)} = \{q_j\}$.  Since the state label of
    $v''$ is also $\{q_j\}$, the claim is easily seen to hold for
    $v''$.  Since no other step of algorithm
    \algo{GeneralizedNextRecursive} or algorithm
    \algo{GeneralizedNext} adds any state to the state label of $v''$,
    this proves the base case of the induction.

    \noindent \emph{Hypothesis}: We assume that the claim is true
    for $r$, where $i \le r < k-1$.

    \noindent \emph{Induction}: Consider the transformation of
    $t_{r+1}$ to $t_{r+2}$.  Let $v'$ be a child of $v$ in some step
    of algorithm \algo{GeneralizedNextRecursive} during this
    transformation.  Suppose further that the edge from $v$ to $v'$ is
    annotated with $j$ and the state label of $v'$ is $S$ in
    this step.  We consider two cases below.
    \begin{itemize}
    \item If $v'$ is present in $t_{r+1}$, then by the argument used
      in the proof of Lemma (\ref{lemmaD}), $v$ must also have been
      present in $t_{r+1}$, with $h_{r+1}(v) = h_{r+2}(v)$ and
      $h_{r+1}(v') = h_{r+2}(v')$.  Therefore, by the definition of
      edge annotations, the edge from $v$ to $v'$ must have been
      annotated with $j$ in $t_{r+1}$ as well.  Step (\ref{Nxt1}) of
      algorithm \algo{GeneralizedNext} ensures that the state label of
      $v'$ is initialized to $\delta(\lambda{r+1}(v), \alpha_{r+1})$
      during the transformation of $t_{r+1}$ to $t_{r+2}$.  This,
      along with the inductive hypothesis, and the facts that
      $h_{r+1}(v) = h_{r+2}(v)$ and the edge annotations from $v$ to
      $v'$ are the same in $t_{r+1}$ and in $t_{r+2}$, imply that the
      claim holds for $v'$ after the initialization step during the
      transformation of $t_{r+1}$ to $t_{r+2}$.  Since no other step
      of algorithm \algo{GeneralizedNextRecursive} or algorithm
      \algo{GeneralizedNext} adds any state to the state label of
      $v'$, this proves the inductive claim for $v'$.

    \item If $v'$ is not present in $t_{r+1}$, it must have been
      created as a child of $v$ in Step (\ref{GNR2}) or in Step
      (\ref{GNR3a}) of algorithm \algo{GeneralizedNextRecursive}
      during the transformation of $t_{r+1}$ to $t_{r+2}$.  Since $i <
      r+1 < k$ (by the condition in our inductive hypothesis), we know
      that $v$ is a non-leaf node in $t_{r+1}$.  Therefore, $v'$ could
      not have been created in Step (\ref{GNR2}) of algorithm
      \algo{GeneralizedNextRecursive} (this step requires $v$ to be a
      leaf node in $t_{r+1}(v)$).  Hence, $v'$ must have been created
      in Step (\ref{GNR3a}).

      From the pseudocode of algorithm
      \algo{GeneralizedNextRecursive}, we note that when $v'$ is
      created as a child of $v$ in Step (\ref{GNR3a}), the state label
      of $v'$ is set to $\{q_{j_x}\}$, where $j_x$ is the annotation
      of the edge from $v$ to an already existing child $v_x$, and
      $q_{j_x}$ is in the state label of $v_x$ at the time of creation
      of $v'$.  In addition, the annotation of the new edge from $v$
      to $v'$ is set to $l = \max(\{0\} \cup (h_{r+2}(v) \cap \{1, 2,
      \ldots j_x-1\}))$.  Since $v$ is a non-leaf node in $t_{r+1}$,
      the child $v_x$ itself could not have been created in Step
      (\ref{GNR2}) of algorithm \algo{GeneralizedNextRecursive} during
      the transformation of $t_{r+1}$ to $t_{r+2}$.  It could not have
      been created in Step (\ref{GNR3a}) of algorithm
      \algo{GeneralizedNextRecursive} either, since Step (\ref{GNR32})
      of algorithm \algo{GeneralizedNextRecursive} iterates over the
      children of $v$ existing prior to execution of Step
      (\ref{GNR3}).  Therefore, the child $v_x$ of $v$ must be present
      in $t_{r+1}$.

      Since $v$ and $v_x$ are present in both $t_{r+1}$ and in the
      intermediate \CGS tree at the time of creation of $v'$, the hope
      sets of $v$ and $v_x$, and the annotation of the edge from $v$
      to $v_x$ must be the same in $t_{r+1}$ and in the intermediate
      \CGS tree.  This implies that the edge from $v$ to $v_x$ is
      annotated with $j_x$ in $t_{r+1}$.  By virtue of Step
      (\ref{Nxt1}) of algorithm \algo{GeneralizedNext}, we also know
      that there is a state $q' \in \lambda_{r+1}(v_x)$ such that
      $q_{j_x} \in \delta(q', \alpha_{r+1})$.  This, along with the
      inductive hypothesis, and the facts that $h_{r+1}(v) =
      h_{r+2}(v)$ and the annotation of the new edge from $v$ to $v'$
      is $l = \max(\{0\} \cup (h_{r+2}(v) \cap \{1, 2, \ldots
      j_x-1\}))$, imply that for every run $\psi$ of $\aA$ on
      $\alpha(i, r+1)$ such that $\psi(z-i) \in \lambda_z(v)$ for $z
      \in \{i, \ldots r\}$ and $\psi(r+1-i) = q_{j_x}$, all states in
      $\{q_n, q_{n-1}, \ldots, q_{l + 1}\} \cap Q_{h_{r+2}(v)}$ are
      visited.

      From the pseudocode of algorithm
      \algo{GeneralizedNextRecursive}, no step other than Step
      (\ref{GNR3a}) adds any state to the state label of $v'$ after it
      is created in Step (\ref{GNR3a}).  Therefore, $v'$ has at most
      one state, $q_{j_x}$, in its state label in any intermediate
      step of algorithm \algo{GeneralizedNextRecursive} during the
      transformation of $t_{r+1}$ to $t_{r+2}$.  We have already
      considered the case of $q_{j_x}$ in the state label of $v'$
      above.  Hence, this proves the inductive claim for $v'$ and
      also completes the proof of Claim 1.
    \end{itemize}

    To complete the proof of Lemma (\ref{claim1}-\ref{claim1c}), we
    note from Lemma (\ref{claim1}-ref{claim1a}) that $v$ is ``reset''
    during the transformation of $t_{k-1}$ to $t_k$.  Therefore, from
    Step (\ref{GNR6}) of algorithm \algo{GeneralizedNextRecursive},
    $v$ must have had at least one child with non-empty state label
    prior to being ``reset''.  In addition, the annotations of all
    edges from $v$ to its children with non-empty state labels must
    have been $0$ prior to the resetting of $v$.  It then follows from
    Claim 1 that for every run $\psi$ of $\aA$ such that $\psi(z-i)
    \in \lambda_z(v)$ for all $z \in \{i, k-1\}$ and $\psi(k-i)$ is in
    the state label of some child of $v$ prior to it being reset, all
    states in $\{q_n, \ldots q_1\} \cap Q_{h_i(v)} = Q_{h_i(v)}$ are
    visited in $\psi$.  

    This does not prove Lemma (\ref{claim1}-\ref{claim1c}) yet, since
    we must show the above result for $\psi(k-i) \in \lambda_{k}(v)$.
    We have seen earlier, in the proof of Lemma (\ref{lemmaC}), that
    the state label of a node $v$ may temporarily contain states that
    are not in the state labels of any of its children after
    intermediate steps of algorithm \algo{GeneralizedNextRecursive}.
    However, we also saw in the same proof that all such states are
    eventually removed from the state label of $v$ after all recursive
    invokations of \algo{GeneralizedNextRecursive} have returned.
    Therefore, proving the claim of Lemma (\ref{claim1}-\ref{claim1c})
    for $\psi(k-i)$ in the state labels of children of $v$ prior to
    $v$ being ``reset'' proves Lemma (\ref{claim1}-\ref{claim1c})
    itself. 
\end{enumerate}
\qed


\begin{lemma}\label{claim2} Let $\alpha$ be an $\omega$-word and 
let $\rho = t_0 t_1 \ldots$ be the unique run of $\dD$ on $\alpha$.
For every $i \ge 0$ and for every node $v$ in $t_i$, $\lambda_i(v)
\subseteq Q_{h_i(v)}$.
\end{lemma}

\noindent {\bf Proof:} We will prove this claim by contradiction.
Suppose there exists an $i \ge 0$ and a node $v$ in $t_i$ such that
$q_{l} \in \lambda_i(v)$ although $l \notin h_i(v)$.  Clearly, $v$
cannot be the root, $r_i$, of $t_i$, since $h_i(r_i) (= h_0[r_0] =
[n])$ contains the indices of all states of $\aA$.  Therefore, $v$
must have a parent, say $u$, in $t_i$.  Recalling that $t_0$ has only
a single node (i.e., $r_0$) without any parent, we can immediately
infer that $i > 0$.  In other words, there exists a \CGS tree
$t_{i-1}$ such that $t_i$ is obtained by applying algorithm
\algo{GeneralizedNext} to $t_{i-1}$.

From the pseudocode of algorithm \algo{GeneralizedNextRecursive}, we
observe that during the transformation of $t_{i-1}$ to $t_i$, the only
nodes in $t_i$ on which the recursive algorithm \algo{GeneralizedNextRecursive} is
not recursively invoked are those that are generated in Step
(\ref{GNR3a}).  Furthermore, every node generated in Step
(\ref{GNR3a}) is either deleted or survives as a leaf in the
transformation of $t_i$ to $t_{i+1}$.  Since node $u$ is a non-leaf
node in $t_i$, algorithm \algo{GeneralizedNextRecursive} must have
been invoked with $u$ as the root of the \CGS subtree to be
transformed, during the transformation of $t_{i-1}$ to $t_i$.

Let $j$ be the annotation of the edge from $u$ to $v$ in $t_i$.  There
are two possibilities that we consider separately below.
\begin{itemize}
\item Suppose $v$ is created during the transformation of $t_{i-1}$ to
  $t_i$.  This can happen either in Step (\ref{GNR2}) or in Step
  (\ref{GNR3a}) of the recursive invokation of algorithm
  \algo{GeneralizedNextRecursive} with $u$ as the root of the \CGS
  subtree to be transformed.  
  \begin{itemize}
    \item If $v$ is created in Step (\ref{GNR3a}), it follows from the
      pseudocode of algorithm \algo{GeneralizedNextRecursive} that
      $\lambda_i(v) = \{q_l\}$, where $l (> 0)$ is the annotation of
      an edge from $u$ to an already existing child, say $v''$, of
      $u$.  In addition, $h_i(v)$ is set to $h_i(u) \setminus
      \{\max\left((h_i(u) \cup \{0\}) \cap \{0, 1, 2, \ldots,
      l-1\}\right)\}$.  By the definition of edge annotations, $l \in
      h_i(u) \setminus h_i(v'')$ and hence $l \in h_i(u)$.  It then
      follows that $l \in h_i(u) \setminus \{\max\left((h_i(u) \cup
      \{0\}) \cap \{0, 1, 2, \ldots, l-1\}\right)\} = h_i(v)$ as well.
      Therefore, $\lambda_i(v) \subseteq Q_{h_i(v)}$.  Since no other
      step of algorithm \algo{GeneralizedNextRecursive} adds any state
      to $\lambda_i(v)$ subsequently, we have $\lambda_i(v) \subseteq
      Q_{h_i(v)}$.  This gives us a contradiction!
    \item If $v$ is created in Step (\ref{GNR2}), then Step
      (\ref{GNR3a}) must subsequently be executed in the same
      recursive invokation of \algo{GeneralizedNextRecursive} with $u$
      as the root of the \CGS subtree to be transformed.  This is
      similar to the case considered below wherein $v$ exists in
      $t_{i-1}$, and Step (\ref{GNR3a}) is executed in the recursive
      invokation of \algo{GeneralizedNextRecursive} with $u$ as the
      root of the \CGS subtree to be transformed.
  \end{itemize}
  \item Suppose $v$ exists in $t_{i-1}$.  It follows that the parent,
    $u$, of $v$ must also exist in $t_{i-1}$.  Consider Step
    (\ref{GNR3a}) in the recursive invokation of algorithm
    \algo{GeneralizedNextRecursive} with $u$ as the root of the \CGS
    subtree, during the transformation of $t_{i-1}$ to $t_i$.  We
    have two sub-cases to consider.
    \begin{itemize}
    \item If $j = l$, a new child, say $v'$, of $u$ is been created in
      Step (\ref{GNR3a}), the state label of $v'$ is set to $\{q_l\}$
      and the edge from $u$ to $v'$ is annotated with an index $< l$.
      This implies that in Step (\ref{GNR4}) of algorithm
      \algo{GeneralizedNextRecursive}, $q_l$ is removed from the state
      label of $v$.  Since no other step of algorithm
      \algo{GeneralizedNextRecursive} adds states to $\lambda_i(v)$
      subsequently, it follows that $q_l \notin \lambda_i(v)$.
      This gives a contradiction!
    \item Suppose $j \neq l$.  Since $l$ is also not in $h_i(v)$, it
      follows that in Step (\ref{GNR3b}) of the recursive invokation
      of \algo{GeneralizedNextRecursive} with $u$ as the root of the
      \CGS subtree to be transformed, $q_l$ is removed from
      $\lambda_i(v)$.  By the same argument used above, $q_l$ cannot
      be subsequently added to $\lambda_i(v)$.  Hence, $q_l \notin
      \lambda_k(v)$ -- a contradiction again!
    \end{itemize}
\end{itemize}

We have therefore shown that there is no $i \ge 0$ and no node $v$ in
$t_i$ such that $q_{l} \in \lambda_i(v)$ and $l \notin h_i(v)$.
\qed

Armed with the above properties of \CGS trees encountered along a run
of $\dD$, we will now show that the languages accepted by $\dD$ and
$\aA$ are the same.  As before, let $\alpha$ be an $\omega$-word in
$L(\dD)$ and let $\rho = t_0 t_1 \ldots$ be the unique run of $\dD$ on
$\alpha$.  By definition of the acceptance condition for $\dD$, there
exists an even index $2a+2$, where $0 \le 2a+2 < 2m-1$, such that \CGS
trees from the parity acceptance set $F_{2a+2}$ are seen infinitely
often along $\rho$, while \CGS trees from all parity acceptance sets
$F_y$, where $0 \le y < 2a+2$, are seen only finitely often along
$\rho$.  Let $i^*$ be the smallest index $(\ge 0)$ such that all \CGS
trees $t_i$ for $i > i^*$ are outside $\bigcup_{0 \le y < 2a+2} F_y$.
The following lemma describes important properties of the suffix $t_i,
t_{i+1}, \ldots$ of $\rho$, where $i > i^*$.

\begin{lemma}\label{greentogreen}
Let $i$ and $i'$ be indices such that (i) $0 \le i^* < i < i'$, (ii)
both $t_i$ and $t_{i'}$ are in $F_{2a+2}$, and (iii) $t_z \notin
F_{2a+2}$ for all $z \in \{i+1, \ldots i'-1\}$.  Then there exists a
node $v$ such that the following hold.
\begin{enumerate}
\item $v$ is present in $t_z$ for all $z \in \{i, i+1, \ldots i'\}$.
  In addition, $M_z(v) = a+2$ and $h_z(v) = h_i(v) \neq \emptyset$ for
  all $z \in \{i, i+1, \ldots i'\}$.
  
\item $v$ is a non-leaf node in $t_z$, for all $z \in \{i+1, \ldots
  i'-1\}$.

\item For every state $q' \in \lambda_{i'}(v)$, there is some state $q
  \in \lambda_i(v)$ such that there is a run of $\aA$ from $q$ to $q'$
  on $\alpha(i, i'-1)$ that visits all and only states in
  $Q_{h_i(v)}$.
\end{enumerate}
\end{lemma}

\noindent {\bf Proof:}
\begin{enumerate}
\item Since both $t_i$ and $t_{i'}$ are in $F_{2a+2}$, it follows from
  the definition of even-indexed parity acceptance sets that $f_i =
  f_{i'} = a+2$.  Also, since $0 \le 2a+2 < 2m-1$, we have $1 \le a+2
  \le m$.  Therefore, by Lemma (\ref{lemmaB}), both $t_i$ and $t_{i'}$
  contain a leaf node with name $a+2$ and with a non-empty hope set.

  Since $i^* < i < i'$, it follows from the definition of $i^*$ that
  for all $z \in \{i, i+1, \ldots i'\}$, the \CGS tree $t_z$ is not in
  $\bigcup_{0 \le x < 2a+2} F_{x}$.  Recalling the definitions of
  $F_{x}$ for odd and even indices $x$, we see that this implies $e_z
  > a+2$ for all $z \in \{i, i+1, \ldots i'\}$.  Hence no node with
  name $\leq a+2$ is removed in the process of transforming $t_i$ to
  $t_{i+1}$, $t_{i+1}$ to $t_{i+2}$, and so on until $t_{i'}$ is
  obtained.  Therefore, the node $v$ with name $a+2$ in $t_i$
  continues to be a part of all $t_z$, where $i \le z \le i'$.  Since
  $e_z > a+2$, the name-compaction step of algorithm
  \algo{GeneralizedNext} keeps the name of node $v$, i.e, $a+2$,
  unchanged in all of $t_z$.  Hence, node $v$ is present in $t_z$ and
  $M_z(v) = a + 2$, for all $z \in \{i, i+1, \ldots i'\}$.
  Furthermore, since $h_i(v) \neq \emptyset$ and since $v$ is not
  deleted in the sequence of transformations from $t_i$ to $t_{i'}$,
  it follows that $h_z(v) = h_i(v) \neq \emptyset$, for $i \leq z \leq
  i'$.

\item Consider an index $z$ such that $i < z < i'$.  If $v$ was a
  non-leaf node in $t_{z-1}$, then it starts off as a non-leaf node
  with at least one child having a non-empty state label when
  algorithm \algo{GeneralizedNextRecursive} is invoked on $t_{z-1}$ to
  transform it to $t_z$.  If $v$ was a leaf node in $t_{z-1}$ (as is
  the case when $z = i+1$, for example), then since $h_{z-1}(v) \neq
  \emptyset$ (by Lemma (\ref{greentogreen}-$1$) above), Step
  (\ref{GNR2}) of algorithm \algo{GeneralizedNextRecursive}) ensures
  that $v$ becomes a non-leaf node with at least one child having a
  non-empty state label in an intermediate step during the
  transformation of $t_{z-1}$ to $t_z$.  Thus, in either case, $v$
  becomes a non-leaf node with at least one child having a non-empty
  state label in some intermediate step of algorithm
  \algo{GeneralizedNextRecursive}.

  In order for $v$ to subsequently become a leaf node in $t_z$, all
  its children must be deleted.  Deletion of nodes can only happen in
  Step (\ref{GNR5}) or Step (\ref{GNR6}) of algorithm
  \algo{GeneralizedNextRecursive}.  We show that none of these steps
  can delete all children of $v$ in $t_z$.
  \begin {itemize}
    \item Since $v$ stays back in $t_z$ (by Lemma
      (\ref{greentogreen}-$1$) above), if the leaves of $v$ are
      deleted in Step (\ref{GNR6}) of algorithm
      \algo{GeneralizedNextRecursive}, $v$ must be ``reset'' and
      $M_z(v) = a+2$ must be added to $U$ (since $P_\phi(Q_{h_z(v)}) =
      P_\phi(Q_{h_i(v)}) = \true$) in Step (\ref{GNR6}).  Therefore,
      $f_z$ must be set to a value no larger than $a+2$ in Step
      (\ref{Nxt4}) of algorithm \algo{GeneralizedNext}.  Since $e_z >
      a+2$ (as shown in the proof of Lemma (\ref{greentogreen})-$1$),
      this would imply that $t_z \in F_x$, where $0 \le x \le 2a+2$.
      Recalling the definition of $i^*$, this contradicts the fact
      that $z > i > i^*$.
    \item If all leaves of $v$ are deleted in Step (\ref{GNR5}) of
      algorithm \algo{GeneralizedNextRecursive}, then the union of
      state labels of the children of $v$ must be empty at some
      intermediate step of algorithm \algo{GeneralizedNextRecursive}.
      We have seen above in the proof of Lemma (\ref{lemmaC}) that the
      state label of a node is eventually no larger than the union of
      state labels of its children at any intermediate step of
      algorithm \algo{GeneralizedNextRecursive}.  Therefore, if all
      leaves of $v$ are deleted in Step (\ref{GNR5}) of algorithm
      \algo{GeneralizedNextRecursive}, the state label of $v$ must
      eventually become empty in $t_z$.  However, $v$ must then be
      deleted from $t_z$ by Step (\ref{GNR5}) of algorithm
      \algo{GeneralizedNextRecursive}.  This contradicts Lemma
      (\ref{greentogreen}-$1$) proved above.
  \end{itemize}
  Therefore, $v$ must be a non-leaf node in $t_z$.

\item Lemma (\ref{greentogreen}-$3$) is an immediate consequence of
  Lemmas (\ref{greentogreen}-$1$), (\ref{greentogreen}-$2$),
  (\ref{claim1}-\ref{claim1b}), (\ref{claim1}-\ref{claim1c}) and
  (\ref{claim2}).

\end{enumerate}
\qed



\begin{lemma}\label{lemma1}
$L(\dD) \subseteq L(\aA)$.
\end{lemma}

\noindent {\bf Proof:} We will prove this lemma by constructing a
finitely branching infinite tree $K$ along the lines of Safra's proof
of correctness of his NSW determinization construction, and by showing
the existence of an infinite accepting path of $\aA$ in this tree.

The vertices of $K$ are elements of $\{\mathbf{r}\} \cup (Q \times
\mathbb{N})$, where $\mathbf{r}$ is a special vertex representing the
root of $K$.  For every $q_0 \in Q_0$, we draw an edge from
$\mathbf{r}$ to $(q_0, 0)$.  As defined earlier, let $i^*$ be the
minimum index after which no \CGS tree from $F_{x}$, for $x < 2a+2$,
is visited in the sequence $t_0, t_1, \ldots$.  Let $i_1$ be the
smallest index greater than $i^*$ such that $f_{i_1} = a+2$, and let
$v$ be the node in $t_{i_1}$ identified in Lemma
(\ref{greentogreen}-$1$).  From Lemma (\ref{greentogreen}-$1$), we
know that $M_{i}(v) = a+2$ and $h_i(v) = h_{i_1}(v) \neq \emptyset$
for all $i \geq i_1$.  For every state $q$ in $\lambda_{i_1}(v)$ we
add a vertex $(q, i_1)$ to the tree $K$.  For every such state $q$,
Proposition (\ref{initialpath}) tells us that there is a state $q_0
\in Q_0$ such that there is a run of $\aA$ from $q_0$ to $q$ on
$\alpha(0, i_1-1)$.  We add an edge from $(q_0, 0)$ to $(q, i_1)$ in
tree $K$ for every such $q_0 \in Q_0$ and $q \in \lambda_{i_1}(v)$.
Subsequently, we extend the tree $K$ inductively as follows. Given a
tree with a leaf $(q_k, i_z)$, where $q_k \in \lambda_{i_z}(v)$ and
$i_z \geq i_1$ is such that $f_{i_z} = a+2$, we find the smallest
$i_{z+1} > i_z$ such that $f_{i_{z+1}} = a+2$.  Since \CGS trees in
$F_{2a+2}$ are encountered infinitely often in $t_0, t_1, \ldots$ (by
the acceptance condition of $\dD$), such an $i_{z+1}$ always exists.
For every state $q' \in \lambda_{i_{z+1}}(v)$, we now add a vertex
$(q', i_{z+1})$ to the tree $K$.  By Lemma (\ref{greentogreen}-$3$),
there is a state $q$ in $\lambda_{i_z}(v)$ such that there is a run of
$\aA$ from $q$ to $q$ on $\alpha(i_z, i_{z+1}-1)$ that visits all and
only states in $Q_{h_{i_1}(v)}$.  For every such $q' \in
\lambda_{i_{z+1}}(v)$ and $q \in \lambda_{i_z(v)}$, we add an edge
from $(q, i_z)$ to $(q', i_{z+1})$ to extend the tree $K$.  It is
easy to see that $K$ is an infinite tree with the branching of each
node $(q, i_z)$ restricted by the cardinality of
$\lambda_{i_{z+1}}(v)$, i.e. $|Q|$.  Therefore, it follows from
K\"onig's lemma that there is an infinite path in $K$. 

From Proposition (\ref{initialpath}), every edge $((q_0, 0), (q',
i_1))$ corresponds to a run of $\aA$ on $\alpha(0, i_1 - 1)$ that
starts at $q_0$ and ends at $q'$. From Lemma (\ref{greentogreen}-$3$),
every edge $((q, i_z), (q',i_{z+1}))$ for $z \ge 1$ corresponds to a
run of $\aA$ on $\alpha(i_z, i_{z+1}-1)$ that starts at $q$ and ends
at $q'$ and visits all and only states in $Q_{h_{i_1}(v)}$.
Therefore, the infinite path in $K$ identified above corresponds to a
run $\rho$ of $\aA$ that starts from some $q_0 \in Q_0$ and eventually
visits all and only states in $Q_{h_{i_1}}(v)$.  In other words,
$\infi(\rho) = Q_{h_{i_1}(v)}$.  Furthermore, since $f_{i_1} = a+2$
and $M_{i_1}(v) = a+2$, we must have $P_\phi(Q_{h_{i_1}(v)}) =
\true$.  In other words, $\infi(\rho) \models \phi$, and hence $\rho$
is an accepting run of $\aA$. This implies $\alpha \in L(\aA)$.  \qed


\begin{lemma}\label{lemma2}
$L(\aA) \subseteq L(\dD)$.
\end{lemma}


\noindent {\bf Proof: } Consider an $\omega$-word $\alpha \in L(\aA)$.
Let $\psi = q_{k_0}, q_{k_1}, q_{k_2} \ldots$ be an accepting run of
$\aA$ on $\alpha$, and let $\rho = t_0, t_1, t_2 \ldots$ be the unique
run of $\dD$ on $\alpha$, where $t_i$ is the \CGS tree $(N_i, M_i,
r_i, p_i, l_i, h_i, f_i, e_i)$.  Consider the transformation of $t_i$
to $t_{i+1}$ by algorithm \algo{GeneralizedNext}.  Step (\ref{Nxt1})
of algorithm \algo{GeneralizedNext} updates the state label of $r_i$
to $\delta(\lambda_i(r_i), \alpha_i)$.  Subsequently, no step of
algorithm \algo{GeneralizedNext} or algorithm
\algo{GeneralizedNextRecursive} deletes any state from the state label
of $r_i$, deletes $r_i$, or adds $r_i$ as the child of any other node.
It therefore follows from an easy inductive argument that the root
$r_i$ of $t_i$ eventually survives as the root $r_{i+1}$ of $t_{i+1}$,
for all $i \geq 0$.  Since $M_0(r_0) = 1$ and $h_0(r_0) = [n]$, and
since every node in $t_i$ that is not deleted in transforming $t_i$ to
$t_{i+1}$ retains its name and hope set in $t_{i+1}$, we have
$M_{i+1}(r_{i+1}) = 1, h_{i+1}(r_{i_1}) = [n]$ and $e_{i+1} > 1$ for
all $i \geq 0$.  Also, by definition, $e_0 = m+1 > 1$.  Therefore,
$e_i > 1$ for all $i \geq 0$.

Let $J$ be the set of indices of all states in $\infi(\psi)$, i.e., $J
= \{j \mid q_{j} \in \infi(\psi)\}$.  Let $i_1$ be the smallest index
such that for all $i > i_1$, we have $q_{k_i} \in \infi(\psi)$.  We
wish to identify those nodes $v$ in $t_z$ that have $\psi(z) \in
\lambda_z(v)$, for all $z \ge i_l + 1$.  In other words, we wish to
identify nodes in the sequence of \CGS trees $t_{i_l + 1}, t_{i_l +
  2}, \ldots$ that track the run $\psi$ of $\aA$ from position $i_l +
1$ onwards.  

We have already seen above that $r_{0}$ survives as the root node in
all \CGS trees in $\rho$.  We also know that $\lambda_0(r_0) = Q_0$,
by definition.  Since Step (\ref{Nxt1}) of algorithm
\algo{GeneralizedNext} updates $\lambda_{i+1}(r_{i+1})$ to
$\delta(\lambda_i(r_i), \alpha_i)$ for all $i \ge 0$, and since no
subsequent step during the transformation of $t_i$ to $t_{i+1}$
deletes any state from the state label of the root $r_{i+1}$, it
follows from an easy inductive argument that $\psi(z) \in
\lambda_z(r_z)$, for all $z \ge 0$.

Now suppose the root node becomes a leaf infinitely often in
$\rho(i_1+1, \infty)$.  Let $j$ and $j'$ be arbitrary indices such
that $i_l + 1 \le j < j'$, and the root node is a leaf in $t_j$ and
$t_j'$, but not in any $t_z$, for $j < z < j'$.  Since we also know
that $h_{i}(r_i) = [n] \neq \emptyset$ for all $i \ge 0$, it follows
from Lemma (\ref{claim1}-\ref{claim1c}) and Lemma (\ref{claim2}) that
the set of states visited in $\psi(j, j')$ is exactly $Q_{h_i(r_i)} =
Q_{[n]}$.  By repeating the same argument for all successive pairs of
indices $j, j'$ such that $i_l + 1 \le j < j'$, and the root node is a
leaf in $t_j$ and $t_j'$, but not in any $t_z$ in between, we get
$\infi(\psi) = Q_{h_i(r_i)}$, for every $i > i_1$.  Since $\psi$ is an
accepting run of $\aA$, we also know that $P_\phi(\infi(\psi)) =
\true$.  This implies that $P_\phi(Q_{h_i(v)}) = \true$ for all those
indices $i > i_l$ where $r_i$ becomes a leaf node in $\rho(i_l + 1,
\infty)$.  By Lemma (\ref{claim1}-\ref{claim1a}), we know that $r_i$
is ``reset'' in these steps as well.  Hence $r_i$ is added to the set
$U$ in Step (\ref{GNR6}) of algorithm \algo{GeneralizedNextRecursive}
during the transformation of $t_{i-1}$ to $t_i$ for each such $i$.
Since the root has the smallest name ($M_i(r_i) = 1$), the component
$f_i$ of the \CGS tree $t_i$ is set to $1$ infinitely often, while
$e_i > 1$. Hence the set $F_0$ is visited infinitely often and $w \in
L(\dD)$.


If the root node becomes a leaf finitely often, there is an index $i_2
> i_1$ such that the root node is a non-leaf node in all $t_z$ for $z
> i_2$.  By Lemma (\ref{lemmaC}), we know that for all $z > i_2$,
every state in $\lambda_z(r_z)$ is also in $\lambda_z(v)$ for some
child $v$ of $r_z$.  Since $\psi(z) \in \lambda_z(r_z)$ for all $z \ge
0$, it follows that for all $z > i_2$, there is a child $v$ of $r_z$
such that $\psi(z) \in \lambda_z(v)$.  Now consider the transformation
of $t_z$ to $t_{z+1}$ for $z > i_2$, and let $v_z$ be the node in
$t_z$ such that $\psi(z) \in \lambda_z(v_z)$.  Step (\ref{Nxt1}) of
algorithm \algo{GeneralizeNext} initializes the state label of $v_z$
with $\delta(\lambda_z(v_z), \alpha_z)$, thereby placing $\psi(z+1)$
in the state label of $v_z$.  Subsequently, if $\psi(z+1)$ is moved
out of the state label of $v_z$, either Step (\ref{GNR3b}) or Step
(\ref{GNR4}) of algorithm \algo{GeneralizedNextRecursive} must be
responsible for this.  However, if $\psi(z+1)$ is removed from the
state label of $v_z$ in Step (\ref{GNR3b}), from the argument used in
the proof of Lemma (\ref{lemmaC}), we know that $\psi(z+1)$ must
eventually be removed from the state label of the parent of $v_z$ in
$t_{z+1}$, i.e. from the state label of $r_{z+1}$.  This is a
contradiction, since $\psi(z) \in \lambda_z(r_z)$ for all $z \ge 0$.
Therefore, if $\psi(z+1)$ is removed from the state label of $v_z$,
Step (\ref{GNR4}) of algorithm \algo{GeneralizedNextRecursive} must be
responsible for the removal.  From the pseudocode of
\algo{GeneralizedNextRecursive}, we now observe that if $v_{z+1}$ is
the new node containing $\psi(z+1)$ in its state label in $t_{z+1}$,
then either $M_{z+1}(v_{z+1}) < M_{z+1}(v_z) = M_z(v_z)$ or the
annotation of the edge from $r_{z+1}$ to $v_{z+1}$ in $t_{z+1}$ is
lesser than the annotation of the edge from $r_{z+1}$ to $v_z$ in
$t_{z+1}$.  Since both $r_z (= r_{z+1})$ and $v_z$ existed in $t_z$,
the annotation of the edge from $r_{z+1}$ to $v_z$ in $t_{z+1}$ must
be the same as the annotation of the edge from $r_z$ to $v_z$ in
$t_z$.  Therefore, if the child of the root that tracks $\psi$ changes
from $t_z$ to $t_{z+1}$, then either the name of the node reduces or
the annotation of the edge from the root to this node reduces during
the transformation from $t_z$ to $t_{z+1}$.  Since neither the name
nor the annotation can decrease infinitely, there must be an index
$i_3 > i_2$ such that for all $z > i_3$, the child of the root that
contains $\psi(z)$ in its state label has the same name and the same
annotation of the edge from the root to this child.  In other words,
if $v_z$ and $v_{z+1}$ are children of the root in $t_z$ and $t_{z+1}$
respectively such that $\psi(z) \in \lambda_z(v_Z)$ and $\psi(z+1) \in
\lambda_{z+1}(v_{z+1})$, then $M_z(v_z) = M_{z+1}(v_{z+1})$ and
$h_z(v_z) = h_{z+1}(v_{z+1})$.

If possible, let $v_z$ and $v_{z+1}$ be distinct nodes.  As seen
above, Step (\ref{GNR4}) of algorithm \algo{GeneralizedNextRecursive}
is responsible for moving $\psi(z+1)$ from the state label of $v_z$ to
that of $v_{z+1}$ during the transformation of $t_z$ to $t_{z+1}$.
From the pseudocode of algorithm \algo{GeneralizedNextRecursive}, we
note that either the annotation of the edge from the root to $v_{z+1}$
must be less than the annotation of the edge from the root to $v_z$,
or the name of $v_{z+1}$ must be less than the name of $v_z$ at the
time of execution of Step (\ref{GNR4}).  Since the name of $v_{z+1}$
can only reduce further during the name-compaction step and since the
annotation of the edge from the root to $v_{z+1}$ cannot change
subsequently in any step of algorithm \algo{GeneralizedNextRecursive}
or algorithm \algo{GeneralizedNext}, we cannot have both the names and
the annotations of the edges from the root identical for $v_z$ in
$t_z$ and for $v_{z+1}$ in $t_{z+1}$.  Since $z > i_3$, this gives us
a contradiction!  Therefore, $v_z$ is the same node as $v_{z+1}$ for
all $z > i_3$.  Since $M_z(v_z)$ also stays unchanged for all $z >
i_3$, no node with name $< M_z(v_z)$ is deleted during the
transformation of $t_z$ to $t_{z+1}$, for $z > i_3$.  This implies
that $e_z > M_z(v_z)$ for all $z > i_3$.

We now claim that $h_z(v_z) \neq \emptyset$ for all $z > i_3$.  To see
why this is so, suppose $h_z(v_z) = \emptyset$ for some $z > i_3$ and
let $j$ be the annotation of the edge from $r_z$ to $v_z$ in $t_z$.
Consider Step (\ref{GNR32}) of the recursive invokation of algorithm
\algo{GeneralizedNextRecursive} with the parent of $v_z$, i.e. $r_z$,
as the root of the \CGS subtree to be transformed.  Let $q_l$ be a
state in the state label of $v_z$ when Step (\ref{GNR32}) is executed.
If $l = j$, then Step (\ref{GNR3a}) creates a new sibling $v'$ of
$v_z$, sets the state label of $v'$ to $\{q_l\}$ and sets the
annotation of the edge from $r_z$ to $v'$ to an index $< l$.  Since no
further step removes the state label of the newly created leaf $v'$,
state $q_l$ gets removed from the state label of $v_z$ in Step
(\ref{GNR4}) of algorithm \algo{GeneralizedNextRecursive}.  If, on the
other hand, $l \neq j$, then since $h_z(v_z)$ is assumed to be
$\emptyset$, Step (\ref{GNR3b}) removes $q_l$ from the state label of
$v_z$.  Thus, in either case, no state eventually remains in the state
label of $v_z$ in $t_z$ if $h_z(v_z) = \emptyset$.  This implies that $v_z$
is deleted from $t_z$ in Step (\ref{GNR5}) -- a contradiction!
Therefore, we must have $h_z(v_z) \neq \emptyset$ for all $z > i_3$. 

We now consider the case where the node $v_z$ becomes a leaf
infinitely often in $\rho(i_3+1, \infty)$.  By using the same argument
as used above when the root becomes a leaf infinitely often, we find
that for every $z > i_3$ such that $v_z$ is a leaf in $t_z$, the node
$v_z$ is added to the set $U$ in Step (\ref{GNR6}) of algorithm
\algo{GeneralizedNextRecursive} during the transformation of $t_{z-1}$
to $t_{z}$.  Therefore, $f_z < M_z(v_z)$ for all $z > i_3$.  We have
also seen above that $e_z > M_z(v_z)$ for all $z > i_3$.  This implies
that a parity acceptance set $F_x$ with an even index $x$ is visited
infinitely often by the run $\rho$ of $\dD$.  Hence $w \in L(\dD)$.

If $v_z$ becomes a leaf only finitely often in $\rho(i_3+1, \infty)$,
we can repeat the same argument as used above and show that there is
an index $i_4 > i_3$ and a child $v'$ of $v_z$ such that (i) $v'$ is
present in $t_i$, (ii) $\psi(i) \in \lambda_i(v')$, (iii) $h_i(v') =
h_{i+1}(v') \neq \emptyset$, and (iv) $M_i(v') = M_{i+1}(v')$, for all
$i > i_4$.  Since all \CGS trees $t_i$ have height $\le n$ (as argued
in Section (\ref{sec:inf-set-size})), by continuing the above
argument, we find that there must exist an even index $x$ such that
$F_x$ is visited infinitely often by $\rho$.  In other words, $w \in
L(\dD)$.
\qed

\begin{theorem}
$L(D)=L(A)$
\end{theorem}


\noindent {\bf Proof:} Follows from Lemmas (\ref{lemma1}) and
(\ref{lemma2}). \qed


\section{Complexity}
\label{sec:inf-set-size}



\begin{theorem}\label{count-thm}
Given an automaton $\aA$ with $n$ states, the deterministic parity
automaton $\dD$ constructed above has at most $n^{O(n^2)}$ states
and $O(n^2)$ parity acceptance sets.
\end{theorem}

\noindent {\bf Proof:}
The computation for the number of states of the automaton $\dD$ is
similar to that done by Piterman for his NSW to DPW construction
\cite{piterman}.  Since every state of $\dD$ is a \CGS tree over
$\aA$, we will count the total number of \CGS trees over $\aA$ below,
assuming $n = |Q|$ and $m = n^2 + n + 1$.


The salient steps in counting the number of \CGS trees over $\aA$
are as follows.
\begin{itemize}
\item Since the state labels of leaves in a \CGS tree are pair-wise
  disjoint, and since every leaf has a nonempty state label, there can
  be at most $n$ leaves.

\item If we collapse the vertices at the head and tail of every
  $0$-annotated edge in a \CGS tree, we will get a tree with no
  $0$-annotated edges.  Since the hope set of the root is always $[n]$
  and since the hope set of a child in the collapsed tree misses
  exactly one index from the hope set of its parent, the height of the
  collapsed tree can be at most $n$.  This, along with the fact that
  there are at most $n$ leaves, implies that there are at most $n^2 +
  1$ nodes in the collapsed tree.

\item To count the nodes that were removed due to the collapsing
  operation described above, we note that each node in the original
  \CGS tree must have a path (possibly of zero length) to a leaf such
  that each edge along this path has a non-$0$ annotation.  Hence, if
  $u$ and $v$ are nodes such that the edge from the parent of $u$ to
  $u$ and that from the parent of $v$ to $v$ are both annotated with
  $0$, the path of non-$0$ annotated edges from $u$ to a leaf cannot
  overlap with the corresponding path from $v$ to a leaf.  Therefore,
  there can be atmost $n$ nodes in a \CGS tree such that the edges
  from the respective parents to these nodes are annotated with $0$.
  This implies that the total number of nodes in a \CGS tree can be
  atmost $m = n^2 + n + 1$.

\item By construction, the parent of a node always has a smaller name
  than the node. Thus the parenthood relation can be represented by a
  sequence of at most $m-1$ names where the $i^{th}$ name is a value
  in $\{1, \ldots i-1\}$.  For a tree with $k$ nodes, the there are at
  most $\le (k-2)!$ such sequences of length $k-1$. Considering all
  trees with number of nodes in $\{1, \ldots m\}$, there are at most
  $\Sigma_{k=1}^m (k-2)!$, i.e. $\leq (m-1)!$ such sequences.  Hence,
  there are at most as many named trees where children have larger
  names than their respective parents.

\item The state label of a node is given by the union of state labels
  of leaves in the sub-tree rooted at that node.  In addition, the
  labels of leaves are pairwise disjoint. Therefore, the state labels
  of all nodes in a tree can be obtained by associating each
  $\aA$-state with the leaf that contains it in its state label.
  Since leaves in a tree may not be named with the first few
  contiguous names, we sort the leaves by names and then use a mapping
  from $\aA$-states to positions of leaves in this name-sorted order.
  If an $\aA$-state doesn't appear in any leaf, we associate the
  position $0$ with it.  Thus, the number of state labelings of a
  named tree is at most the number of mappings $Q \rightarrow \{0, 1,
  \ldots n\}$, i.e.  $\le (n+1)^n$.


\item The (hope set) annotation of a node is represented using edge
  annotations as follows. Suppose the hope set of a node $v$ is $h(v)$
  and that of its child $v'$ is $h(v')$.  Then the edge from $v$ to
  $v'$ is annotated with $h(v) \setminus h(v')$, if $h(v') \subset
  h(v)$, and with $0$ if $h(v')=h(v)$.  By properties of \CGS trees,
  $h(v') \subseteq h(v)$ and $|h(v') \setminus h(v)| \leq 1$.
  Therefore, the edge annotation is a unique element in $[n] \cup
  \{0\}$.  Similarly, the hope set for every node is uniquely
  determined if the annotations of all edges are given.  Specifically,
  the hope set of a node is simply $[n]$ sans the annotations on edges
  along the path from the root to this node.  Therefore, it is
  sufficient to count the number of edge annotation functions to
  obtain the count of hope set annotations of nodes.  Each edge can be
  identified by the name of the node it points to.  The total number
  of edge annotation functions is then easily seen to be the number of
  functions $[m] \rightarrow [n] \cup\{0\}$.  This is bounded above by
  $(n+1)^m$.

\item For the acceptance condition, we need to know the value of $e$
  when $e \leq f$, and the value of $f$ when $f < e$. Thus we need to
  keep track of at most $2m$ values.
\end{itemize}

Combining the above counts, the total number of \CGS trees over $\aA$
is at most
$$(m-1)! \cdot (n+1)^{n+m} \cdot (2m) = n^{O(n^2)}$$.  
The number of parity acceptance sets is $2m = 2\cdot(n^2 + n + 1) =
O(n^2)$. \qed

\section{An improved upper bound for $\omega$-automata}
\label{sec:large-acc-sets}

The determinization construction proposed above gives a DPW starting
from a variety of different non-deterministic automata, all of which
have an acceptance condition based on infinity sets.  By Theorem
(\ref{count-thm}), the number of states of the DPW is at most
$n^{O(n^2)}$ or $2^{O(n^2\log n)}$, while the number of sets in the
parity acceptance condition is at most $O(n^2)$, where $n$ is the
number of states of the original automaton $\aA$.  This bound also
holds when the input automaton is a pairs automaton viz. a Streett or
a Rabin automaton. This is significant since the size of the output
DPW, both in terms of number of states and acceptance pairs, is
independent of the number of pairs of the input pairs automaton. This
is different from the case of Safra's determinization construction for
NSW\cite{safra06}\cite{2001automata}\cite{piterman}, where the output
DRW/DPW has at most $2^{O(nh\log(nh))}$ states and $O(nh)$ pairs,
where $n$ and $h$ are the count of states and pairs, respectively, of
the input NSW.

This naturally leads us to ask if $2^{O(n^2\log n)}$ is a better bound
than $2^{O(nh\log(nh))}$ for determinization of NSW/NRW. The answer to
this question is not immediately obvious and requires us to show that
there are indeed examples of NSW/NRW with $O(n)$ states and $h$ pairs
for which Safra's and Piterman's NSW determinization construction will
end up constructing automata with state count worse than $2^{O(n^2\log
  n)}$. In the following, we present a class of such automata. In the
case when $h \geq n^k$, where $k>1$, this immediately implies an
improved worst case complexity bound on NSW/NRW determinization.


\begin{theorem}
\label{familyimprovedbound}

There exists a family $\mathsf{A_S}$ of NSW where each NSW $\aA_S \in
\mathsf{A_S}$ has $3n+1$ states and $2^n+1$ accepting pairs for which the
Safra-Schwoon (Piterman) construction constructs a DRW (DPW) with
$2^{\Omega(n^3)}$ states, while our construction (algorithm
\algo{GeneralizedNext}) constructs a DRW/DPW with $2^{O(n^2\log n)}$ states.

\end{theorem}

The proof of Theorem (\ref{familyimprovedbound}) is given in Subsection
(\ref{familyconstruction}) by demonstrating the construction of an automaton
from the family $\mathsf{A_S}$.
 
To begin with, a strategy to generate more than $2^{O(n^2\log n)}$ states for
the DRW/DPW using Safra's/Piterman's construction is established. The input NSW
for such a strategy has $O(n)$ states and $h=2^n$ pairs.  One way to generate a
sufficiently large number of $(Q,H)$-trees (as used in Schwoon's exposition of
Safra's construction) is to obtain different permutations of the edge labels on
a path from a leaf to the root, and then repeat this for all paths in the tree.
We shall follow the construction of Schwoon\cite{2001automata} described in
algorithms \algo{SafraNext} and \algo{SafraNextRecursive} (see Subsection
(\ref{safransw})) for NSW determinization.

\begin{figure}[ht]
\begin{minipage}[b]{0.48\linewidth}
\begin{center}
\includegraphics[scale=0.53]{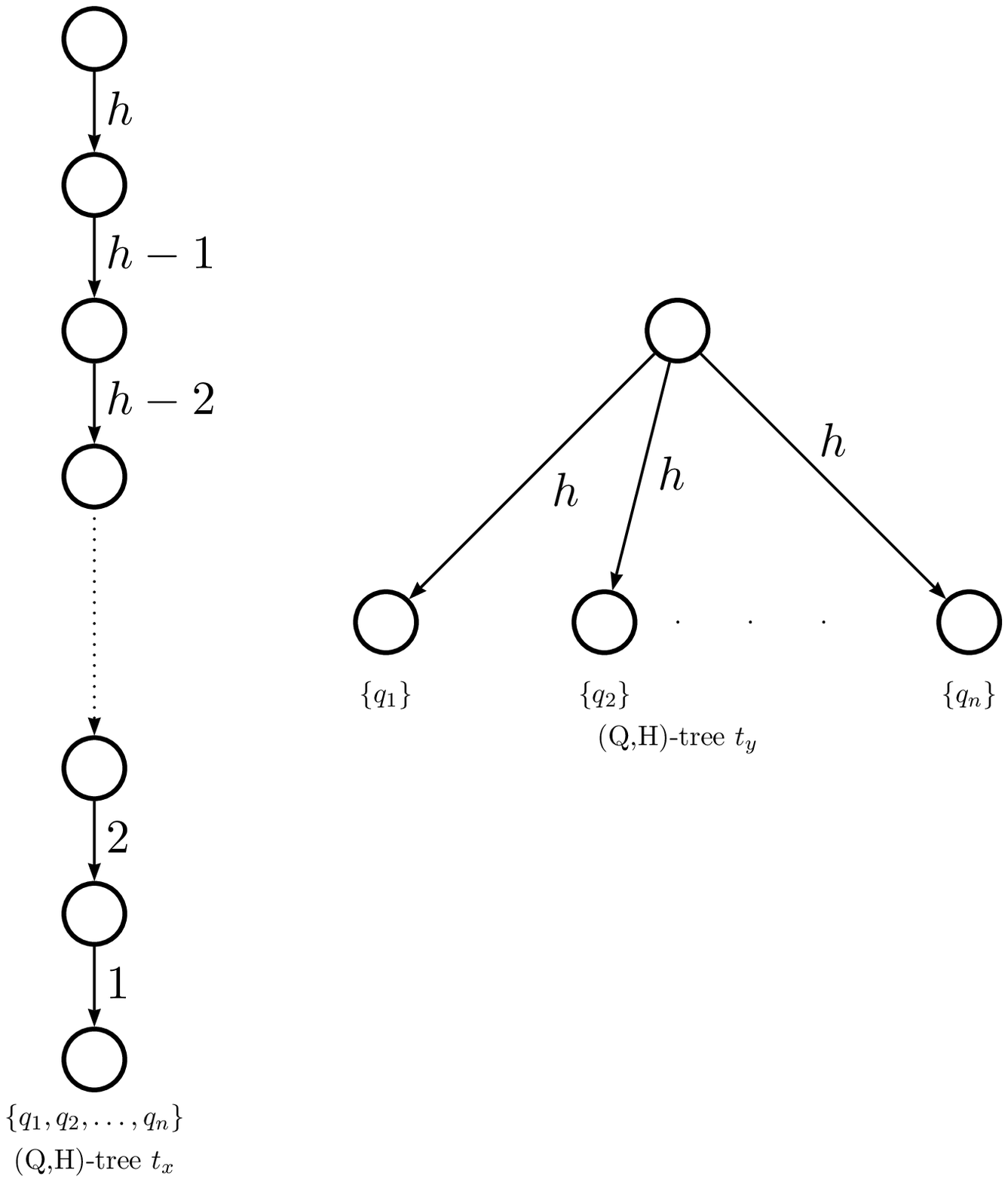}
\end{center}
\label{counterex1}
\end{minipage}
\hspace{5mm}
\begin{minipage}[b]{0.48\linewidth}
\begin{center}
\includegraphics[scale=0.53]{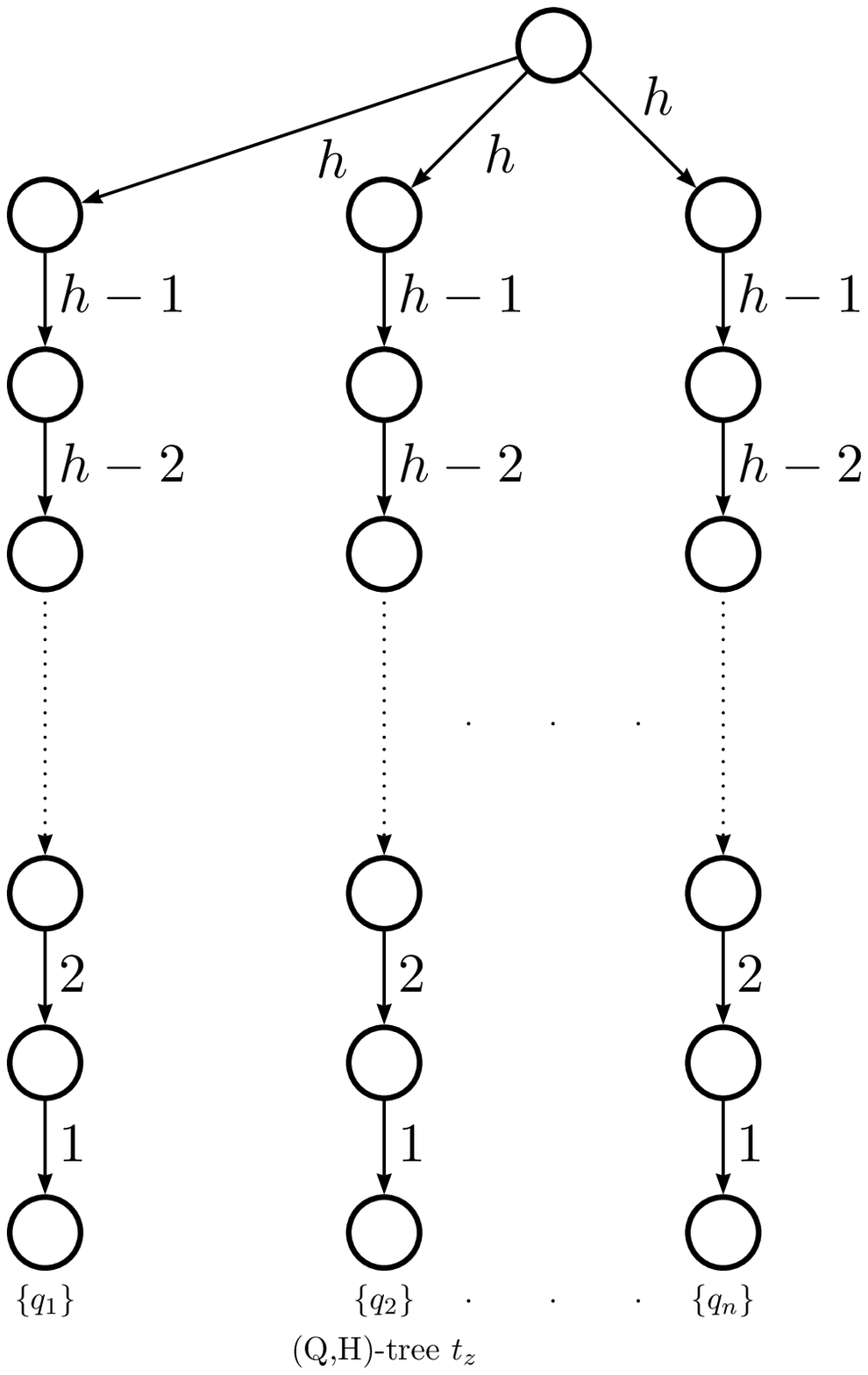}
\end{center}
\label{counterex2}
\end{minipage}
\caption{Steps in construction of counter-example}
\label{counterex12}
\end{figure}

Figure (\ref{counterex12}) shows three possible $(Q, H)$-trees, $t_x,t_y$ and
$t_z$ that can be generated using the Safra-Schwoon construction in algorithm
\algo{SafraNextRecursive} starting from the initial tree $t_0$, where $t_0$ is
the \CGS tree with a single (root) node $r_0$, with $\lambda(r_0)= Q_0$, $M(r_0)
= 1$, $h(r_0)=[n]$ and for $t_0$ we have $e = f = m+1$.

The first tree $t_x$ is not hard to generate, since Steps (\ref{safra1}) and
(\ref{safra2}) recursively extend a $(Q, H)$-tree at its leaves. If the Streett
state label of the leaf node in the first tree $t_x$ is $\{q_1, q_2, \ldots,
q_n\}$ and $q_i \in F_h$ for all $i \in [n]$, then in Step (\ref{safra3b1}) a
new node is created for each such $q_i \in F_h$ with the edge from the root node
to the newly created node annotated $h$, giving the second tree $t_y$.  An
application of Steps (\ref{safra1}) and (\ref{safra2}) will result in the
extension of the second tree $t_y$ at its leaves giving the third tree $t_z$.
For each Streett state $q_i$, $i \in [n]$ that appears in the label of a leaf
node in the third tree $t_z$, the path from the leaf to the node is disjoint
from every other path in the tree. Each such disjoint path has exactly the same
edge annotations.  Note that since the number of leaves in a $(Q, H)$-tree can
never be more than the total number of Streett states, we cannot expect to get
more than $n$ disjoint paths from a leaf to the root.  The challenge now is to
permute the edge annotations giving a large number of $(Q,H)$-trees.



\begin{figure}
\begin{center}
\includegraphics[scale=0.53]{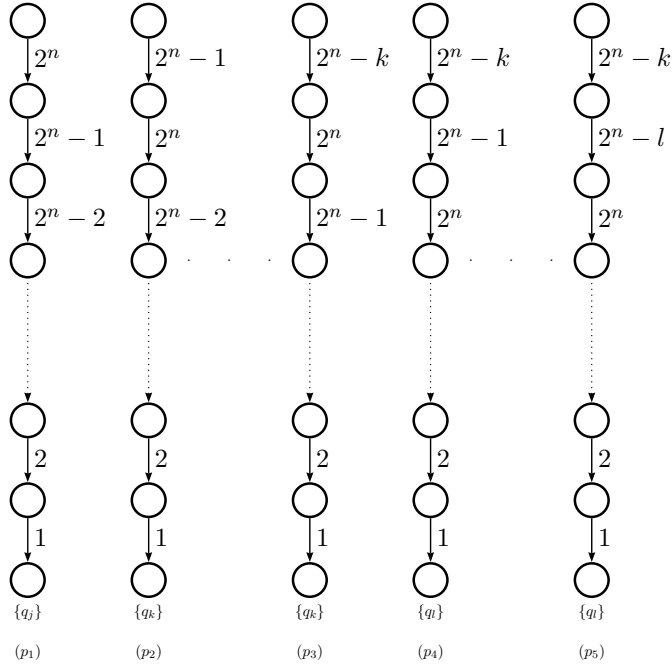}
\end{center}
\caption{Example transformation of $(Q, H)$-trees}
\label{counterex3}
\end{figure}

Since, the maximum length of a disjoint path in a $(Q,H)$-tree depends on the
number of pairs of the NSW, one would like to start with an NSW with as many
pairs as possible. Suppose, we start out with $h=2^n$ pairs in the NSW. A
permutation of $2^n$ edge annotations would give us $(2^n)!$ possible trees with
just one branch and ${((2^n)!)}^n$ trees with all $n$ disjoint branches. With
only $n$ states in the NSW and $2^n$ pairs, it is clear that one or more Streett
states will be replicated across pairs.  This replication of Streett states is a
potential problem as the example in Figure (\ref{counterex3}) shows.  

Figure (\ref{counterex3}) shows different edge annotations for a path of length
$h$ (with no $0$ edges) in a $(Q,H)$-tree. The different edge annotations are
obtained as the Streett state label at the leaf changes. We assume that $h=2^n$.
It is not hard to obtain the edge annotations along $(p_2)$ from the edge
annotations along $(p_1)$. In this transformation only the edge annotation of
the first edge in $(p_1)$ changes from $2^n$ to $2^n-1$. 
This is possible if there is a state $q_k$ in the leaf label that is also in the
pair $E_{2^n}$ of the pair $(E_{2^n},F_{2^n})$. This causes the entire path to
be replaced by a pair of nodes - the root node with exactly one child.  The edge
between the root and its child node is annotated $2^n-1$. This path is again
extended by Steps (\ref{safra1}) and (\ref{safra2}) of algorithm
\algo{SafraNextRecursive}. We see that repeated application of this change
allows us to change the edge annotations of $(p_2)$ to those shown along
$(p_3)$, where the edge annotation on the edge from the root to the first child
node is $2^n-k$. Note that this requires that the NSW has a path from $q_k$ back
to itself on some letter or word segment. Once the first edge annotation is
fixed we can apply a similar set of transformations using some other state $q_l$
to fix the second edge annotation to $2^n-l$.  But, this immediately implies
that the state $q_l$ cannot be in $E_{2^n-k}$ or $F_{2^n-k}$ since that would
either change the annotation of the first edge to $2^n-{k+1}$ or reset the path
back to the third path $(p_3)$ shown in the figure. Hence, every time we fix the edge
annotation for an edge it constrains the possible pairs that a Streett state can
belong to. With only $n$ states and $2^n$ pairs, we are soon forced to repeat
Streett states across pairs in our example NSW.  This in turn forces already
fixed edge annotations to change, defeating our purpose.  Thus, generating
arbitrary permutations of $2^n$ pair indices along paths in a $(Q,H)$-tree is
extremely hard with an NSW with just $n$ states. 

We then ask if $2^n$ is too many pairs and try to see if $n$ or $n^2$ or some
number of pairs polynomial in $n$ allows us to achieve our objective of
obtaining arbitrary permutations of edge annotations.  But, with $n^k$ pairs in
the NSW, for some constant $k$, even if obtaining arbitrary permutations of edge
annotations is possible, we can obtain at most $(n^k)!$ permutations along a
path and hence ${((n^k)!)}^n$ $(Q, H)$-trees using all the paths. But,
${((n^k)!)}^n$ is $2^{O(n^2\log n)}$, which matches the bound given by our
construction and does not serve our purpose. 

We now show a solution to the above dilemma. We start out with $h=2^n$ pairs in
the NSW, but we partition the $2^n$ pairs into $\lfloor \frac{2^n}{n} \rfloor$
\emph{blocks} of $n$ pairs each.  Hence $B_1=\langle (L_{2^n},U_{2^n}),
(L_{2^n-1},U_{2^n-1}), \ldots, (L_{2^n-(n-1)},U_{2^n-(n-1)}) \rangle$ is the
first block, $B_2=\langle (L_{2^n-(n)},$ $U_{2^n-(n)}),
(L_{2^n-(n+1)},U_{2^n-(n+1)}), \ldots, (L_{2^n-(2n-1)},U_{2^n-(2n-1)}) \rangle$
is the second block and so on. If $\lfloor \frac{2^n}{n} \rfloor=k$, then the
last or $k^{th}$ block is $B_k=(L_{2^n-((k-1)n)},U_{2^n-((k-1)n)}),$
$(L_{2^n-((k-1)n+1)},$ $U_{2^n-((k-1)n+1)}), \ldots,
(L_{2^n-(kn-1)},U_{2^n-(kn-1)})$.  Instead of trying to generate arbitrary
permutations of $2^n$ pair indices we try to generate permutations of only $n$
pair indices, but with the following properties for a permutation $\langle j_1,
j_2, \ldots, j_n \rangle$, where $j_i \in [h]$ for all $i \in \{1,2, \ldots,
n\}$.  

\begin{itemize}

\item We pick $k=\lfloor\frac{2^n}{n}\rfloor$ blocks starting with the last
block $B_k$ and picking successively lower numbered blocks $B_{k-1}, B_{k-2},
\ldots$.

\item From each block we pick exactly one pair index. For example if we pick the
$i^{th}$ pair in block $B_j$ then pair is
$(L_{2^(j-1)n+(i-1)},U_{2^(j-1)n+(i-1)})$. We call this pair index
${\mathsf{idx}}_j^i$. 

\item If pair index ${\mathsf{idx}}_j^i$ is already picked from block $j$, then
we do not pick ${\mathsf{idx}}_l^i$ for $l \neq j$, for every pair of blocks
$B_j$ and $B_l$ that are picked. 

\end{itemize}

This system of picking elements of the permutation not only allows us to permute
only $n$ elements along every path from a leaf to the root, but also allows us
to choose from $2^n$ Streett pairs and at the same time have only $O(n)$ states
for the example NSW. We shall see later that this method ends up generating more
than $2^{O(n^2\log n)}$ $(Q, H)$-trees. We shall call a permutation that
satisfies the conditions described above as a \emph{block permutation} of size
$n$. An example of a NSW with $O(n)$ states and $2^{n}$ pairs for which the
corresponding DPW constructed using the Safra/Piterman construction has more
than $2^{O(n^2\log n)}$ states is given below. 
 
\subsection{An example showing improved worst case bounds}
\label{familyconstruction}

Consider the the NSW ${\aA}_s=(\Sigma, Q^s,q_0^s,\delta^s,\mathcal{T})$ defined
as follows. The NSW $\aA_S$ is an automaton in the family $\mathsf{A_S}$
described in Theorem (\ref{familyimprovedbound}).

\begin{itemize}

\item $Q^s$ is the state set containing $3n+1$ states $\{q_0\} \cup
\{q_{0,\bot}, q_{1,\bot}, \ldots, q_{n-1,\bot}\} \cup$ \{ $q_{0,s}, q_{1,s},
\ldots,$ $q_{n-1,s}$ \} $\cup \{q_{0,\top}, q_{1,\top}, \ldots, q_{n-1,\top}\}$.
States of the form $q_{i,\bot}, q_{i,s}, q_{i,\top}$ are called $\bot$-states,
$s$-states and $\top$-states respectively.

\item $q_0$ is the initial state.

\item $\Sigma$ is the alphabet $\{a_0\} \cup \{a_{x,s} \mid x \in \{0,1,2,
\ldots,n-1\}\} \cup \{a_0, \ldots, a_{n-1}\} \cup \{a_{\bot}\}$.

\item The transitions for the automaton are defined as follows

\begin{enumerate}

\item \label{ex1} $\delta^s(q_0,a_0)=\{q_{(0,\top)}, q_{(1,\top)}, \ldots,
q_{(n-1,\top)}\}$

\item \label{ex2} $\delta^s(q_{(i,\top)}, a_\bot)=q_{(i,\top)}$ for all $i \in \{0,1,\ldots,n-1\}$

\item\label{ex3} $\delta^s(q_{(i,\top)},a_{i})=q_{(i,s)}$ for all $i \in \{0,1,\ldots,n-1\}$


\item \label{ex5}$\delta^s(q_{(i,s)},a_{(j,s)})=q_{(j,s)}$ for all $i,j \in \{0,1,\ldots,n-1\}$

\item \label{ex6}$\delta^s(q_{(i,\top)},a_{(j,s)})=q_{(i,\top)}$ for all $i,j \in \{0,1,\ldots,n-1\}$

\item \label{ex7}$\delta^s(q_{(i,s)},a_{\bot})=q_{(i,\bot)}$ for all $i \in \{0,1,\ldots,n-1\}$

\item \label{ex8}$\delta^s(q_{(i,\bot)},a_{(j,s)})=q_{(i,\bot)}$ for all $i,j \in \{0,1,\ldots,n-1\}$

\item \label{ex9}$\delta^s(q_{(i,\bot)},a_{\bot})=q_{(i,\bot)}$ for all $i \in \{0,1,\ldots,n-1\}$

\item \label{ex10}$\delta^s(q_{(i,\bot)},a_{j})=q_{(i,\bot)}$ for all $i,j \in \{0,1,\ldots,n-1\}$

\end{enumerate}

\item There are $2^n+1$ Streett pairs $\mathcal{T}=\{(E_{-1},F_{-1}), (E_1,F_1),
(E_2,F_2), \ldots, (E_{2^n},F_{2^n})\}$, where $F_i, E_i \subseteq Q^s$, for all
$i \in \{-1,0,1, \ldots, 2^n\}$ satisfying the following constraints

\begin{enumerate}

\item $\{q_{(0,\top)}, \ldots, q_{(n-1,\top)}\} \subseteq F_{2^n}$ and 
$q_{(i,\top)} \notin E_{2^n}$ for all $i \in \{0,1, \ldots, n-1\}$.

\item $q_{(i,\bot)} \notin E_{j}$ for all $i \in \{-1,0,1, \ldots, n-1\}$ and for
all $j \in \{1,2, \ldots, 2^n\}$.
 
\item $q_{(i,\bot)} \notin F_{j}$ for all $i \in \{0,1, \ldots, n-1\}$ and for
all $j \in \{1,2, \ldots, 2^n\}$.

\item $\{q_{(0,\bot)}, \ldots, q_{(n-1,\bot)}\} \subseteq F_{-1}$ 

\item $\{q_{(i,s)}\}=E_{2^n-rn-i}$ for all $r \in \{0,1,\ldots,k-1\}$ and 
$q_{(i,s)} \notin F_{2^n-rn-j}$ for all $j \in \{0,1,\ldots,n-1\}$ and $j \neq
i$.

\item $\{q_{(i,s)}\}\notin F_{(2^n-rn-t)}$ for all $r \in \{0,1, \ldots,k-1\}$
and for all $t \in \{0,1,\ldots,n-1\}$.

\end{enumerate}

\end{itemize}

\begin{figure}
\begin{center}
\includegraphics[scale=0.56]{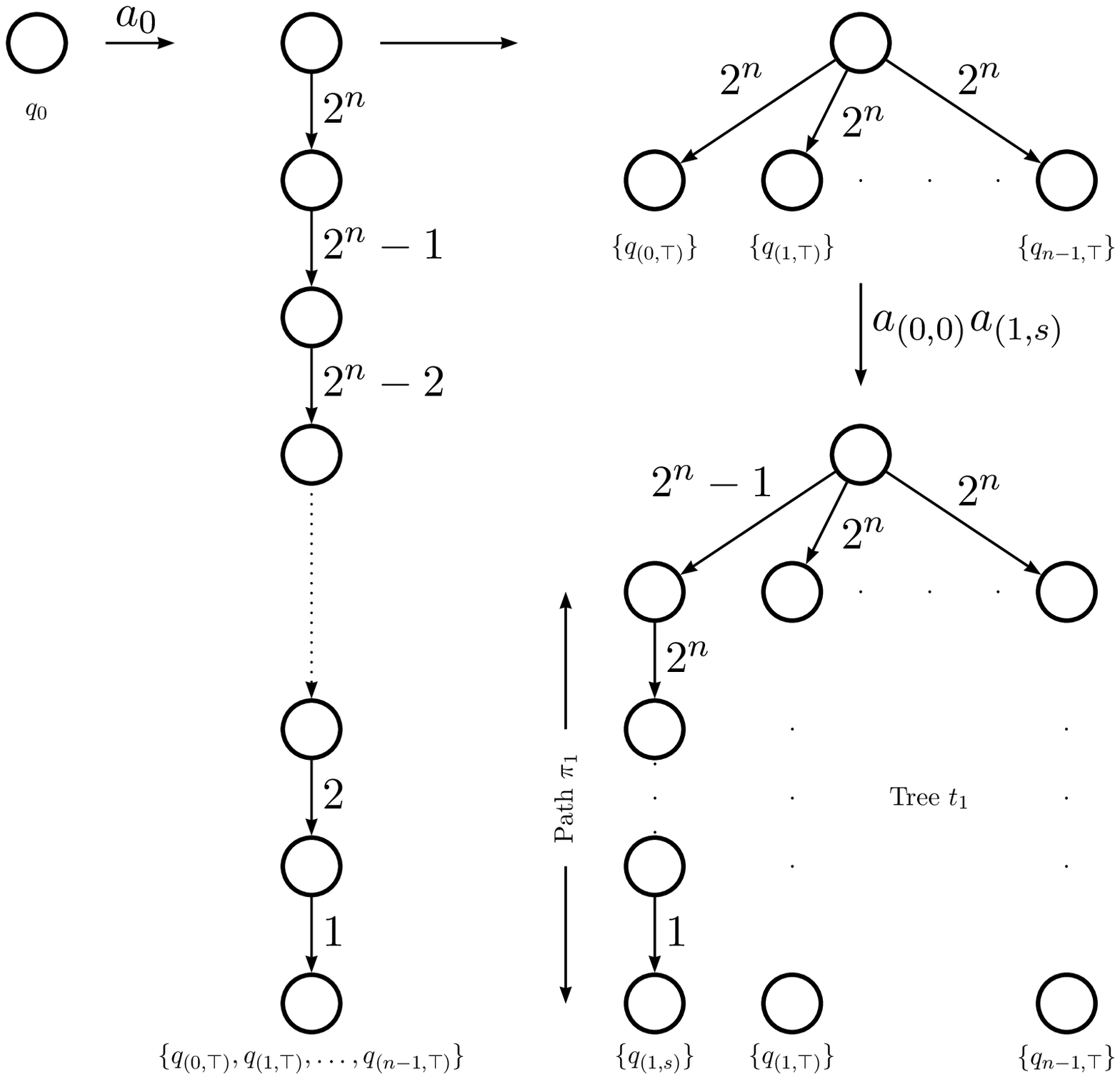}
\end{center}
\caption{Example transformation of $(Q, H)$-trees}
\label{counterex4}
\end{figure}



\begin{figure}[ht]
\begin{minipage}[b]{0.48\linewidth}
\begin{center}
\includegraphics[scale=0.53]{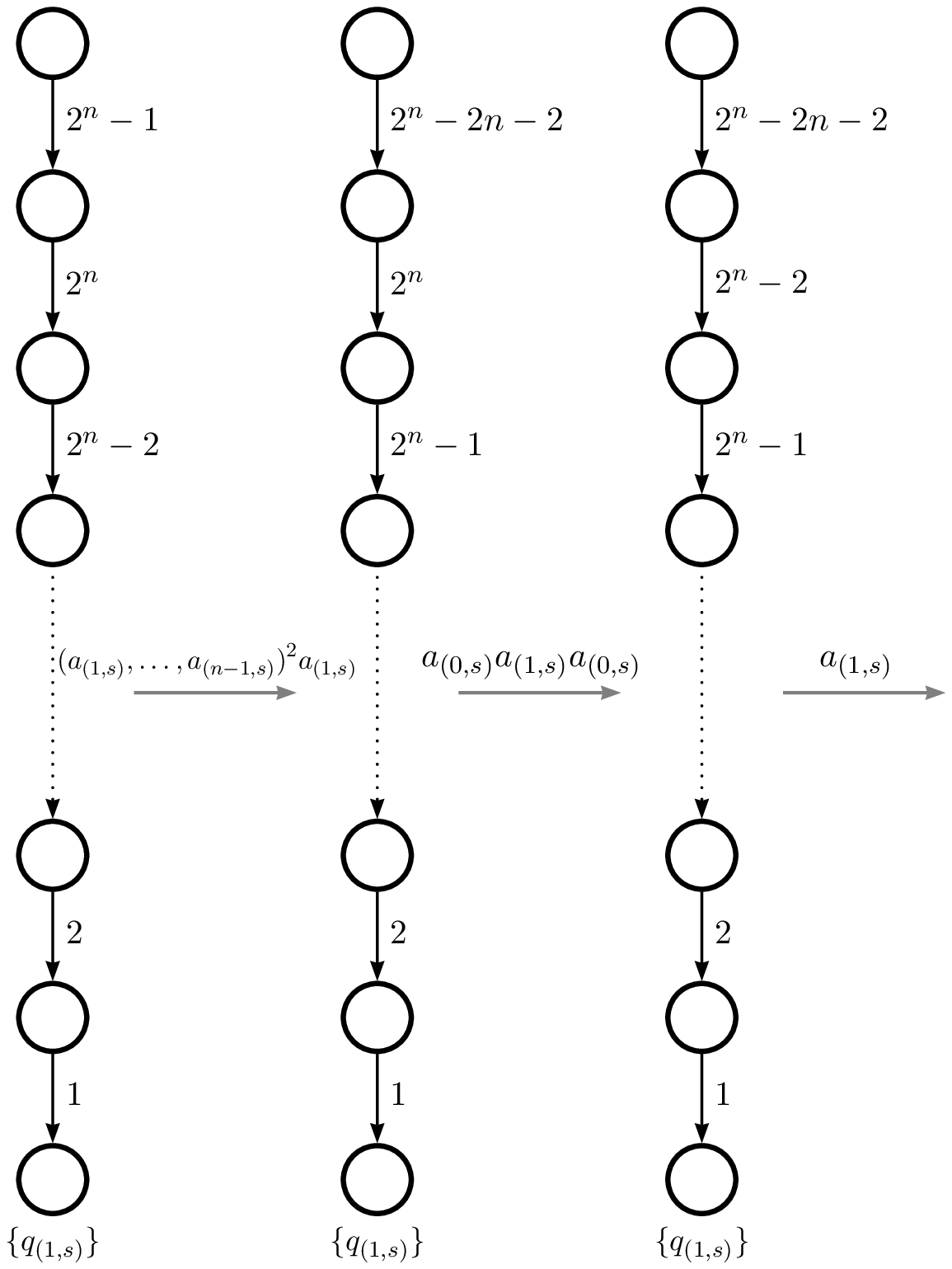}
\end{center}
\label{counterex5}
\end{minipage}
\hspace{5mm}
\begin{minipage}[b]{0.48\linewidth}
\begin{center}
\includegraphics[scale=0.53]{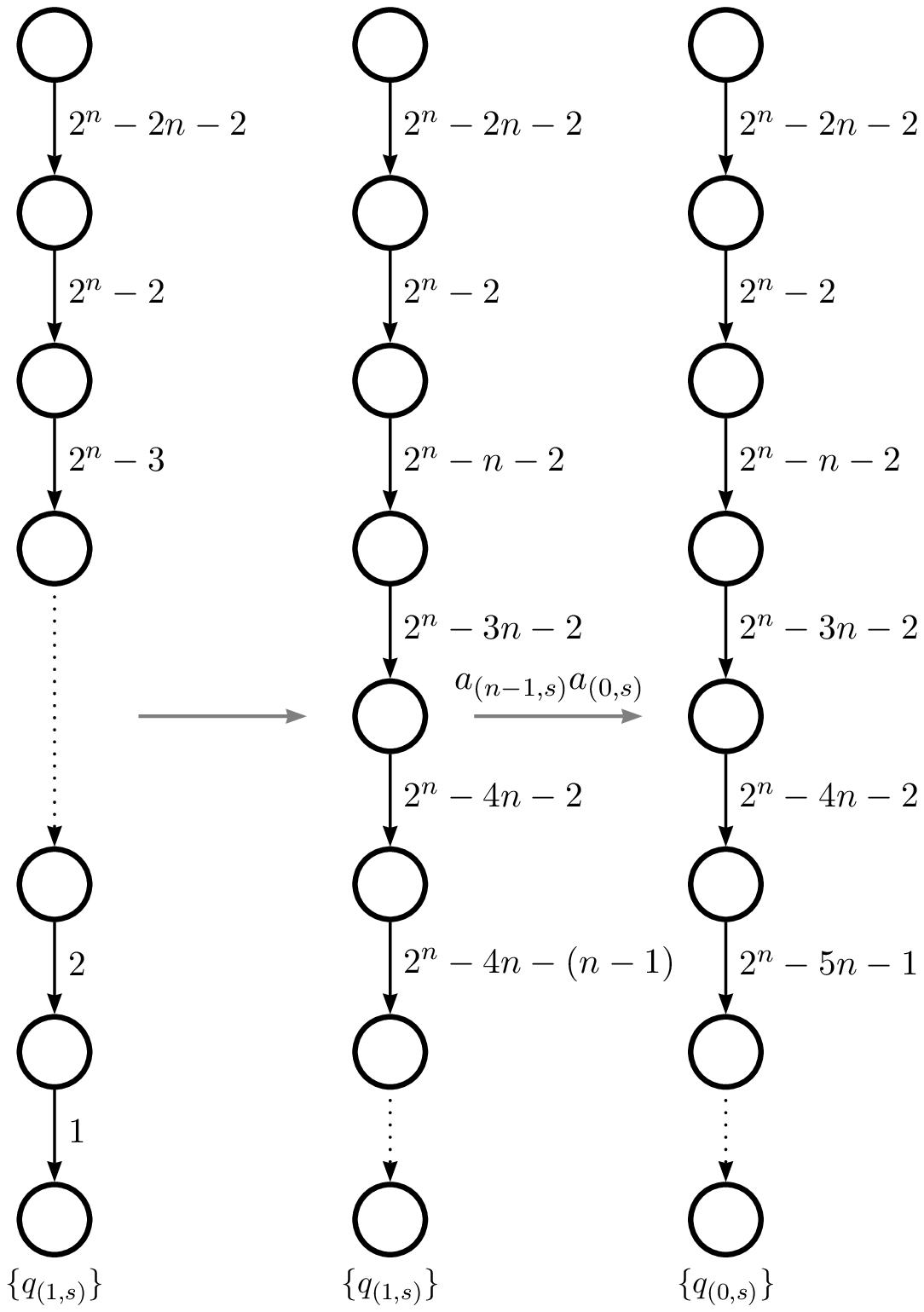}
\end{center}
\label{counterex6}
\end{minipage}
\caption{Example transformation of path $\pi_1$}
\label{counterex56}
\end{figure}

As discussed earlier our goal is to permute $n$ pair indices chosen carefully
from different blocks. For example let $B_1=\langle 2^n-2n-1, 2^n-5n-1, 2^n,
2^n-3n-4, \ldots\rangle$ be a block permutation of size $n$. Our goal is to
start with an arbitrary assignment of edge annotations along a path in a $(Q,
H)$-tree and obtain the permutation $B_1$ along that path. We do not insist that
the elements of $B_1$ appear along successive edges along the path, but we
insist that they appear along the path in the same order as they appear in
$B_1$.

\begin{figure}[ht]
\begin{minipage}[b]{0.48\linewidth}
\begin{center}
\includegraphics[scale=0.53]{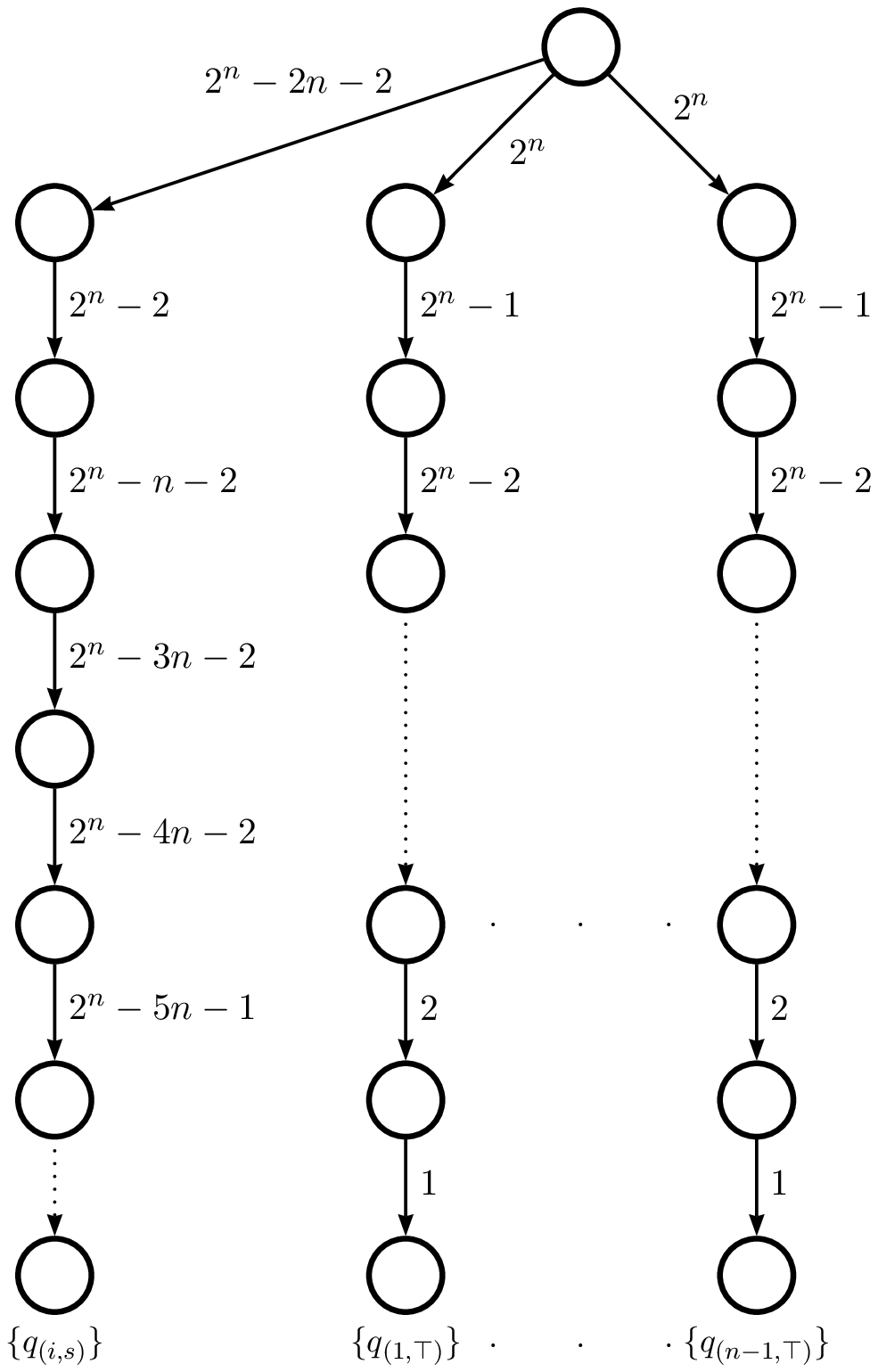}
\end{center}
\label{counterex7}
\end{minipage}
\hspace{5mm}
\begin{minipage}[b]{0.48\linewidth}
\begin{center}
\includegraphics[scale=0.53]{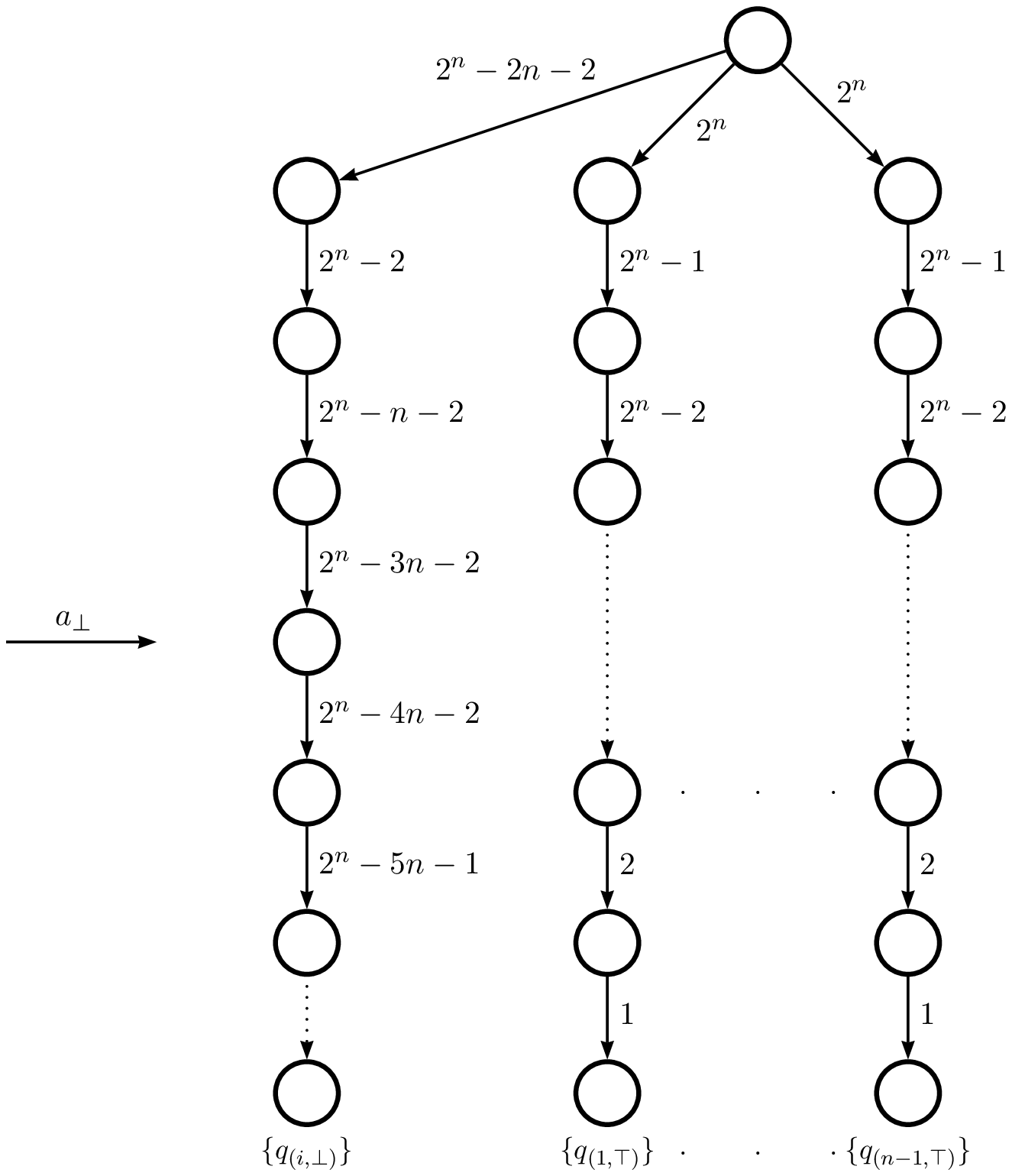}
\end{center}
\label{counterex8}
\end{minipage}
\caption{Example transformation of path $\pi_1$}
\label{counterex78}
\end{figure}

Figures (\ref{counterex4}), (\ref{counterex56}) and (\ref{counterex78})
demonstrate the main steps in the process of generating the required
permutations of pair indices for the example automaton. In Figure
(\ref{counterex4}), starting from the initial $(Q, H)$-tree consisting of just
the root node, we obtain the tree extended at the root and with Streett state
label $\{ q_{(0,\top)}, q_{(1,\top)}, \ldots, q_{(n-1,\top)}\}$ using the
transition from $q_0$ on letter $a_0$ and Steps (\ref{safra1}) and
(\ref{safra2}) of the Safra-Schwoon construction. This single path changes to
the branched tree in which the root has $n$ children with the edge to each child
annotated $2^n$ and the $i^{th}$ child has Streett state label $q_{(i-1,\top)}$.
Using a sequence of transitions on the letters $a_{(0,0)}$ and $a_{(1,s)}$ we
obtain the final tree that has $n$ leaves and $n$ disjoint paths, one from each
leaf to the root node. 

Note that the letter $a_{(0,0)}$ causes only state $q_{(0,\top)}$ to change to
the next state $q_{(0,s)}$, while Streett state labels for all other leaves
remain unchanged. This results in the edge annotation between the root and the
leftmost child to change to $2^n-1$. On reading the letter $a_{(1,s)}$, state
$q_{(0,s)}$ changes to $q_{(1,s)}$ giving us the tree $t_1$ in the figure. Note
that $t_1$ is only an intermediate tree and will evolve through different steps
of the Safra-Schwoon algorithm.  We observe that by changing the Streett label
of just one path at a time we can systematically generate permutations of edge
annotations one path at a time.  This will be our general strategy henceforth
and we shall see how a path $\pi_1$ in tree $t_1$ evolves with succeeding steps. 

The $\top$-states can be thought of as the \emph{source} states of every
path transformation. We change a $\top$-state to an $s$-state only along the
path whose edge annotations we need to modify.  

Figure (\ref{counterex56}) shows the transformations of path $\pi_1$ in order to
obtain the block permutation $B_1$ in order along the edges in $\pi_1$. It is
straightforward to obtain the first element $2^n-2n-2$ along the first edge.
All it requires is successive applications of letter $a_{(1,s)}$ to
$a_{(n-1,s)}$ follows by $a_{(1,s)}$.  We now try and change the other edge
annotations keeping the first edge annotation fixed. On reading the letter
$a_{(0,s)},a_{(1,s)}$ we change the second edge annotation to $2^{n}-2$. Here,
we need to be careful, since an application of $a_{(2,s)}$ at this point will
change $2^{n}-2$ to $2^{n}-3$ but it will also change $2^n-2n-2$ to $2^n-2n-3$,
because of the way the Streett pairs are organised. Hence, we defer the
application of $a_{(2,s)}$ and instead apply letter $a_{(0,s)}$ again, which
changes $2^n$ to $2^n-1$. Now an application of $a_{(1,s)}$ will change $2^n-1$
to $2^n-3$, since $2^n-2$ already appears on the edge above. Using this general
strategy of deferring the application of a letter if it changes an edge
annotation that is already on an edge above and part of $B_1$, we can obtain the
required block permutation $B_1$ along path $\pi_1$. Note that it is possible
that all elements $2^n-rn-1$, for all $r \in \{1,2, \ldots, k\}$, where $k$ is
the number of blocks may appear between the first element $2^n-2n-2$ and the
second element $2^n-5n-1$ of $B_1$ in order. 

Once all elements of $B_1$ appears along $\pi_1$, we ``seal'' path $\pi_1$, by
applying the letter $a_\bot$, which affects only $q_{(i,s)}$ at the leaf of
$\pi_1$ and does not affect the $\top$-states on the other paths. After this the
state $(q_{(i,\bot)})$ and hence the edge annotations for $\pi_1$ do not ever
change. We now apply $a_{(0,1)}$ to change $q_{(1,\top)}$ to $q_{(0,s)}$ at the
leaf of the second path. We then use our usual strategy discussed above to
obtain another block permutation along that path. Continuing this way we can
obtain arbitrary block permutations of size $n$ along every path in $(Q, H)$
trees. 

Although, we consider only special types of $(Q, H)$-trees, where the paths of
the trees are disjoint from one another, we shall show that this is sufficient
to generate enough trees to go beyond the $2^{O(n^2\log n)}$ upper bound given
by our construction. 

There are $k=\lfloor \frac{2^n}{n} \rfloor$ blocks of Streett pairs, with $n$
elements in each block. Note that if $2^n \text{ mod } n \neq 0$ i.e $n$ is not
a power of $2$, then some pairs may not appear in any block, but this does not
affect our construction. Also, the pair $(E_{-1},F_{-1})$ is not considered at
all and serves only as a placeholder for the $\bot$-states.  Consider a block
permutation $B=\langle 2^n-a_0n, 2^n-a_1n-1, 2^n-a_2n-2, \ldots,
2^n-a_{n-1}n-n-1\rangle$, where $a_1, \ldots, a_{n-1} \in \{1, \ldots, k\}$.
Each element $2^n-a_in-j$, for all $i,j \in \{0, \ldots, n-1\}$ can be chosen
from one of $k$ blocks. There are $n!$ ways of ordering the blocks themselves.
Hence there are $n! \times k^n$ ways of choosing a block permutation in each
branch. Since, we consider $(Q, H)$-trees that always have $n$ disjoint
branches/paths there are ${(n! \times k^n)}^n$ ways of choosing block
permutations in all branches. But, ${(n! \times k^n)}^n={(n!)}^n \times
k^{n^2}$. Since $k=\frac{2^n}{n}$ and Stirling's approximation gives us
$n!=\Omega({(\frac{n}{e})}^n)$, this is equal to
$\Omega(\frac{n^{n^2}}{e^{n^2}}) \times (\frac{2^{n^3}}{n^{n^2}})$ or
$\Omega(\frac{2^{n^3}}{e^{n^2}})$, which is $2^{\Omega(n^3)}$. Hence, the
Safra-Schwoon construction generates $2^{\Omega(n^3)}$ $(Q, H)$-trees, which are
states of the DRW, while our construction gives a bound of $2^{O(n^2\log n)}$ on
the number of states of the constructed DPW/DRW. Since, the bounds for the
Safra-Schwoon construction are obtained by counting (Q,H)-trees without names,
the same bounds work when constructing a DPW from an NSW using compact
(Q,H)-trees as described by Piterman\cite{piterman}.

Hence, its has been effectively demonstrated that our construction for
determinization of $\omega$-automata using generalized witness sets, results in
an improved worst case complexity bound for NSW determinization when the number
of pairs of the NSW is $h=2^n$. Since, our construction constructs deterministic
parity automata and complementing parity automata is trivial, the same arguments
can be used to show an improved upper bound for NRW determinization. 

In the following we show another interesting consequence of our construction. We
show a new lower bound on the number of states of any $\omega$-automaton
accepting a given $\omega$-regular language. Interestingly, this lower bound on
the number of states is a function of the Rabin index of the $\omega$-regular
language.

\section{A new lower bound for $\omega$-automata}


demonstrate a new lower bound on the number of states of any
$\omega$-automaton that uses an acceptance condition based on infinity
sets to accept a given $\omega$-regular language $L$. Interestingly,
this lower bound is a function of the Rabin index of the
$\omega$-regular language. The Rabin index of an $\omega$-regular
language is defined as follows.

\begin{definition}[Rabin Index]
\label{ri}
Let $\mathcal{L}(k)$ be the set of all $\omega$-regular languages that are
accepted by DRW with $k$ or less number of pairs. For any $\omega$-regular
language $L$ the smallest $k$ such that $L \in \mathcal{L}(k)$ is called the
Rabin index of $L$.  

\end{definition}

Wagner \cite{wagner} and Kaminski\cite{kaminski} showed that the Rabin
index is a property of an $\omega$-regular language and not of the
deterministic pairs automaton accepting the given language.  They also
provided a characterization of the Rabin index in terms of structural
properties of deterministic automata accepting a given
$\omega$-regular language.  We provide below a lower bound on the
number of states of any $\omega$-automaton that uses an acceptance
condition based on infinity sets and accepts an $\omega$-regular language
with a given Rabin index.

\begin{theorem}
Given an $\omega$-regular language $L$ with Rabin index $k$, any
$\omega$-automaton (deterministic or non-deterministic) that uses an
acceptance condition based on infinity sets and accepts $L$ must have
at least $\sqrt{k} - 1$ states.
\end{theorem}
\begin{proof}
\noindent {\bf Proof :} Let $\aA$ be an $\omega$-automaton with $n$
states that uses an acceptance condition based on infinity sets and
accepts $L$.  Using the construction of Section (\ref{sec:inf-set}),
we can obtain an equivalent DPW with at most $n^{O(n^2)}$ states and
$2\cdot(n^2 + n +1)$ parity acceptance sets.  This DPW can be
interpreted as an equivalent DRW with the same number of states and at
most $n^2 + n + 1$ Rabin acceptance pairs.  By definition of Rabin
index we must have $n^2 + n + 1 \geq k$.  It follows that $n \geq
\sqrt{k} - 1$. \qed
\end{proof}

\section{Conclusion}

In this paper, we presented a new construction for determinization
of $\omega$-automata whose acceptance condition is based on the notion
of infinity sets. We extended the Safra/Piterman construction for NSW
determinization using the concept of generalized witness sets to
construct an equivalent DPW.  We demonstrated, by way of an example,
that there are families of NSW with $O(n)$ states and $2^n$ pairs for
which our construction gives a DPW with better worst case complexity
bounds than the Safra/Piterman construction.  Effectively, we have
improved the worst case complexity for NSW/NRW determinization.  Also,
there is no known direct determinization procedure for NMW; every
known procedure uses an indirect method by first translating the NMW
to either an NSW or an NBW and then using determinization on it. Our
method provides a direct determinization construction for NMW. As an
easy corollary of our construction, we demonstrate a new lower bound
on the number of states of an $\omega$-automaton accepting a given
$\omega$-regular language, as a function of the Rabin index of the
language.

\bibliographystyle{plain}
\bibliography{tcs-infset}


\end{document}

%% file: macros.tex
\usepackage{epsfig}
\usepackage[T1]{fontenc}
\usepackage{amsmath}
\usepackage{amssymb}
\usepackage{graphicx}
\usepackage{wrapfig}
\usepackage{ifthen}
\usepackage[vlined,ruled]{algorithm2e}
\usepackage{algorithmic}
\usepackage{latexsym}
\usepackage{appendix}
\usepackage{epsfig}
\usepackage{scalefnt}

\newtheorem{theorem}{Theorem}

\newtheorem{definition}[theorem]{Definition}

\newtheorem{lemma}[theorem]{Lemma}
\newtheorem{proposition}[theorem]{Proposition}

\newtheorem{proof}[theorem]{Proof}

\newcommand{\buchi}{B\"{u}chi}

\newcommand{\infi}{\ensuremath{\mathit{inf}}}

\newcommand{\true}{\ensuremath{\mathsf{True}}}
\newcommand{\aA}{\ensuremath{\mathcal{A}}}
 \newcommand{\dD}{\ensuremath{\mathcal{D}}}

 \newcommand{\aS}{\ensuremath{\mathcal{A}_S}}
 \newcommand{\ar}{\ensuremath{\mathcal{A}_R}}

 \newcommand{\CGS}{\ensuremath{\textsf{CGS }}}

\newcommand{\killit}[1]{}

\newcommand{\algo}[1]{\ensuremath{\textsf{{#1}}}}

\newcommand{\algoblock}[4]{{{
				     {
                                     \vspace{5mm}
                                     \hrule 
				     \vspace{2mm} 
                                     \noindent \textsf{\bfseries Algorithm : {#1}}
				     \vspace{2mm} 
                                     \hrule 
				     \vspace{2mm}
                                     \hrule 
				     \vspace{2mm}
                                     \noindent \textsf{ {\bfseries Input:} {#2}}\\
                                     \noindent \textsf{ {\bfseries Output:} {#3}}
				     \vspace{2mm}
				     \hrule
				     \vspace{2mm}
                                     {#4} 
				     \vspace{2mm}
				     \hrule
				     \vspace{2mm}
                                     \noindent \textsf{\bfseries End Algorithm : {#1}}}
				     \vspace{2mm}
                                     \hrule 
				     \vspace{2mm}
                          }}
                         }